\pdfoutput=1
\documentclass[
    aps,
    pra,
    reprint,
    superscriptaddress,
    footinbib,
    longbibliography
]{revtex4-2}

\usepackage{hyperref}

\makeatletter

\newif\ifequalcontrib@sw
\newcommand{\equalcontrib}{\equalcontrib@swtrue}

\def\doauthor#1#2#3{%
  \equalcontrib@swfalse
  \ignorespaces#1\unskip\@listcomma
  \begingroup
    #3%
    \ifequalcontrib@sw
      \frontmatter@footnote{These authors contributed equally to this work.}%
    \fi
  \@if@empty{#2}{%
    \endgroup{}{}%
  }{%
    \endgroup{\comma@space}{}\frontmatter@footnote{#2}%
  }%
  \space\@listand
}

\makeatother
\usepackage{graphicx}
\usepackage{subfigure}
\usepackage{amsmath}
\usepackage{amssymb}
\usepackage{dsfont}
\usepackage{braket}
\usepackage[normalem]{ulem}
\usepackage{siunitx}
\usepackage{wasysym}
\usepackage{hyperref}
\usepackage{lipsum} 
\usepackage{booktabs}

\usepackage{float} 

\usepackage{esvect}

\usepackage{dcolumn}
\usepackage{bm}

\usepackage{color} 

\newcommand{\spindown}{\lvert\downarrow\rangle}
\newcommand{\spinup}{\lvert\uparrow\rangle}
\newcommand{\FMdetuning}{\ensuremath{\varepsilon_{12}}}

\usepackage{hyperref}
\hypersetup{
colorlinks=true,final=true,
        linkcolor=blue,
        citecolor=blue,
        filecolor=blue,
        urlcolor=blue,
}

\newcommand{\bmag}{\ensuremath{|\mathrm{\textbf{B}}|}}
\newcommand{\bfield}{\ensuremath{\mathrm{\textbf{B}}}}
\newcommand{\dephasing}{\ensuremath{T\mathrm{^*_{2}}}}
\newcommand{\echo}{\ensuremath{T\mathrm{^{Hahn}_{2}}}}
\newcommand{\qubitf}{\ensuremath{f_\mathrm{FM}}}

\newcommand{\virtone}{\ensuremath{\overline{V}_{\mathrm{P1}}}}
\newcommand{\virttwo}{\ensuremath{\overline{V}_{\mathrm{P2}}}}
\newcommand{\virtthree}{\ensuremath{\overline{V}_{\mathrm{P3}}}}
\newcommand{\virtBR}{\ensuremath{\overline{V}_{\mathrm{B23}}}}
\newcommand{\virtBL}{\ensuremath{\overline{V}_{\mathrm{B12}}}}

\newcommand{\bone}{\ensuremath{\hat{\bm{b}}_1}}
\newcommand{\btwo}{\ensuremath{\hat{\bm{b}}_2}}
\newcommand{\bthree}{\ensuremath{\hat{\bm{b}}_3}}

\newcommand{\Adrive}{A_\mathrm{d, \mathrm{P2}}}

\newcommand{\relaxation}{\ensuremath{T_1}}

\begin{document} 
\title{Coherent and ultra-low-power EDSR with a flopping-mode spin qubit in germanium}

\author{Alexei Orekhov\equalcontrib}
\affiliation{Hybrid Quantum Circuits Laboratory, Institute of Physics, École Polytéchnique Fédérale de Lausanne (EPFL), 1015 Lausanne, Switzerland}
\affiliation{Center for Quantum Science and Engineering, École Polytéchnique Fédérale de Lausanne (EPFL), 1015 Lausanne, Switzerland}

\author{Wonjin Jang\equalcontrib}
\email{wonjin.jang@epfl.ch}
\affiliation{Hybrid Quantum Circuits Laboratory, Institute of Physics, École Polytéchnique Fédérale de Lausanne (EPFL), 1015 Lausanne, Switzerland}
\affiliation{Center for Quantum Science and Engineering, École Polytéchnique Fédérale de Lausanne (EPFL), 1015 Lausanne, Switzerland}

\author{Pan Zhang}
\affiliation{Hybrid Quantum Circuits Laboratory, Institute of Physics, École Polytéchnique Fédérale de Lausanne (EPFL), 1015 Lausanne, Switzerland}
\affiliation{Center for Quantum Science and Engineering, École Polytéchnique Fédérale de Lausanne (EPFL), 1015 Lausanne, Switzerland}
\affiliation{
QuTech and Kavli Institute of Nanoscience, Delft University of Technology,
PO Box 5046, 2600 GA Delft, The Netherlands}

\author{Konstantinos Tsoukalas}

\affiliation{Hybrid Quantum Circuits Laboratory, Institute of Physics, École Polytéchnique Fédérale de Lausanne (EPFL), 1015 Lausanne, Switzerland}
\affiliation{Center for Quantum Science and Engineering, École Polytéchnique Fédérale de Lausanne (EPFL), 1015 Lausanne, Switzerland}

\author{Fabian Oppliger}

\affiliation{Hybrid Quantum Circuits Laboratory, Institute of Physics, École Polytéchnique Fédérale de Lausanne (EPFL), 1015 Lausanne, Switzerland}
\affiliation{Center for Quantum Science and Engineering, École Polytéchnique Fédérale de Lausanne (EPFL), 1015 Lausanne, Switzerland}

\author{Franco De Palma}

\affiliation{Hybrid Quantum Circuits Laboratory, Institute of Physics, École Polytéchnique Fédérale de Lausanne (EPFL), 1015 Lausanne, Switzerland}
\affiliation{Center for Quantum Science and Engineering, École Polytéchnique Fédérale de Lausanne (EPFL), 1015 Lausanne, Switzerland}

\author{Elena Acinapura}
\affiliation{Hybrid Quantum Circuits Laboratory, Institute of Physics, École Polytéchnique Fédérale de Lausanne (EPFL), 1015 Lausanne, Switzerland}
\affiliation{Center for Quantum Science and Engineering, École Polytéchnique Fédérale de Lausanne (EPFL), 1015 Lausanne, Switzerland}

\author{Younghun Ryu}
\affiliation{Hybrid Quantum Circuits Laboratory, Institute of Physics, École Polytéchnique Fédérale de Lausanne (EPFL), 1015 Lausanne, Switzerland}
\affiliation{Center for Quantum Science and Engineering, École Polytéchnique Fédérale de Lausanne (EPFL), 1015 Lausanne, Switzerland}

\author{Inga Seidler}
\affiliation{
IBM Research Europe – Zurich, Säumerstrasse 4, 8803 Rüschlikon, Switzerland}

\author{Lisa Sommer}
\affiliation{
IBM Research Europe – Zurich, Säumerstrasse 4, 8803 Rüschlikon, Switzerland}

\author{Leonardo Massai}
\affiliation{
IBM Research Europe – Zurich, Säumerstrasse 4, 8803 Rüschlikon, Switzerland}

\author{Felix J. Schupp}
\affiliation{
IBM Research Europe – Zurich, Säumerstrasse 4, 8803 Rüschlikon, Switzerland}

\author{Matthias Mergenthaler}
\affiliation{
IBM Research Europe – Zurich, Säumerstrasse 4, 8803 Rüschlikon, Switzerland}

\author{Stefano Bosco}
\affiliation{
QuTech and Kavli Institute of Nanoscience, Delft University of Technology,
PO Box 5046, 2600 GA Delft, The Netherlands}

\author{Patrick Harvey-Collard}
\affiliation{
IBM Research Europe – Zurich, Säumerstrasse 4, 8803 Rüschlikon, Switzerland}

\author{Pasquale Scarlino}
\email{pasquale.scarlino@epfl.ch}
\affiliation{Hybrid Quantum Circuits Laboratory, Institute of Physics, École Polytéchnique Fédérale de Lausanne (EPFL), 1015 Lausanne, Switzerland}
\affiliation{Center for Quantum Science and Engineering, École Polytéchnique Fédérale de Lausanne (EPFL), 1015 Lausanne, Switzerland}



\begin{abstract}

Hole spin qubits in semiconductor quantum dots (QDs) enable high-fidelity all-electric control, but conventional electric dipole spin resonance (EDSR) can require substantial rf drive power at the low magnetic fields that are favorable for qubit coherence and readout. In planar Ge hole spin qubits, this can reach \qty{-27}{dBm} at the device, posing challenges for scalable architectures due to heating and crosstalk. Here, we demonstrate a flopping-mode (FM) qubit in Ge, where a single spin is delocalized in a double QD, combining first-order protection against charge noise with exceptionally efficient electric driving. By mapping out coherence sweet-spots as a function of magnetic field orientation we achieve $T_2^*=\qty{1.4}{\micro\second}$, $T_2^{\mathrm{Hahn}}=\qty{11.5}{\micro\second}$, $T^{\phi, \rm CPMG32}_2=\qty{130}{\micro\second}$, and $T_1=\qty{226}{\micro\second}$, and a  single-qubit gate fidelity of up to \qty{99.76}{\percent} for a gate time $t_{X\pi} = \qty{88}{\nano\second}$. Importantly, these results are obtained at a nearly in-plane magnetic field of \qty{5}{\milli\tesla} using only  $\qty{-52}{dBm}$ drive power at the device. We further find that qubit relaxation in this regime is consistent with a two-photon Orbach process, providing a route for further optimization. Our results demonstrate that FM-EDSR supports ultra-low-power, high-fidelity single-qubit operations, improvements that could benefit scalable hole-spin-based architectures and hybrid spin-photon interfaces.

\end{abstract}

\maketitle

\vspace{0.5cm}

\section*{Introduction} \label{sec:intro}
\begin{figure*}[th]
	\centering
    \includegraphics[width=\textwidth]{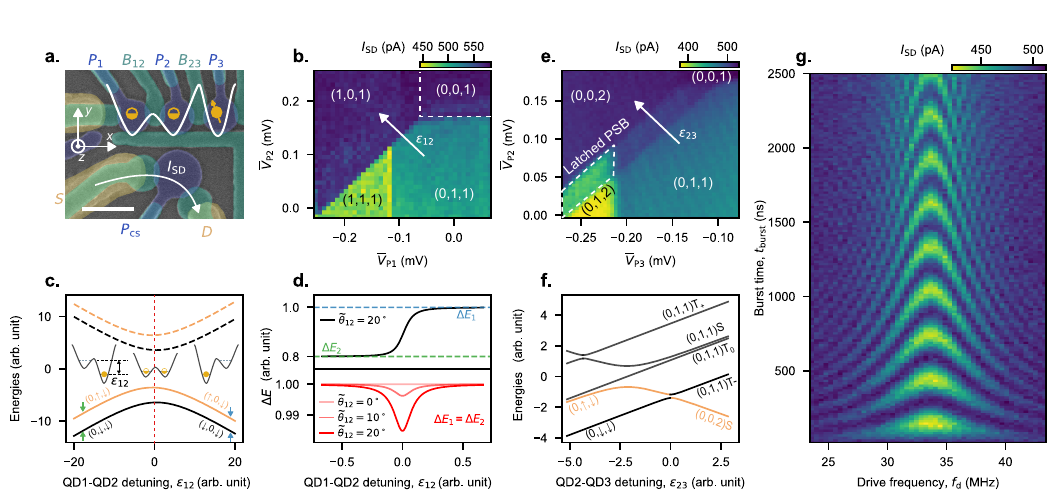}
	\caption{
    \textbf{a.} False colored scanning electron micrograph of a QD device identical to the one utilized in this work. Plungers $P_i$ are used to tune the chemical potential of QD$_i$, and barriers $B_{ij}$ are used to control tunnel coupling between QD$_i$ and QD$_j$. $S$ and $D$ indicate the source and drain ohmic contacts of the SHT charge sensor. The dc current $I_\mathrm{SD}$ through these contacts is used to charge-sense the triple QD. 
    \textbf{b.} Pulsed DQD charge stability diagram measured as a function of the virtualized gate voltages $\overline{V}_\mathrm{P1}$ and $\overline{V}_\mathrm{P2}$, near (1,0,1)-(0,1,1) charge transition. 
    The DQD energy detuning $\FMdetuning$ is tuned along the direction illustrated by the white arrow. 
    \textbf{c.} Eigenenergies as a function of energy detuning $\varepsilon_{12}$ near the (1,0,1)-(0,1,1) charge transition calculated using the Hamiltonian in Supplementary~\autoref{sec:FM hamiltonian} while assuming a finite magnetic field. 
    Solid black and orange lines denote spin-$\downarrow$ and spin-$\uparrow$ charge ground states respectively. Dashed black and orange lines indicate spin-$\downarrow$ and spin-$\uparrow$ charge excited states. Blue (green) arrows represent the Zeeman splitting of the hole in  QD$_{1 (2)}$. The FM spin qubit is defined at the charge symmetry point, denoted by the red dashed line. 
    \textbf{d.} Energy difference between the black and orange solid branches shown in \textbf{c.}, calculated by assuming different (top panel) and equal (bottom panel) Zeeman energies in QD$_1$ and QD$_2$ for different values of Larmor vector misalignment $\tilde\theta_{12}$. Blue and green dashed lines in the top panel illustrate the Zeeman energy of the hole in QD$_1$ and QD$_2$.
    \textbf{e.} Pulsed DQD charge stability diagram (see text) measured as a function of the virtualized gate voltages $\overline{V}_\mathrm{P2}$ and $\overline{V}_\mathrm{P3}$, near the (0,1,1)-(0,0,2) charge transition. The DQD energy detuning $\varepsilon_{23}$ is tuned along the direction illustrated by the white arrow. The region in (0,1,2) enclosed by the white dashed lines corresponds to the latched-PSB readout window. 
    \textbf{f.} Calculated eigenenergies as a function of energy detuning $\varepsilon_{23}$ near the (0,1,1)-(0,0,2) charge transition. S, $\mathrm{T}_0$, $\mathrm{T}_+$ and $\mathrm{T}_-$ indicate spin singlet and triplet states. 
    \textbf{g.} Rabi chevron pattern of the FM spin qubit measured at a detuning sweet-spot for \bfield{} orientation \bthree{} (see section Anisotropic Coherence) with rf drive frequency $f_\mathrm{d}$, drive amplitude \qty{550}{\micro\volt} on \virttwo{} and drive duration $t_\mathrm{burst}$. Here, $|\mathbf{B}| = \qty{5}{\milli\tesla}$ and $t_c/h=\qty{13}{\giga\hertz}$. 
	}
    \label{fig:fig1}
\end{figure*}


Hole spin qubits in planar strained Ge/SiGe heterostructures have emerged as a promising platform for quantum computing and simulation~\cite{hendrickxFourqubit2021, scappucciGermanium2021, zhang_universal_2025, morozova_observation_2026}. 
Their highly anisotropic and voltage-dependent $g$-tensor enables fully electrical spin control via $g$-tensor magnetic resonance ($g$-TMR), yielding single-qubit gate fidelities in excess of $99.9\%$ \cite{lawrie_simultaneous_2023, hendrickxSweetspot2024, dijkemaSimultaneous2026, ademi2026, wang_operating_2024}. 
Recent progress in coherence optimization \cite{hendrickxSweetspot2024, yu_optimising_2026}, high-fidelity two-qubit gates \cite{wang_operating_2024}, robust spin readout \cite{kelly_identifying_2025} and coherent shuttling \cite{ademi2026} has relied on operation in low in-plane magnetic fields below \qty{100}{\milli\tesla}, where charge noise sensitivity and the spin relaxation rate are reduced. The same low-field regime, however, decreases the Rabi driving efficiency and typically requires drive powers around $P_{\rm drive}\approx\qty{-27}{dBm}$ (or equivalently \qty{10}{\milli\volt} drive voltage) at the device for practical gate speeds ~\cite{ademi2026, dijkemaSimultaneous2026, john_robust_2025}. 
In semiconductor spin qubit platforms, a large driving power has been associated with heating \cite{undseth_hotter_2023}, readout fidelity degradation \cite{kelly_capacitive_2023, eggli_coupling_2025} and cross-talk \cite{undseth_nonlinear_2023, john_robust_2025}, making power-efficient control increasingly important as qubit arrays grow  \cite{abraham_digitally_2026, smet_spin_2026, wang_operating_2024, john_robust_2025, tsoukalas2026, nguyen_degenerate_2026}.

Several approaches can be employed to improve the power-efficiency of single-qubit gates.
For instance, it has been observed that Rabi driving is moderately improved for certain hole occupations \cite{john_robust_2025}. 
Alternatively, a large $g$-tensor variability in neighboring quantum dots can be utilized to perform high-fidelity hopping gates via baseband pulsing, which reduces dielectric losses~\cite{wang_operating_2024, unseld_baseband_2025}. Furthermore, both EDSR and baseband gates can be extended to shuttling links where the qubit probes larger magnetic field or $g$-tensor gradients \cite{bosco2024, smet_spin_2026, ademi2026}.
Another approach is to operate in the flopping-mode regime, where a single charge carrier is delocalized across a double quantum dot (DQD) \cite{crootFloppingmode2020, benito2019, mutter2021}. At the charge symmetry point, the increased electric dipole can enhance spin--electric coupling by orders of magnitude. This has enabled strong spin--photon coupling for electrons in Si/SiGe \cite{samkharadzeStrong2018, miCoherent2018, dijkema_cavity-mediated_2025}, and for holes in Si nanowires \cite{yu_strong_2023, noirot_coherence_2026}, with theory predicting fast, low-power and high-fidelity single-qubit gates \cite{youngBenchmarking2025, kinikar2026, teske_flopping-mode_2023, hu2023}. The enhanced dipole, however, presents a trade-off: the same charge hybridization that strengthens electrical control can also increase charge-noise sensitivity and relaxation. Whether strong transverse spin–electric coupling can coexist with weak longitudinal susceptibility and long relaxation times therefore remains a key question for high-fidelity operation.

In this work, we demonstrate that these requirements can be simultaneously satisfied in a flopping-mode (FM) hole-spin qubit in planar Ge. Pauli spin blockade (PSB)-based readout using an ancillary quantum dot (QD) spin enables operation at low magnetic fields, where the spin coherence is improved. 
We study the coherence anisotropy with respect to the magnetic field orientation and identify dephasing sweet-spots where the Zeeman energies in the two QDs are degenerate \cite{hendrickxSweetspot2024, bassiOptimal2026}. At these sweet-spots, we observe that charge noise is suppressed to first order, and that second order charge noise coupling, together with the Overhauser field fluctuation, is limiting the longitudinal coherence time of the qubit. Furthermore, when operating at the sweet-spots, we observe a Rabi driving efficiency normalized by the qubit frequency, exceeding \qty{0.15}{\per\milli\volt}, a two to three orders of magnitude improvement over $g$-TMR drive of single-QD Loss-DiVincenzo (LD) spin qubits in Ge \cite{john_robust_2025, ademi2026, dijkemaSimultaneous2026, tsoukalas_resonant_2025}. Dynamical decoupling extends the coherence time to $T_2^{\phi,\mathrm{CPMG}32} = \qty{130}{\micro\second}$, and randomized benchmarking yields average physical-gate fidelities of up to 99.76(1)\% with drive powers below \qty{-51}{dBm} (0.6 mV drive voltage) at the device. 
We further characterize spin relaxation as a function of magnetic-field orientation, detuning and tunnel coupling, finding behaviour consistent with a second-order Orbach process induced by lattice phonons or thermal photons \cite{tahan_relaxation_2014}. Finally, a comparison with both an LD qubit in the same device and recent spin qubit experiments in Ge places the FM qubit in a favourable power--fidelity regime.

\subsection*{Flopping-mode qubit}\label{sec:device}

\autoref{fig:fig1}a displays a false-colored scanning electron micrograph of a device identical to the one used in this work, defined on a planar Ge/SiGe heterostructure grown by reverse-grading on a Si substrate \cite{massai_impact_2024}. 
Holes are accumulated in a \qty{20}{\nano\meter} strained Ge quantum well residing \qty{47}{\nano\meter} below the surface and contacted with annealed PtGeSi ohmic contacts \cite{massai_spin_2026}. 
We make use of two gate layers, with barriers and screening gates in the first layer, and plungers in the second layer. 
The three uppermost plunger gates $P_{i}$, with $i=1,2,3$ are primarily used to define a triple-QD, while the barriers $B_{12}$ and $B_{23}$ are used to control the inter-dot tunnel couplings. 
The larger QD under $P_{\mathrm{cs}}$ is used as a single-hole transistor. 
It is probed in transport by monitoring the dc-current $I_\mathrm{SD}$ through the source ($S$) and drain ($D$) ohmic contacts (Supplementary~\autoref{sec:Experimental setup}), allowing for sensitive detection of the triple-QD charge occupation down to the single-hole level. 
An external magnetic field (\bfield{}) is applied using a three-axis vector magnet, with the coordinate frame shown in \autoref{fig:fig1}a.    

The dc voltage setpoint is set to the middle of the (0,1,1) charge region, where ($n_1$, $n_2$, $n_3$) denotes the charge configuration in the triple-QD, and $n_i$ designates the number of holes in QD$_i$ under plunger $P_i$. 
All gate voltages are virtualized to compensate for cross-capacitance, with $\overline{V}_i$ denoting the pulsed virtual gate voltage of gate $i$ (see  Supplementary \autoref{sec:vmatrix} for more details). 
Charge stability diagrams are measured by pulsing the virtual plunger gates to each point and reading out $I_\mathrm{SD}$ after a settling time. 
The pulsed-gate charge stability diagram of the DQD, defined under $P_1$ and $P_2$, is shown in \autoref{fig:fig1}b. 
It reveals the inter-dot transition between (0,1,1) and (1,0,1), which is the operating point for the FM qubit. We define the energy detuning $\FMdetuning=\mu_2-\mu_1$, where $\mu_i$ is the chemical potential of  QD$_i$. The corresponding detuning direction in gate-voltage space is shown as a white arrow in \autoref{fig:fig1}b. 

The energy diagram of the leftmost DQD in the presence of a finite \bfield{} is presented in \autoref{fig:fig1}c, as a function of \FMdetuning{}.
Black and orange solid lines represent the charge ground states having spin-down ($\spindown$) and spin-up ($\spinup$) respectively, where the splitting between the two states is primarily given by the Zeeman energy ($E_{\rm z}$) of the hole.
Similarly, black and orange dashed lines correspond to the charge-excited  $\spindown$ and $\spinup$ states. For large detunings $|\FMdetuning{}|\gg t_c$, where $t_c$ is the tunnel coupling, the eigenstates correspond to the localized charge states $|L\rangle = |(1,0,1)\rangle$ and $|R\rangle = |(0,1,1)\rangle$. At the charge symmetry point ($\FMdetuning = 0$), the charge eigenstates are the symmetric (bonding, $\ket{+} = (|L\rangle + |R\rangle)/\sqrt{2}$) and antisymmetric (antibonding, $\ket{-} = (|L\rangle - |R\rangle)/\sqrt{2}$) states, separated by an energy $2t_c$ \cite{hayashi2003b, petersson2010}. 
The FM qubit is encoded in the Zeeman-split bonding states, $\lvert 0\rangle = \lvert+,\downarrow\rangle$ and $\lvert 1\rangle = \lvert+,\uparrow\rangle$~\cite{benito2019}.  
At $\FMdetuning = 0$, the charge density is delocalized across the DQD, implying an enhanced charge dipole moment, and therefore faster electric dipole spin resonance (EDSR) drive compared to that of an LD qubit \cite{crootFloppingmode2020, benito2019}. 

Importantly, the Larmor vector, and therefore $E_{\rm z}$ can be site-dependent due to $g$-tensor variability between QDs \cite{hendrickxSweetspot2024, seidler2025}. The energy splitting between the Zeeman-split charge ground states $\Delta E$ at $\FMdetuning=0$ is given by \cite{vanriggelen-doelman2024, wang_operating_2024}:
\begin{equation*}
\Delta E(\varepsilon_{12}=0)
=
\frac{1}{2}\sqrt{(E_\mathrm{z,1}+E_\mathrm{z,2})^2
-4E_\mathrm{z,1}E_\mathrm{z,2}\sin^2(\tilde{\theta}_{12}/2)},
\end{equation*}
where $E_{\mathrm{z},i}\ll 2t_c$ is the Zeeman energy in QD$_{i}$ and $\tilde{\theta}_{12}$ is a rotation angle taking into account site-dependent Larmor vector directions and spin-flip tunneling induced by spin-orbit interaction (SOI) (see Supplementary~\autoref{sec:FM hamiltonian}) \cite{seidler2025, massai_engineering_2026}.  
The top panel of \autoref{fig:fig1}d illustrates $\Delta E(\varepsilon_{12})$, for the case where the two QDs exhibit very different $E_\mathrm{z}$.
Here, $\Delta E$ approaches $E_\mathrm{z,1(2)}$ for $\varepsilon_{12} \gg 0$ ($\varepsilon_{12} \ll 0$) denoted by a blue (green) dashed line.
In contrast, the bottom panel presents $\Delta E(\varepsilon_{12})$ for the case when the Larmor vectors of the two QDs have comparable magnitudes ($E_\mathrm{z,1} \approx E_\mathrm{z,2}$) and $\tilde{\theta}_{12}$ is finite. 
In this regime, $\Delta E(\varepsilon_{12})$ exhibits a local minimum at $\varepsilon_{12}=0$, corresponding to a coherence sweet-spot where the spin is first-order insensitive to detuning charge noise (i.e., $\partial \Delta E/\partial\varepsilon_{12}=0$). 
We emphasize that the depth of the $\Delta E$ minimum relative to $E_\mathrm{z,1(2)}$ is independent of $t_c$, in contrast to the regime $E_z \approx 2t_c$ often considered in the context of strong spin-photon coupling to electron spins \cite{miCoherent2018, samkharadzeStrong2018}. Consequently, the operating point considered here is simultaneously a detuning and tunnel-coupling sweet-spot. Since the $g$-tensors of Ge QDs are inherently anisotropic, electrically tunable, and site dependent \cite{terrazos_theory_2021, scappucciGermanium2021, valvo2025, seidler2025, mauro2025, massai_engineering_2026}, the conditions $E_\mathrm{z,1}\approx E_\mathrm{z,2}$ together with a finite $\tilde{\theta}_{12}$ can be systematically realized by choosing an appropriate magnetic-field orientation, as discussed later.

Readout of the FM qubit is performed using latched Pauli spin blockade (PSB) \cite{kelly_identifying_2025, harvey-collard_high-fidelity_2018}, employing the spin ancilla in QD$_3$. 
\autoref{fig:fig1}e presents the pulsed charge stability diagram of the DQD formed under $P_2$ and $P_3$, near the (0,0,2)--(0,1,1) charge transition, where the latched PSB window is revealed by a pulsed-gate measurement (Supplementary~\autoref{sec:Readout and initialization}).
The relevant energy diagram is shown in \autoref{fig:fig1}f, as a function of the DQD energy detuning $\varepsilon_{23} = \mu_2 - \mu_3$. 
To initialize the FM qubit, a (0,0,2)S singlet is first prepared by ramping into the (0,0,2) charge region and waiting for the spin to relax, followed by adiabatic mapping to (0,$\downarrow$,$\downarrow$)~\cite{kelly_identifying_2025}. 
The exchange between QD$_2$ and QD$_3$ is then turned off by applying a positive voltage pulse on barrier \virtBR{}. 
This is followed by a negative pulse on \virtBL{} to increase $t_c$. We then ramp to the charge symmetry point between (1,0,1) and (0,1,1) ($\FMdetuning = 0$), which is the operating point of the qubit (for more details see Supplementary~\autoref{sec:Readout and initialization}). 

An example chevron pattern of the FM qubit operated at the first-order detuning sweet-spot is shown in \autoref{fig:fig1}g, obtained by radio-frequency (rf) drive of the virtual voltage \virttwo{} with the frequency $f_\mathrm{d}$, drive duration $t_\mathrm{burst}$, and drive amplitude $\Adrive{} = \qty{550}{\micro\volt}$ on \virttwo{}. We extract a Rabi frequency $f_\mathrm{Rabi} = \qty{3.5}{\mega\hertz}$ and qubit frequency $f_{\rm FM} = \qty{33.7}{\mega\hertz}$. 
The detailed configuration of this measurement is discussed next.

\subsection*{Anisotropic coherence}\label{sec:coherence_vs_angle}

\begin{figure*}[t]
    \includegraphics[width=\textwidth]{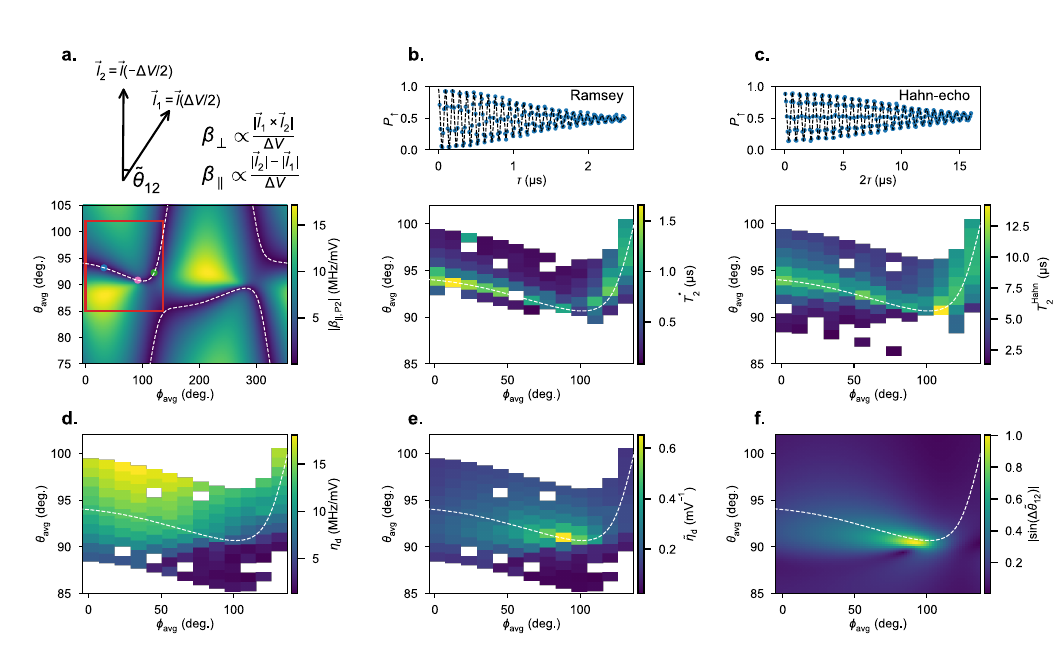}
	\caption{\textbf{a.} Top panel: Schematic showing the definition of longitudinal ($\beta_{||}$ ) and transverse ($\beta_{\perp}$) spin-electric susceptibility (LSES and TSES), with FM qubit Larmor vector $\textbf{\textit{l}}$ evaluated at two (incrementally small) voltage detunings of $\Delta V = \Delta\overline{V}_{\mathrm{P}i}$ from the charge symmetry point.  Bottom panel: Absolute value of the LSES, $|\beta_{||, \mathrm{P2}}|$, with respect to \virttwo{}. Blue, pink and green dots define the field directions, \bone{}, \btwo{} and \bthree{} respectively, corresponding to experimentally measured zero-detuning sweet-spots. \textbf{b.} Top panel: Example Ramsey decay measurement at \bfield{} orientation $\hat{\bm{b}}_{1}$, yielding $\dephasing{}=\qty{1.4}{\micro\second}$. Bottom panel: \dephasing{} measured as a function of \bfield{}, with spherical angles defined in the average $g$-tensor frame. \textbf{c.} Top panel: Example Hahn echo measurement at \bfield{} orientation $\hat{\bm{b}}_{1}$, showing $\echo = \qty{11.2}{\micro\second}$. Bottom panel: \echo{} measured as a function of \bfield{}-field spherical angles. \textbf{d, e.} Rabi driving efficiency ($\eta_{\rm d}$) and normalized Rabi driving efficiency ($\tilde{\eta}_{\rm d}$) as a function of \bfield{} direction. \textbf{f.} Calculated $|\sin(\Delta \tilde{\theta}_{12})|$ with $\Delta \tilde{\theta}_{12}$ denoting the change in angle of the effective FM qubit Larmor vector for $\Delta V = \Delta \virttwo{}=\qty{250}{\micro\volt}$. 
    All measurements taken at $\FMdetuning{}=0$, with fixed $\bmag{}=\qty{5}{\milli\tesla}$ and $t_c/h = \qty{13}{\giga\hertz}$.
    }
    \label{fig:fig3}
\end{figure*}

To accurately describe the dynamics of a single spin confined in a DQD, we consider a model Hamiltonian that incorporates the $g$-tensors of the spin in QD$_1$ ($g_1$) and QD$_2$ ($g_2$), their dependence on gate voltage, and contributions from spin-flip tunneling due to SOI.
To this end, we characterize the $g_1$ and $g_2$ respectively by measuring the Zeeman splitting of the spin in each QD as a function of the \bfield{} direction (see Supplementary~\autoref{sec:gtensor_SOI}) \cite{crippa_electrical_2018, hendrickxSweetspot2024}.
We also characterize the gate-voltage dependence of each $g$-tensor using a modified Hahn echo sequence \cite{hendrickxSweetspot2024}, and measure the energy splitting at $\FMdetuning = 0$ as a function of \bfield{} direction, from which we estimate a spin-flip tunneling angle $\theta_{\rm SO}=\qty{14.7\pm0.6}{\degree}$, similar to values extracted from the exchange interaction in other works~\cite{massai_engineering_2026, seidler2025} (see Supplementary~\autoref{sec:gtensor_SOI}).
Based on this model, we calculate the effective Larmor vector of the FM qubit $\textbf{\textit{l}}(\virtone{},\virttwo{})$, near $\FMdetuning = 0$, assuming linear gate-voltage dependence of the $g$-tensor \cite{wang_operating_2024}. 
This is used to characterize the longitudinal spin-electric susceptibility \cite{hendrickxSweetspot2024, bassiOptimal2026} (LSES) of the FM qubit at $\FMdetuning=0$ with respect to $\overline{V}_{\mathrm{P}i}$, given by $\beta_{||, \mathrm{P}i} = \frac{d\textbf{\textit{l}}}{d\overline{V}_{\mathrm{P}i}} \cdot \textbf{\textit{l}}/(h |\textbf{\textit{l}}|)$, as illustrated in the top panel of \autoref{fig:fig3}a.
The LSES measures the sensitivity of the qubit frequency to gate voltage, which implies that minimizing $|\beta_{||, \mathrm{P}i}|$ is beneficial for suppressing charge-noise-induced dephasing.

The bottom panel in \autoref{fig:fig3}a presents the calculated $\mathrm{|\beta_{||, P2}|}$, as a function of the magnetic field orientation (see Supplementary~\ref{sec:LSES and TSES extraction} for $\beta_{||, \mathrm{P1}}$).
Here, $\phi_\mathrm{avg}$ and $\theta_\mathrm{avg}$ are spherical coordinates, used to denote the \bfield{} direction in the average $g$-tensor frame of $g_1$ and $g_2$, with $\theta_\mathrm{avg}=\qty{90}{\degree}$ corresponding to the $xy$ (sample) plane of the average frame (see Supplementary \autoref{fig:fig2}).
Importantly, the white dashed curve in \autoref{fig:fig3}a marks the contours $\beta_{||, \mathrm{P2}} = 0$, or LSES sweet-lines \cite{bassiOptimal2026}, where the Zeeman energies of the two QDs are degenerate at $\FMdetuning{}=0$, resulting in a first-order detuning sweet-spot. 
We note that the geometry of the FM qubit sweet-line resembles that of the LD qubits in QD$_1$ and QD$_2$, with a sweet-spot generally occurring at small out-of-plane \bfield{} tilts (see Supplementary~\autoref{fig:Supple_LSES_TSES}).   


To confirm the noise suppression at the sweet-line, we measure the longitudinal qubit coherence times $T_2^*$ and $T_2^\mathrm{Hahn}$ as a function of $\phi_\mathrm{avg}$ and $\theta_\mathrm{avg}$ at $\FMdetuning{}=0$, around the region denoted by the red square in \autoref{fig:fig3}a, for fixed $\bmag{}=\qty{5}{\milli\tesla}$ and $t_c/h=\qty{13}{\giga\hertz}$ (see Supplementary~\autoref{sec:Lever arm extraction}).
At each field direction, the control pulses and $\FMdetuning = 0$ are calibrated as detailed in Supplementary~\autoref{sec:Detuning and sweet-spot calibration}.
Representative coherence decay traces are shown in the top panels of \autoref{fig:fig3}b,c for the \bfield{} orientation \bone{} (blue point on the sweet-line in \autoref{fig:fig3}a), yielding $T_2^*$ = $\qty{1.4}{\micro\second}$ and $T_2^{\mathrm{Hahn}} = \qty{11.2}{\micro\second}$ from a fit to a generalized exponential decay model (see Supplementary~\autoref{eq:decay fit}). The full maps in \autoref{fig:fig3}b,c show that the coherence maxima track the calculated LSES sweet-line over magnetic field orientation, directly demonstrating first-order suppression of charge-noise sensitivity. 
Along the sweet-line, \dephasing{} reaches up to $\qty{1.5}{\micro\second}$ and \echo{} about $\qty{15}{\micro\second}$. \dephasing{} reduces rapidly away from the sweet-line, whereas Hahn echo coherence is more robust, especially at larger out-of-plane angles. 


Importantly, suppressing the longitudinal susceptibility does not compromise electrical control. 
We define the Rabi driving efficiency $\eta_{\rm d}=f_{\rm Rabi}/\Adrive{}$ and the normalized Rabi driving efficiency $\tilde{\eta}_\mathrm{d}=f_{\rm Rabi}/(\Adrive{}\qubitf{})$. 
From the same chevrons used to obtain the coherence maps (\autoref{fig:fig3}b,c), we extract $\eta_d > 5$~MHz/mV and $\tilde{\eta}_\mathrm{d}> 0.15$~Hz/(Hz\,mV) along the sweet-line (\autoref{fig:fig3}d, e). 
$\tilde{\eta}_\mathrm{d}$ is two to three orders of magnitude larger than values reported for $g$-TMR-driven LD qubits \cite{dijkemaSimultaneous2026, john_robust_2025, tsoukalas_resonant_2025}. 
Noting that the power of the qubit drive tone approximately scales as $V_{\rm drive}^2$ \cite{wang_operating_2024, undseth_hotter_2023}, the significantly enhanced $\eta_d$ and $\tilde{\eta}_d$ of the FM qubit facilitate minimally dissipative qubit operations, as will be clear in the comparison with LD spin qubits later in \autoref{fig:figRB}.


The origin of the large transverse spin-electric susceptibility (TSES), defined as $\beta_{\perp,\mathrm{P}i} = |\frac{d\textbf{\textit{l}}}{d\overline{V}_{\mathrm{P}i}} \times \textbf{\textit{l}}|/(h |\textbf{\textit{l}}|)$,  is revealed by the relative orientation of the local Larmor vectors. 
At the sweet-line their magnitudes are nearly matched, whereas their directions remain misaligned. 
Resonant gate modulation therefore changes primarily the direction of the effective Larmor vector, while leaving its magnitude comparatively insensitive, analogous to iso-Zeeman EDSR \cite{crippa_electrical_2018, carballido_compromise-free_2025, geyer_-situ_2025}. 
In this regime, $\beta_{\perp,\mathrm{P}i}$ scales approximately as $f_{\rm FM}|\sin (\Delta  \tilde{\theta}_{12})|$, where $\Delta \tilde{\theta}_{12}$ is the rotation of the effective Larmor vector with respect to the change of a gate-voltage by $\Delta \virttwo{}$ (see top panel in \autoref{fig:fig3}a).
The calculated $|\sin(\Delta \tilde{\theta}_{12})|$ map in \autoref{fig:fig3}f closely follows the measured angular dependence of $\tilde{\eta}_d$ in \autoref{fig:fig3}e. 
Thus, the same inter-dot Larmor-vector misalignment that provides strong transverse electrical control can coexist with longitudinal noise protection. 
The key feature of the FM operating point is therefore that the longitudinal and transverse electrical susceptibilities can be engineered separately: matching the Larmor-vector magnitudes suppresses dephasing, while their finite relative angle preserves strong electrical driving.

For the remainder of the study, we focus on three representative \bfield{} directions on the sweet-line (see \autoref{fig:fig3}a) denoted by unit vectors \bone{} (blue), \btwo{} (pink) and \bthree{} (green). Within the angular range in \autoref{fig:fig3}b-f, $\bone{}$ = ($\phi_{\rm avg}$,$\theta_{\rm avg}$) = ($32.36^{\circ}$,$93.16^{\circ}$) maximizes $f_{\rm Rabi}$, $\btwo{}$ = ($91.80^{\circ}$,$90.83^{\circ}$) maximizes $f_{\rm Rabi}/f_{\rm FM}$, and $\bthree{}$ = ($120.6^{\circ}$,$92.25^{\circ}$) gives the lowest Rabi frequency. We also fix $\bmag{}=\qty{5}{\milli\tesla}$ and $t_c/h=\qty{13}{\giga\hertz}$ unless stated otherwise. 

\subsection*{Coherence scaling}\label{sec:B field scaling}

\begin{figure}[!t]
    \includegraphics[width=\columnwidth]{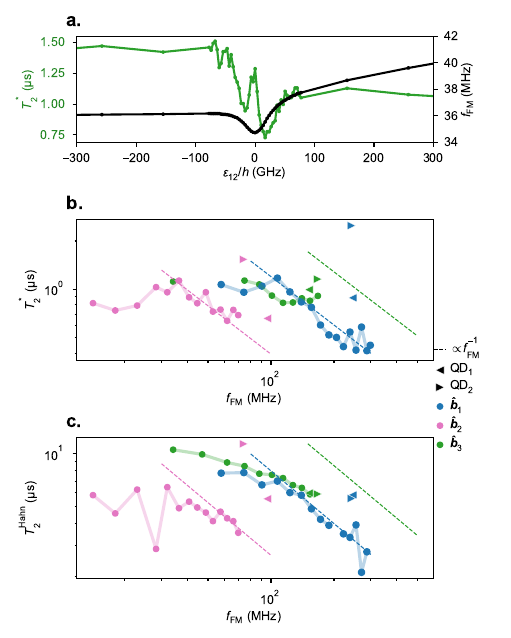}
	\caption{
    \textbf{a.} FM qubit frequency $f_{\mathrm{FM}}$ (black) and $T^*_2$ (green) as a function of $\FMdetuning$, extracted from the same Ramsey measurement, with $|\textbf{B}|=\qty{5}{\milli\tesla}$ and $t_c/h=\qty{13}{\giga\hertz}$.
    \textbf{b. (c.)} $T_2^*$ ($T_2^\mathrm{Hahn}$) as a function of $f_\mathrm{FM}$ for the \bfield{} orientations \bone{} (blue), \btwo{} (pink), and \bthree{} (green), and $t_c/h=\qty{13}{\giga\hertz}$. Each dashed line is proportional to $1/\qubitf{}$ with prefactors related by the extracted second derivatives of the qubit spectra (see text). The LD qubit coherence times in QD$_{1(2)}$ are measured at the center of (1,0,1) and (0,1,1) represented with left (right) pointing triangles. 
    }
    \label{fig:fig4}
\end{figure}

\begin{figure*}[t]
	\centering
\includegraphics[width=\textwidth]{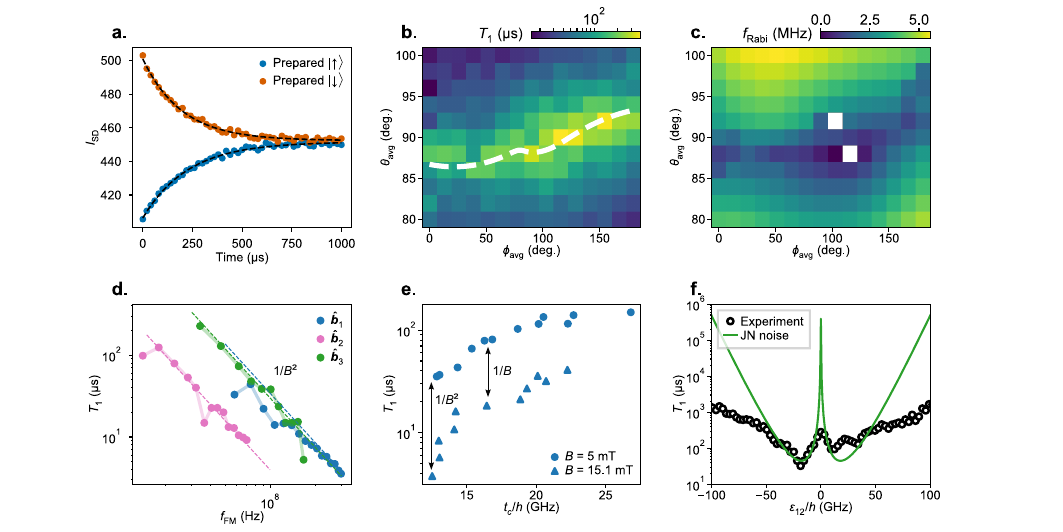}
	\caption{\textbf{a.} Spin relaxation for \bfield{} direction $\hat{\bm{b}}_{3}$, $\bmag{} = \qty{5}{\milli\tesla}$, $t_c/h = \qty{13}{\giga\hertz}$. Measured for spin-up and spin-down initializations. \textbf{b.} $\relaxation{}$ dependence on \bfield{} direction for spin-down initialization. Dashed line shows analytical $T_1$ sweet-spot (see Supplementary~\autoref{sec:second-order relaxation process}). \textbf{c.} Rabi frequency dependence on \bfield{} direction with $\Adrive{}=\qty{600}{\micro \volt}$. \textbf{d.} $\relaxation{}$ as a function of qubit frequency at zero detuning, measured for directions $\hat{\bm{b}}_{1,2,3}$. Dashed lines are obtained as $T_1=\tilde{\eta}_\mathrm{d}^{-2} k \qubitf{}^{-2}$, where $k$ is a fixed constant for all three \bfield{} directions. \textbf{e.} Scaling of $T_1$ with tunnel coupling $t_c$ for \bone{} and two different magnetic field magnitudes $\bmag{}=\qty{5}{\milli\tesla}$ and $\qty{15.1}{\milli\tesla}$. At $t_c/h=\qty{13}{\giga\hertz}$ the \bfield{} scaling of $T_1$ follows $1/\bmag{}^2$, consistent with \textbf{d.}, while for higher $t_c$ it appears to satisfy $1/\bmag{}$. \textbf{f.}  Scaling of relaxation time with energy detuning \FMdetuning{}. Green curve corresponds to the calculated $T_1$, assuming a Johnson-Nyquist (JN) noise with a characteristic impedance of \qty{250}{\ohm} and temperature of \qty{300}{\milli\kelvin}. 
    }
    \label{fig:fig5}
\end{figure*}

We next examine the mechanisms limiting qubit coherence at the detuning sweet-spot. 
To first explicitly demonstrate the charge detuning noise protection at the sweet-spot, we present \qubitf{} and \dephasing{} as a function of $\FMdetuning$ for \bthree{} in \autoref{fig:fig4}a, using a Ramsey sequence with an interleaved detuning pulse. 
As expected, \qubitf{} exhibits a minimum at $\FMdetuning = 0$ (i.e. $\partial{f_\mathrm{FM}}/\partial{\FMdetuning} = 0$).
Accordingly, \dephasing{} exhibits local minima at the flanks (where $|\partial{f_\mathrm{FM}}/\partial{\FMdetuning}|$ is maximal), and a maximum at $\FMdetuning = 0$, clearly demonstrating the protection against detuning noise. 
We note that the \dephasing{} at $\FMdetuning = 0$ remains comparable to that of the LD qubits in the deeply detuned single-QD regimes. 
This suggests that \dephasing{} of the qubit both in the FM and LD regimes is mainly limited by hyperfine noise for the given $\bfield{}$ \cite{hendrickxSweetspot2024, stehouwer2025, zeng_high-fidelity_2026}.

In Supplementary \autoref{sec:Flopping-mode qubit noise spectrum} we present an analysis of the noise power spectral density (PSD) of the FM qubit extracted using CPMG and Ramsey measurements for \bthree{} \cite{rojas-arias2025, bluhm_dephasing_2011}. Sweet-spot operation suppresses the  noise by about an order of magnitude compared to flank operation at frequencies above \qty{100}{\kilo\hertz}. We further observe an extension of the longitudinal coherence to $T^{\phi, \rm CPMG}_2 = \qty{130\pm3}{\micro\second}$ with 32 refocusing pulses.

The scaling of the coherence times with \bmag{} provides additional insights about the origin of decoherence mechanisms \cite{hendrickxSweetspot2024}. 
We vary \bmag{} in the range \qty{5}{\milli\tesla} to \qty{25}{\milli\tesla} for the \bfield{} orientations $\bone{}$, $\btwo{}$ and $\bthree{}$ and present the measured \dephasing{} and \echo{} as a function of \qubitf{} in \autoref{fig:fig4}b, and c. 
For relatively low qubit frequencies (\qubitf{}~$<$ 100, 40, and 200~MHz for $\bone{}$, $\btwo{}$ and $\bthree{}$ respectively) the coherence plateaus or decreases with lower \qubitf{}, which we attribute to the finite broadening of the hyperfine line approaching the Larmor frequency of the $^{73}$Ge isotope \cite{hendrickxSweetspot2024, bluhm_dephasing_2011}. We also note that $T_1$ is not limiting the coherence times in this experiment, with the corresponding values later discussed in \autoref{fig:fig5}d. 

At higher frequencies, the coherence times \dephasing{} and \echo{} appear to follow $T_2 \propto 1/\qubitf{}$ for $\bone{}$. 
This is consistent with the trend observed for charge-noise limited coherence times in planar Ge LD qubits \cite{hendrickxSweetspot2024, wang_operating_2024}. The coherence times at the FM sweet-spot are expected to follow $T_2 =  1/(\sigma_{\varepsilon}^2\chi f_{\mathrm{FM}})$, for a value $\chi = |\frac{\partial^2\qubitf{}}{\partial \varepsilon^2}|/ f_\mathrm{FM}$ that is assumed to be frequency-independent, and an effective detuning noise amplitude $\sigma_{\varepsilon}$ \cite{noirot_coherence_2026, benito2019}. 
We extract $\chi$ from the measured qubit spectra at specific \bmag{} values, which allows for a comparison between charge-noise limited coherence times for the respective \bfield{} orientations. 
Assuming constant $\sigma_{\varepsilon}^* $ and $\sigma_{\varepsilon}^\mathrm{Hahn}$, same for the different \bfield{} orientations (visually fitted), we present the calculated $1/((\sigma^*_{\varepsilon})^2\chi f_{\mathrm{FM}})$  and $1/((\sigma^{\mathrm{Hahn}}_{\varepsilon})^2\chi f_{\mathrm{FM}})$ in \autoref{fig:fig4}b and c (dashed lines). 
Apparently, for $\bone{}$ and $\btwo{}$, the measured coherence times asymptotically approach the charge-noise-limited line for higher qubit frequencies.
For the orientation $\bthree{}$ the model predicts a later onset of charge-noise-limited decay, beyond the measured range. This suggests that the qubit coherence in this regime is instead hyperfine-limited, consistent with the previous observation in \autoref{fig:fig4}a. 

We additionally investigate the coherence of the LD qubits in QD$_1$ and  QD$_2$ by evaluating \dephasing{} and \echo{} at the nominal center of the (1,0,1) and (0,1,1) charge regions. 
The measurements are taken for \bone{}, \btwo{} and \bthree{} with $\bmag{}=\qty{20}{\milli\tesla}$, with the results shown as triangles in \autoref{fig:fig4}b,c. 
For the cases where charge noise is expected to limit the FM qubit coherence (\bone{} and \btwo{} orientations), the LD qubit coherence time is significantly larger than that of  the FM qubit. 
However, for \bthree{} the coherence of the FM qubit and the LD qubits are almost identical, further supporting our hypothesis that the FM qubit is limited by hyperfine interactions in this regime.

\subsection*{Spin relaxation}\label{sec:T1_characterization}

To understand the physical process limiting the spin lifetime, we measure the relaxation time $T_1$ as a function of \bfield{}, \FMdetuning{} and $t_c$. \autoref{fig:fig5}a shows the measured $\relaxation{}$ decay at $\FMdetuning{}=0$ for $\hat{\bm{b}}_3$, $\bmag{} = \qty{5}{\milli\tesla}$ and $t_c/h = \qty{13}{\giga\hertz}$. 
Here, we prepare either a $\ket{\uparrow}$ or a $\ket{\downarrow}$ state, and measure its population decay as a function of time as represented by the orange (initialized in $\ket{\uparrow}$) and blue (initialized in $\ket{\downarrow}$) datapoints.
They both decay into the maximally mixed state with a similar decay time of $T_1 \approx \qty{226}{\micro\second}$. 

In Supplementary \autoref{sec:T1 theory}, we present a comprehensive treatment of $T_1$ relaxation of the FM qubit taking into account thermal photons (e.g. from gate lines and insufficient packaging), lattice phonons and charge noise. 
By absorbing a photon or a phonon close to the charge transition frequency $2t_\mathrm{c}/h$, an FM qubit can be excited from the ground ($\lvert + \rangle$) to the excited ($\lvert - \rangle$) charge state manifold. 
Then, this excited charge state can relax back to the FM qubit subspace (i.e. the ground charge state) by emitting a phonon or a photon in a mechanism known as the Orbach process \cite{tahan_relaxation_2014}. 
Because each of the absorption and emission events can be accompanied by a spin flip due to the effective transverse magnetic field gradient, the second-order process can result in the observed decay into the mixed state.
We note that the first-order process involving direct emission or absorption of phonons at the qubit frequency is negligible in our regime $\qubitf{} \sim 10-100$~MHz due to the reduced density of states of the environmental bath.
In our setup, we expect the thermal photons, rather than the lattice phonons, to play the dominant role in the second-order process. This is because lattice phonons are expected to be well-thermalized to the base temperature of the cryostat $T_\mathrm{base} \sim 10$~mK ($k_\mathrm{B}T_\mathrm{base}/h \sim 200~\mathrm{MHz} \ll 2t_\mathrm{c}/h)$.

In \autoref{fig:fig5}b we present the measured \relaxation{} decay as a function of the \bfield{} field direction.
Here, the qubit is prepared in the $\ket{\downarrow}$ state and $\bmag$ is kept constant at \qty{5}{\milli\tesla}. 
The white dashed line corresponds to the calculated contour where $\relaxation{}$ is expected to maximize due to a reduction of the effective transverse magnetic field gradient (see Supplementary~\autoref{sec:T1 theory}). 
It closely follows the maximum of the measured \relaxation{}.
\autoref{fig:fig5}c presents the corresponding Rabi frequency for each field orientation, which is showing an inverse dependence to the measured \relaxation{} \cite{benito2019}.
This confirms that the effective transverse magnetic field gradient, which drives the qubit excitation, also mediates the relaxation process.

\autoref{fig:fig5}d shows \relaxation{} measured for the \bfield{} orientations \bone{}, \btwo{} and \bthree{} (see \autoref{fig:fig3}a), while varying \qubitf{} with the magnetic field strength. Dashed lines correspond to a fit to $T_1=\tilde{\eta}_\mathrm{d}^{-2} k \qubitf{}^{-2}$, where $k$ is a fixed constant and $\tilde{\eta}_\mathrm{d}$ is the normalized Rabi driving efficiency extracted separately for each orientation. We observe a quadratic decay of $T_1$, which is consistent with either the second-order photon (or phonon) process, or first-order coupling to photons, and rules out $1/f$ charge noise (see Supplementary~\autoref{sec:T1 theory}) \cite{noirot_coherence_2026}. 

Since the Orbach process  depends on a finite thermal population of the excited charge state it should be exponentially suppressed for higher values of $t_c$. We measure \relaxation{} as a function of $t_\mathrm{c}/h$ at $\FMdetuning = 0$ at the \bfield{} orientation \bone{} for two different magnetic field strengths $\bmag{} = 5$ and \qty{15.1}{\milli\tesla}. 
The results are shown in \autoref{fig:fig5}e. 
Indeed, we observe that \relaxation{} increases with $t_\mathrm{c}$ for both magnetic field strengths. For relatively low $t_\mathrm{c}/h\lesssim 15$~GHz, the \relaxation{} at $\bmag{} = 5$~mT is approximately nine times larger than that at $\bmag{} = 15.1$~mT, which is consistent with the $1/f^2_{\rm FM}$ trend shown in \autoref{fig:fig5}d.
However, for $t_\mathrm{c}/h \gtrsim 15$~GHz, the \relaxation{} at $\bmag{} = 5$~mT is about three times larger than the value at $\bmag{} = 15.1$~mT, suggesting instead a $1/\qubitf{}$ scaling of the \relaxation{}. 
This trend points towards the Orbach process being suppressed at larger values of $t_\mathrm{c}$, with relaxation instead dominated by charge noise which results in $\relaxation{} \sim 1/\qubitf{}$. 

Next, the \relaxation{} is extracted as a function of $\FMdetuning$ (\autoref{fig:fig5}f) for \bfield{} orientation \bthree{} and $\bmag{} = \qty{5}{\milli\tesla}$. At zero detuning the $T_1$ is enhanced, which is a hallmark of the second-order process (see Supplementary~\autoref{sec:T1 theory}). A theory calculation superimposed on the data (green curve) qualitatively captures the trend. Here, we have assumed a Johnson-Nyquist noise generated by five \qty{50}{\ohm} rf gates acting as an effective \qty{250}{\ohm} noise source to qualitatively capture the trend. The data was then fitted by setting a photonic bath temperature of \qty{300}{\milli\kelvin}. The asymmetry in the data may be due to a two-level fluctuator displacing the inter-dot detuning during the measurement (as discussed in Supplementary \autoref{sec:LD qubit performance}). We note that the model predicts a sharply peaked \relaxation{} exactly at $\FMdetuning = 0$, where the second-order process is negligible. We attribute this discrepancy to the model not considering the low-frequency charge noise, which effectively modulates $\FMdetuning$. Given that the second-order relaxation process rapidly increases away from $\FMdetuning = 0$, a small detuning modulation can effectively reduce the measured \relaxation{}. This suggests that reducing low-frequency charge noise can not only improve longitudinal coherence times of the qubit, but also indirectly enhance \relaxation{}. 

\subsection*{High-fidelity, ultra-low-power EDSR}

\begin{figure}[h]
    \includegraphics[width=\columnwidth]{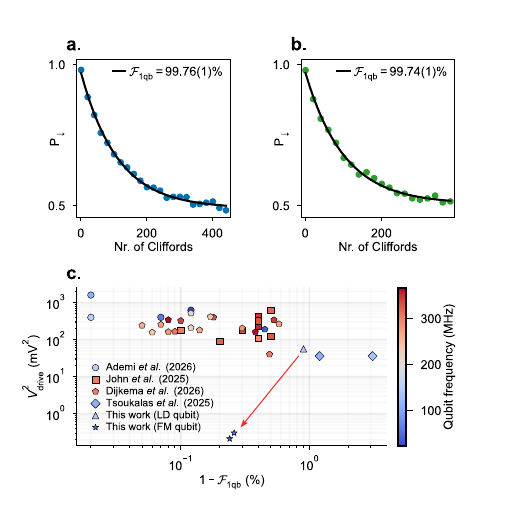}
	\caption{
    \textbf{a.} (\textbf{b.}) Single-qubit Clifford randomized benchmarking experiments for \bfield{} orientations \bone{}  (\bthree{}) for $\bmag{}=\qty{5}{\milli\tesla}$ and $t_c/h=\qty{13}{\giga\hertz}$. The qubit is driven with an rf amplitude $\Adrive{}=\qty{460}{\micro\volt}$ (\qty{550}{\micro\volt}) at the device. The average physical gate fidelity calculated from the exponential fits (black lines) is shown in the insets. Corresponding longitudinal and transverse coherence metrics are summarized in Supplementary~\autoref{tab:coherence}. \textbf{c.} Compiled data from references \cite{ademi2026, john_robust_2025, dijkemaSimultaneous2026, tsoukalas_resonant_2025}, showing the square of the drive amplitude at the device, $V_{\rm drive}^2$ and the physical single-qubit gate infidelity. Stars denote the two measured configurations in \textbf{a.} and \textbf{b.} Triangle denotes an LD qubit measured at the hyperfine sweet-spot in the same device (Supplementary~\autoref{sec:LD qubit performance}). Points are color coded according to qubit frequency. 
    }
    \label{fig:figRB}
\end{figure}

The qubit fidelity is characterized using Clifford randomized benchmarking \cite{magesan_scalable_2011}, for the \bfield{} orientations \bone{} and \bthree{}, $\bmag{} = \qty{5}{\milli\tesla}$ and $t_c/h = \qty{13}{\giga\hertz}$. Decay traces of the average survival probability after the recovery Clifford are shown in \autoref{fig:figRB}a,b. 
Here, the drive amplitudes were fixed to \qty{460}{\micro\volt} and \qty{550}{\micro\volt} resulting in $f_\mathrm{Rabi}$ of \qty{5.7}{\mega\hertz} and \qty{3.5}{\mega\hertz}, at  $\qubitf = \qty{58.8}{\mega\hertz}$ and $\qubitf =\qty{33.7}{\mega\hertz}$ respectively. The drive amplitudes were chosen to satisfy the validity of the rotating wave approximation, while maximizing gate speed.  
The average physical gate fidelities are evaluated to $\mathcal{F}_{\mathrm{1qb}} = 99.76(1)\%$ and $99.74(1)\%$ as outlined in Supplementary~\autoref{sec:Randomized benchmarking}. 
The corresponding qubit coherence metrics are summarized in Supplementary~\autoref{tab:coherence}. 
Our results demonstrate that flopping-mode EDSR is compatible with high-fidelity operations, while simultaneously reducing control power by several orders of magnitude compared to $g$-TMR operation at similar qubit frequencies~\cite{ademi2026, john_robust_2025, tsoukalas_resonant_2025, dijkemaSimultaneous2026}. 

We further illustrate this advantage by explicitly comparing our results with those reported in the literature. 
In \autoref{fig:figRB}c, we plot $V_\mathrm{drive}^2$ as a function of the gate infidelity $1-\mathcal{F}_{\mathrm{1qb}}$, where $V_\mathrm{drive}$ denotes the rf drive amplitude at the device. 
We compile results from recently reported Ge hole spin qubits operated at magnetic fields below \qty{100}{\milli\tesla}, with the field oriented approximately in the sample plane~\cite{ademi2026, john_robust_2025, tsoukalas_resonant_2025, dijkemaSimultaneous2026}. 
Each point is color-coded according to the qubit frequency reported in each study.
The stars, indicated by a red arrow, represent the FM qubit in this work for the two configurations shown in \autoref{fig:figRB}a,b. 
We observe about a two-order-of-magnitude reduction in drive power for the FM qubit compared to typical LD qubits, while maintaining comparable gate performance. 
We also show a comparison to an LD qubit (marked with a triangle) measured in our device in QD$_2$ and driven with the same gate (\virttwo{}) as the FM qubit, with a field of $\bmag{} = \qty{15}{\milli\tesla}$ oriented along the in-plane hyperfine sweet-spot (see Supplementary~\autoref{sec:LD qubit performance}). 
Notably, we observe an improvement by a factor of three in the infidelity of the FM qubit, while drive power is reduced by two orders of magnitude. 
Assuming a linear scaling of $f_\mathrm{Rabi}$ with $f_\mathrm{qubit}$ and drive amplitude, we infer a four-order-of-magnitude reduction in  driving power for the FM qubit compared to an LD qubit, when operated at the same $f_\mathrm{Rabi}$ and $f_\mathrm{qubit}$. 

\section*{Conclusions}\label{sec:discussion}

We have demonstrated a flopping-mode hole-spin qubit in a Ge DQD, showing that it can support high-fidelity EDSR control of a single spin at low magnetic fields, while reducing the required drive power by two orders of magnitude compared to state-of-the-art single-QD hole spin qubits in Ge. 
We systematically map the magnetic field sweet-lines where the Zeeman energies of the two QDs are degenerate and the qubit becomes first-order insensitive to detuning fluctuations.
The coherence measurements on the sweet-lines reveal a crossover from a low-field regime consistent with hyperfine-limited dephasing to a higher-field regime governed by second-order charge-noise coupling. 
The key feature of these operating points is the ability to separately engineer the longitudinal and transverse spin-electric  susceptibilities: matching the Larmor vector magnitudes between the two QDs significantly reduces dephasing, while their finite relative angle
preserves a large transverse spin--electric susceptibility. 
This allows Rabi control at MHz frequencies with sub-millivolt drive amplitudes at the device, while operating at a low magnetic field where the qubit coherence is optimized. 
Randomized benchmarking yields physical-gate fidelities of up to 99.76(1)\%, similar to median values reported for LD qubits in state-of-the-art Ge devices, while maintaining a high gate speed of $t_{X\pi} = \qty{88}{\nano\second}$ \cite{dijkemaSimultaneous2026, john_robust_2025}. 

The relaxation measurements expose a complementary trade-off: the transverse Larmor-vector gradient that enables efficient control also opens a relaxation channel. 
Our modeling indicates that thermal photons are the dominant relaxation mechanism in the present setup. 
Improved control-line filtering, radiation-tight packaging, and larger tunnel couplings could enhance $T_1$ by reducing thermal excitations, while lower charge noise would reduce fluctuations away from the FM operating point ($\FMdetuning{}=0$) thereby allowing the qubit to better exploit the sharp $T_1$ enhancement at this point.

These results suggest a practical operating strategy for larger spin-qubit arrays: the FM regime could be activated for ultra-low-power resonant control, while idling and two-qubit operations could be performed in the LD qubit regime, where longer coherence and spin lifetime may be available. 
Because the power reduction is obtained per driven qubit, its absolute benefit grows with the number of simultaneously operated qubits and directly addresses the aggregate rf-power budget of large semiconductor-qubit arrays. 
Highly uniform neighboring $g$-factor magnitudes would be advantageous for realizing FM sweet spots across an array, while gate-voltage tunability of the $g$-tensors provides an additional degree of freedom for engineering the finite inter-dot Larmor-vector tilt required for efficient EDSR and sweet-spot operation~ \cite{tosato2026, john_robust_2025, nguyen_degenerate_2026}. Isotopically purified heterostructures  may enable FM-qubit operation in out-of-plane fields \cite{zeng_high-fidelity_2026}, where more uniform Larmor vectors are expected and the SOI vector remains perpendicular to the quantization axis, enhancing the drive via spin-flip tunneling.


Beyond scalable control, the large electric dipole of the FM qubit is also attractive for coherent spin--photon interfaces at higher qubit frequencies, and the high driving efficiency opens access to spin dynamics beyond the rotating-wave approximation \cite{zwanenburg_single-qubit_2025, teske_flopping-mode_2023}.

\subsection*{Acknowledgment}\label{sec:acknowledgement}
We acknowledge the staff of the Binnig and Rohrer Nanotechnology Center (BRNC) for their contributions to the sample
fabrication, and Alberto Bordin for providing feedback towards the manuscript. P.S. acknowledges support from the Swiss State Secretariat for Education, Research and Innovation (SERI) under contract number MB22.00081. 
This research was funded in part by NCCR
SPIN, a National Centre of Competence in Research,
funded by the Swiss National Science Foundation (grants
51NF40-180604 and 51NF40-225153) and by the Swiss
National Science Foundation (grant 200021-188752).

\subsection*{Author contributions}\label{sec:author_contrib}

AO and WJ conceptualized the experiment. AO performed the measurements with the support of WJ and KT. AO and WJ analyzed the data. PZ and SB developed the relaxation theory model.  FO, FDP, EA, YR, WJ and AO implemented the experimental set-up. MM and FJS fabricated the device. IS, LM, LS, KT and AO contributed to device testing and development. KT designed the device. AO and WJ wrote the manuscript with input from all authors. PHC and PS supervised the project. 



\bibliographystyle{ieeetr}
\bibliography{references}

@article{hendrickxSweetspot2024,
    title = {Sweet-spot operation of a germanium hole spin qubit with highly anisotropic noise sensitivity},
    volume = {23},
    copyright = {2024 The Author(s)},
    issn = {1476-4660},
    url = {https://www.nature.com/articles/s41563-024-01857-5},
    doi = {10.1038/s41563-024-01857-5},
    language = {en},
    number = {7},
    urldate = {2026-04-16},
    journal = {Nature Materials},
    publisher = {Nature Publishing Group},
    author = {Hendrickx, N. W. and Massai, L. and Mergenthaler, M. and Schupp, F. J. and Paredes, S. and Bedell, S. W. and Salis, G. and Fuhrer, A.},
    month = jul,
    year = {2024},
    pages = {920--927},
}

@article{scappucciGermanium2021,
    title = {The germanium quantum information route},
    volume = {6},
    copyright = {2020 Springer Nature Limited},
    issn = {2058-8437},
    url = {https://www.nature.com/articles/s41578-020-00262-z},
    doi = {10.1038/s41578-020-00262-z},
    language = {en},
    number = {10},
    urldate = {2026-04-16},
    journal = {Nature Reviews Materials},
    publisher = {Nature Publishing Group},
    author = {Scappucci, Giordano and Kloeffel, Christoph and Zwanenburg, Floris A. and Loss, Daniel and Myronov, Maksym and Zhang, Jian-Jun and De Franceschi, Silvano and Katsaros, Georgios and Veldhorst, Menno},
    month = oct,
    year = {2021},
    pages = {926--943},
}

@article{hendrickxFourqubit2021,
    title = {A four-qubit germanium quantum processor},
    volume = {591},
    copyright = {2021 The Author(s), under exclusive licence to Springer Nature Limited},
    issn = {1476-4687},
    url = {https://www.nature.com/articles/s41586-021-03332-6},
    doi = {10.1038/s41586-021-03332-6},
    language = {en},
    number = {7851},
    urldate = {2023-10-04},
    journal = {Nature},
    publisher = {Nature Publishing Group},
    author = {Hendrickx, Nico W. and Lawrie, William I. L. and Russ, Maximilian and van Riggelen, Floor and de Snoo, Sander L. and Schouten, Raymond N. and Sammak, Amir and Scappucci, Giordano and Veldhorst, Menno},
    month = mar,
    year = {2021},
    note = {Number: 7851},
    pages = {580--585},
}

@article{wang_operating_2024,
    title = {Operating semiconductor quantum processors with hopping spins},
    volume = {385},
    issn = {0036-8075, 1095-9203},
    url = {https://www.science.org/doi/10.1126/science.ado5915},
    doi = {10.1126/science.ado5915},
    language = {en},
    number = {6707},
    urldate = {2026-04-30},
    journal = {Science},
    author = {Wang, Chien-An and John, Valentin and Tidjani, Hanifa and Yu, Cécile X. and Ivlev, Alexander S. and Déprez, Corentin and Van Riggelen-Doelman, Floor and Woods, Benjamin D. and Hendrickx, Nico W. and Lawrie, William I. L. and Stehouwer, Lucas E. A. and Oosterhout, Stefan D. and Sammak, Amir and Friesen, Mark and Scappucci, Giordano and De Snoo, Sander L. and Rimbach-Russ, Maximilian and Borsoi, Francesco and Veldhorst, Menno},
    month = jul,
    year = {2024},
    pages = {447--452},
}

@article{bassiOptimal2026,
    title = {Optimal operation of hole spin qubits},
    volume = {22},
    copyright = {2025 The Author(s), under exclusive licence to Springer Nature Limited},
    issn = {1745-2481},
    url = {https://www.nature.com/articles/s41567-025-03106-1},
    doi = {10.1038/s41567-025-03106-1},
    number = {1},
    urldate = {2026-04-17},
    journal = {Nature Physics},
    publisher = {Nature Publishing Group},
    author = {Bassi, M. and Rodríguez-Mena, E. A. and Brun, B. and Zihlmann, S. and Nguyen, T. and Champain, V. and Abadillo-Uriel, J. C. and Bertrand, B. and Niebojewski, H. and Maurand, R. and Niquet, Y.-M. and Jehl, X. and De Franceschi, S. and Schmitt, V.},
    month = jan,
    year = {2026},
    pages = {75--80},
}

@article{samkharadzeStrong2018,
    title = {Strong spin-photon coupling in silicon},
    volume = {359},
    issn = {0036-8075, 1095-9203},
    url = {https://www.science.org/doi/10.1126/science.aar4054},
    doi = {10.1126/science.aar4054},
    language = {en},
    number = {6380},
    urldate = {2026-02-25},
    journal = {Science},
    author = {Samkharadze, N. and Zheng, G. and Kalhor, N. and Brousse, D. and Sammak, A. and Mendes, U. C. and Blais, A. and Scappucci, G. and Vandersypen, L. M. K.},
    month = mar,
    year = {2018},
    pages = {1123--1127},
}

@article{miCoherent2018,
    title = {A coherent spin–photon interface in silicon},
    volume = {555},
    copyright = {2018 Macmillan Publishers Limited, part of Springer Nature. All rights reserved.},
    issn = {1476-4687},
    url = {https://www.nature.com/articles/nature25769},
    doi = {10.1038/nature25769},
    language = {en},
    number = {7698},
    urldate = {2023-10-04},
    journal = {Nature},
    publisher = {Nature Publishing Group},
    author = {Mi, X. and Benito, M. and Putz, S. and Zajac, D. M. and Taylor, J. M. and Burkard, Guido and Petta, J. R.},
    month = mar,
    year = {2018},
    note = {Number: 7698},
    pages = {599--603},
}

@article{crootFloppingmode2020,
    title = {Flopping-mode electric dipole spin resonance},
    volume = {2},
    url = {https://link.aps.org/doi/10.1103/PhysRevResearch.2.012006},
    doi = {10.1103/PhysRevResearch.2.012006},
    number = {1},
    urldate = {2026-04-16},
    journal = {Physical Review Research},
    publisher = {American Physical Society},
    author = {Croot, X. and Mi, X. and Putz, S. and Benito, M. and Borjans, F. and Burkard, G. and Petta, J. R.},
    month = jan,
    year = {2020},
    pages = {012006},
}

@misc{dijkemaSimultaneous2026,
    title = {Simultaneous operation of an 18-qubit modular array in germanium},
    url = {http://arxiv.org/abs/2604.01063},
    doi = {10.48550/arXiv.2604.01063},
    urldate = {2026-04-15},
    publisher = {arXiv},
    author = {Dijkema, J. J. and Zhang, X. and Bardakas, A. and Bouman, D. and Cuzzocrea, A. and Driel, D. van and Girardi, D. and Stehouwer, L. E. A. and Scappucci, G. and Zwerver, A. M. J. and Hendrickx, N. W.},
    month = apr,
    year = {2026},
    note = {arXiv:2604.01063 [cond-mat]},
}

@article{benito2019,
    title = {Electric-field control and noise protection of the flopping-mode spin qubit},
    volume = {100},
    url = {https://link.aps.org/doi/10.1103/PhysRevB.100.125430},
    doi = {10.1103/PhysRevB.100.125430},
    number = {12},
    urldate = {2026-04-15},
    journal = {Physical Review B},
    publisher = {American Physical Society},
    author = {Benito, M. and Croot, X. and Adelsberger, C. and Putz, S. and Mi, X. and Petta, J. R. and Burkard, Guido},
    month = sep,
    year = {2019},
    pages = {125430},
}

@article{mutter2021,
    title = {Natural heavy-hole flopping mode qubit in germanium},
    volume = {3},
    issn = {2643-1564},
    url = {https://link.aps.org/doi/10.1103/PhysRevResearch.3.013194},
    doi = {10.1103/PhysRevResearch.3.013194},
    language = {en},
    number = {1},
    urldate = {2026-05-04},
    journal = {Physical Review Research},
    author = {Mutter, Philipp M. and Burkard, Guido},
    month = feb,
    year = {2021},
    pages = {013194},
}

@misc{ademi2026,
    title = {Distributing entanglement between distant semiconductor qubit registers using a shared-control shuttling link},
    url = {http://arxiv.org/abs/2510.26860},
    doi = {10.48550/arXiv.2510.26860},
    urldate = {2026-05-04},
    publisher = {arXiv},
    author = {Ademi, Zarije and Bassi, Marion and Yu, Cécile X. and Oosterhout, Stefan D. and Matsumoto, Yuta and Snoo, Sander L. de and Sammak, Amir and Vandersypen, Lieven M. K. and Scappucci, Giordano and Déprez, Corentin and Veldhorst, Menno},
    month = jan,
    year = {2026},
    note = {arXiv:2510.26860 
version: 3},
}

@misc{kinikar2026,
    title = {Microscopic modeling of flopping-mode quantum dot spin qubits},
    url = {http://arxiv.org/abs/2604.20510},
    doi = {10.48550/arXiv.2604.20510},
    urldate = {2026-05-04},
    publisher = {arXiv},
    author = {Kinikar, Ashutosh and Levajac, Vukan and Moors, Kristof and Simion, George and Benito, Monica and Soree, Bart},
    month = apr,
    year = {2026},
    note = {arXiv:2604.20510},
}

@article{youngBenchmarking2025,
    title = {Benchmarking low-power flopping-mode spin-qubit fidelities in {Si}/{Si0}.{7Ge0}.3 devices with alloy disorder},
    volume = {24},
    url = {https://link.aps.org/doi/10.1103/j4ww-lt1l},
    doi = {10.1103/j4ww-lt1l},
    number = {6},
    journal = {Phys. Rev. Appl.},
    author = {Young, Steve and Brickson, Mitchell and Petta, Jason R. and Jacobson, N. Tobias},
    month = dec,
    year = {2025},
    pages = {064042},
}

@article{tsoukalas2026,
    title = {A dressed singlet-triplet qubit in germanium},
    volume = {17},
    copyright = {2025 The Author(s)},
    issn = {2041-1723},
    url = {https://www.nature.com/articles/s41467-025-65569-3},
    doi = {10.1038/s41467-025-65569-3},
    language = {en},
    number = {1},
    urldate = {2026-05-04},
    journal = {Nature Communications},
    author = {Tsoukalas, K. and von Lüpke, U. and Orekhov, A. and Hetényi, B. and Seidler, I. and Sommer, L. and Kelly, E. G. and Massai, L. and Aldeghi, M. and Pita-Vidal, M. and Hendrickx, N. W. and Bedell, S. W. and Paredes, S. and Schupp, F. J. and Mergenthaler, M. and Salis, G. and Fuhrer, A. and Harvey-Collard, P.},
    month = jan,
    year = {2026},
    pages = {699},
}

@misc{seidler2025,
    title = {Spatial uniformity of g-tensor and spin-orbit interaction in germanium hole spin qubits},
    url = {http://arxiv.org/abs/2510.03125},
    doi = {10.48550/arXiv.2510.03125},
    urldate = {2026-05-04},
    publisher = {arXiv},
    author = {Seidler, Inga and Hetényi, Bence and Sommer, Lisa and Massai, Leonardo and Tsoukalas, Konstantinos and Kelly, Eoin G. and Orekhov, Alexei and Aldeghi, Michele and Bedell, Stephen W. and Paredes, Stephan and Schupp, Felix J. and Mergenthaler, Matthias and Salis, Gian and Fuhrer, Andreas and Harvey-Collard, Patrick},
    month = oct,
    year = {2025},
    note = {arXiv:2510.03125},
}

@article{john_robust_2025,
    title = {Robust and localised control of a 10-spin qubit array in germanium},
    volume = {16},
    copyright = {2025 The Author(s)},
    issn = {2041-1723},
    url = {https://www.nature.com/articles/s41467-025-65577-3},
    doi = {10.1038/s41467-025-65577-3},
    language = {en},
    number = {1},
    urldate = {2026-05-07},
    journal = {Nature Communications},
    author = {John, Valentin and Yu, Cécile X. and van Straaten, Barnaby and Rodríguez-Mena, Esteban A. and Rodríguez, Mauricio and Oosterhout, Stefan D. and Stehouwer, Lucas E. A. and Scappucci, Giordano and Rimbach-Russ, Maximilian and Bosco, Stefano and Borsoi, Francesco and Niquet, Yann-Michel and Veldhorst, Menno},
    month = nov,
    year = {2025},
    pages = {10560},
}

@article{crippa_electrical_2018,
    title = {Electrical {Spin} {Driving} by g -{Matrix} {Modulation} in {Spin}-{Orbit} {Qubits}},
    volume = {120},
    issn = {0031-9007, 1079-7114},
    url = {https://link.aps.org/doi/10.1103/PhysRevLett.120.137702},
    doi = {10.1103/PhysRevLett.120.137702},
    language = {en},
    number = {13},
    urldate = {2026-05-07},
    journal = {Physical Review Letters},
    author = {Crippa, Alessandro and Maurand, Romain and Bourdet, Léo and Kotekar-Patil, Dharmraj and Amisse, Anthony and Jehl, Xavier and Sanquer, Marc and Laviéville, Romain and Bohuslavskyi, Heorhii and Hutin, Louis and Barraud, Sylvain and Vinet, Maud and Niquet, Yann-Michel and De Franceschi, Silvano},
    month = mar,
    year = {2018},
    pages = {137702},
}

@article{massai_impact_2024,
    title = {Impact of interface traps on charge noise and low-density transport properties in {Ge}/{SiGe} heterostructures},
    volume = {5},
    copyright = {2024 The Author(s)},
    issn = {2662-4443},
    url = {https://www.nature.com/articles/s43246-024-00563-8},
    doi = {10.1038/s43246-024-00563-8},
    language = {en},
    number = {1},
    urldate = {2026-05-08},
    journal = {Communications Materials},
    author = {Massai, Leonardo and Hetényi, Bence and Mergenthaler, Matthias and Schupp, Felix J. and Sommer, Lisa and Paredes, Stephan and Bedell, Stephen W. and Harvey-Collard, Patrick and Salis, Gian and Fuhrer, Andreas and Hendrickx, Nico W.},
    month = aug,
    year = {2024},
    pages = {151},
}

@misc{geyer_-situ_2025,
    title = {In-situ control of hole-spin driving mechanisms},
    url = {http://arxiv.org/abs/2512.19467},
    doi = {10.48550/arXiv.2512.19467},
    urldate = {2026-05-08},
    publisher = {arXiv},
    author = {Geyer, Simon and Eggli, Rafael S. and Santos, Carlos dos and Carballido, Miguel J. and Stano, Peter and Loss, Daniel and Zumbühl, Dominik M. and Warburton, Richard J. and Kuhlmann, Andreas V.},
    month = dec,
    year = {2025},
    note = {arXiv:2512.19467},
}

@article{noirot_coherence_2026,
    title = {Coherence of a hole-spin flopping-mode qubit in a circuit quantum electrodynamics environment},
    copyright = {2026 The Author(s), under exclusive licence to Springer Nature Limited},
    issn = {1745-2481},
    url = {https://www.nature.com/articles/s41567-026-03262-y},
    doi = {10.1038/s41567-026-03262-y},
    language = {en},
    urldate = {2026-05-08},
    journal = {Nature Physics},
    author = {Noirot, Léo and Yu, Cécile X. and Abadillo-Uriel, José C. and Dumur, Etienne and Niebojewski, Heimanu and Bertrand, Benoit and Maurand, Romain and Zihlmann, Simon},
    month = may,
    year = {2026},
    pages = {1--7},
}

@article{hu2023,
    title = {Flopping-mode spin qubit in a {Si}-{MOS} quantum dot},
    volume = {122},
    issn = {0003-6951, 1077-3118},
    url = {https://pubs.aip.org/apl/article/122/13/134002/2881059/Flopping-mode-spin-qubit-in-a-Si-MOS-quantum-dot},
    doi = {10.1063/5.0137259},
    language = {en},
    number = {13},
    urldate = {2026-05-11},
    journal = {Applied Physics Letters},
    author = {Hu, Rui-Zi and Ma, Rong-Long and Ni, Ming and Zhou, Yuan and Chu, Ning and Liao, Wei-Zhu and Kong, Zhen-Zhen and Cao, Gang and Wang, Gui-Lei and Li, Hai-Ou and Guo, Guo-Ping},
    month = mar,
    year = {2023},
    pages = {134002},
}

@article{rojas-arias2025,
    title = {The origins of noise in the {Zeeman} splitting of spin qubits in natural-silicon devices},
    volume = {12},
    copyright = {2025 The Author(s)},
    issn = {2056-6387},
    url = {https://www.nature.com/articles/s41534-025-01150-6},
    doi = {10.1038/s41534-025-01150-6},
    language = {en},
    number = {1},
    urldate = {2026-05-12},
    journal = {npj Quantum Information},
    author = {Rojas-Arias, Juan S. and Kojima, Yohei and Takeda, Kenta and Stano, Peter and Nakajima, Takashi and Yoneda, Jun and Noiri, Akito and Kobayashi, Takashi and Loss, Daniel and Tarucha, Seigo},
    month = dec,
    year = {2025},
    pages = {9},
}

@article{stehouwer2025,
    title = {Exploiting strained epitaxial germanium for scaling low-noise spin qubits at the micrometre scale},
    volume = {24},
    copyright = {2025 The Author(s)},
    issn = {1476-4660},
    url = {https://www.nature.com/articles/s41563-025-02276-w},
    doi = {10.1038/s41563-025-02276-w},
    language = {en},
    number = {12},
    urldate = {2026-05-12},
    journal = {Nature Materials},
    author = {Stehouwer, Lucas E. A. and Yu, Cécile X. and van Straaten, Barnaby and Tosato, Alberto and John, Valentin and Degli Esposti, Davide and Elsayed, Asser and Costa, Davide and Oosterhout, Stefan D. and Hendrickx, Nico W. and Veldhorst, Menno and Borsoi, Francesco and Scappucci, Giordano},
    month = dec,
    year = {2025},
    pages = {1906--1912},
}

@article{unseld_baseband_2025,
    title = {Baseband control of single-electron silicon spin qubits in two dimensions},
    volume = {16},
    copyright = {2025 The Author(s)},
    issn = {2041-1723},
    url = {https://www.nature.com/articles/s41467-025-60351-x},
    doi = {10.1038/s41467-025-60351-x},
    language = {en},
    number = {1},
    urldate = {2026-05-17},
    journal = {Nature Communications},
    author = {Unseld, Florian K. and Undseth, Brennan and Raymenants, Eline and Matsumoto, Yuta and de Snoo, Sander L. and Karwal, Saurabh and Pietx-Casas, Oriol and Ivlev, Alexander S. and Meyer, Marcel and Sammak, Amir and Veldhorst, Menno and Scappucci, Giordano and Vandersypen, Lieven M. K.},
    month = jul,
    year = {2025},
    pages = {5605},
}

@article{bylander2011,
    title = {Noise spectroscopy through dynamical decoupling with a superconducting flux qubit},
    volume = {7},
    copyright = {2011 Springer Nature Limited},
    issn = {1745-2481},
    url = {https://www.nature.com/articles/nphys1994},
    doi = {10.1038/nphys1994},
    language = {en},
    number = {7},
    urldate = {2026-05-20},
    journal = {Nature Physics},
    author = {Bylander, Jonas and Gustavsson, Simon and Yan, Fei and Yoshihara, Fumiki and Harrabi, Khalil and Fitch, George and Cory, David G. and Nakamura, Yasunobu and Tsai, Jaw-Shen and Oliver, William D.},
    month = jul,
    year = {2011},
    pages = {565--570},
}

@article{kelly_capacitive_2023,
    title = {Capacitive crosstalk in gate-based dispersive sensing of spin qubits},
    volume = {123},
    issn = {0003-6951, 1077-3118},
    url = {https://pubs.aip.org/apl/article/123/26/262104/2931555/Capacitive-crosstalk-in-gate-based-dispersive},
    doi = {10.1063/5.0177857},
    language = {en},
    number = {26},
    urldate = {2026-06-08},
    journal = {Applied Physics Letters},
    author = {Kelly, Eoin G. and Orekhov, Alexei and Hendrickx, Nico W. and Mergenthaler, Matthias and Schupp, Felix J. and Paredes, Stephan and Eggli, Rafael S. and Kuhlmann, Andreas V. and Harvey-Collard, Patrick and Fuhrer, Andreas and Salis, Gian},
    month = dec,
    year = {2023},
    pages = {262104},
}

@article{undseth_hotter_2023,
    title = {Hotter is {Easier}: {Unexpected} {Temperature} {Dependence} of {Spin} {Qubit} {Frequencies}},
    volume = {13},
    issn = {2160-3308},
    shorttitle = {Hotter is {Easier}},
    url = {https://link.aps.org/doi/10.1103/PhysRevX.13.041015},
    doi = {10.1103/PhysRevX.13.041015},
    number = {4},
    urldate = {2026-06-08},
    journal = {Physical Review X},
    author = {Undseth, Brennan and Pietx-Casas, Oriol and Raymenants, Eline and Mehmandoost, Mohammad and Madzik, Mateusz T. and Philips, Stephan G.J. and De Snoo, Sander L. and Michalak, David J. and Amitonov, Sergey V. and Tryputen, Larysa and Wuetz, Brian Paquelet and Fezzi, Viviana and Esposti, Davide Degli and Sammak, Amir and Scappucci, Giordano and Vandersypen, Lieven M.K.},
    month = oct,
    year = {2023},
    pages = {041015},
}

@article{undseth_nonlinear_2023,
    title = {Nonlinear {Response} and {Crosstalk} of {Electrically} {Driven} {Silicon} {Spin} {Qubits}},
    volume = {19},
    issn = {2331-7019},
    url = {https://link.aps.org/doi/10.1103/PhysRevApplied.19.044078},
    doi = {10.1103/PhysRevApplied.19.044078},
    language = {en},
    number = {4},
    urldate = {2026-06-11},
    journal = {Physical Review Applied},
    author = {Undseth, Brennan and Xue, Xiao and Mehmandoost, Mohammad and Rimbach-Russ, Maximilian and Eendebak, Pieter T. and Samkharadze, Nodar and Sammak, Amir and Dobrovitski, Viatcheslav V. and Scappucci, Giordano and Vandersypen, Lieven M.K.},
    month = apr,
    year = {2023},
    pages = {044078},
}

@article{eggli_coupling_2025,
    title = {Coupling a high-{Q} resonator to a spin qubit with all-electrical control},
    volume = {7},
    issn = {2643-1564},
    url = {https://link.aps.org/doi/10.1103/PhysRevResearch.7.013197},
    doi = {10.1103/PhysRevResearch.7.013197},
    language = {en},
    number = {1},
    urldate = {2026-06-11},
    journal = {Physical Review Research},
    author = {Eggli, Rafael S. and Patlatiuk, Taras and Kelly, Eoin G. and Orekhov, Alexei and Salis, Gian and Warburton, Richard J. and Zumbühl, Dominik M. and Kuhlmann, Andreas V.},
    month = feb,
    year = {2025},
    pages = {013197},
}

@misc{kelly_identifying_2025,
    title = {Identifying and mitigating errors in hole spin qubit readout},
    url = {https://arxiv.org/abs/2504.06898v1},
    doi={10.48550/arXiv.2504.06898},
    urldate = {2026-06-11},
    journal = {arXiv.org},
    author = {Kelly, Eoin Gerard and Massai, Leonardo and Hetényi, Bence and Pita-Vidal, Marta and Orekhov, Alexei and Carlsson, Cornelius and Seidler, Inga and Tsoukalas, Konstantinos and Sommer, Lisa and Aldeghi, Michele and Bedell, Stephen W. and Paredes, Stephan and Schupp, Felix J. and Mergenthaler, Matthias and Fuhrer, Andreas and Salis, Gian and Harvey-Collard, Patrick},
    month = apr,
    year = {2025},
    note = {arXiv:2504.06898}
}

@article{bluhm_dephasing_2011,
    title = {Dephasing time of {GaAs} electron-spin qubits coupled to a nuclear bath exceeding 200 {$\mu$}s},
    volume = {7},
    copyright = {2010 Springer Nature Limited},
    issn = {1745-2481},
    url = {https://www.nature.com/articles/nphys1856},
    doi = {10.1038/nphys1856},
    language = {en},
    number = {2},
    urldate = {2026-06-12},
    journal = {Nature Physics},
    author = {Bluhm, Hendrik and Foletti, Sandra and Neder, Izhar and Rudner, Mark and Mahalu, Diana and Umansky, Vladimir and Yacoby, Amir},
    month = feb,
    year = {2011},
    pages = {109--113},
}

@misc{tsoukalas_resonant_2025,
    title = {Resonant two-qubit gates for fermionic simulations with spin qubits},
    url = {http://arxiv.org/abs/2507.13781},
    doi = {10.48550/arXiv.2507.13781},
    urldate = {2026-06-17},
    publisher = {arXiv},
    author = {Tsoukalas, Konstantinos and Orekhov, Alexei and Hetényi, Bence and Lüpke, Uwe von and Arunseangroj, Jeth and Seidler, Inga and Sommer, Lisa and Kelly, Eoin G. and Massai, Leonardo and Aldeghi, Michele and Pita-Vidal, Marta and Bedell, Stephen W. and Paredes, Stephan and Schupp, Felix J. and Mergenthaler, Matthias and Salis, Gian and Fuhrer, Andreas and Harvey-Collard, Patrick},
    month = jul,
    year = {2025},
    note = {arXiv:2507.13781},
}

@article{yu_strong_2023,
    title = {Strong coupling between a photon and a hole spin in silicon},
    volume = {18},
    copyright = {2023 The Author(s), under exclusive licence to Springer Nature Limited},
    issn = {1748-3395},
    url = {https://www.nature.com/articles/s41565-023-01332-3},
    doi = {10.1038/s41565-023-01332-3},
    language = {en},
    number = {7},
    urldate = {2026-06-19},
    journal = {Nature Nanotechnology},
    author = {Yu, Cécile X. and Zihlmann, Simon and Abadillo-Uriel, José C. and Michal, Vincent P. and Rambal, Nils and Niebojewski, Heimanu and Bedecarrats, Thomas and Vinet, Maud and Dumur, Etienne and Filippone, Michele and Bertrand, Benoit and De Franceschi, Silvano and Niquet, Yann-Michel and Maurand, Romain},
    month = jul,
    year = {2023},
    pages = {741--746},
}

@article{teske_flopping-mode_2023,
    title = {Flopping-mode electron dipole spin resonance in the strong-driving regime},
    volume = {107},
    issn = {2469-9950, 2469-9969},
    url = {https://link.aps.org/doi/10.1103/PhysRevB.107.035302},
    doi = {10.1103/PhysRevB.107.035302},
    language = {en},
    number = {3},
    urldate = {2026-06-20},
    journal = {Physical Review B},
    author = {Teske, Julian D. and Butt, Friederike and Cerfontaine, Pascal and Burkard, Guido and Bluhm, Hendrik},
    month = jan,
    year = {2023},
    pages = {035302},
}

@thesis{massai_spin_2026,
	title = {Spin Qubits in Germanium: Driving through Noise and Anisotropies},
	rights = {metadata-only},
	url = {https://infoscience.epfl.ch/handle/20.500.14299/262676},
	shorttitle = {Spin Qubits in Germanium},
	institution = {Lausanne, {EPFL}},
	type = {phdthesis},
	author = {Massai, Leonardo},
	urldate = {2026-06-22},
	date = {2026},
	langid = {english},
	doi = {10.5075/EPFL-THESIS-11800},
}

@article{harvey-collard_high-fidelity_2018,
    title = {High-{Fidelity} {Single}-{Shot} {Readout} for a {Spin} {Qubit} via an {Enhanced} {Latching} {Mechanism}},
    volume = {8},
    issn = {2160-3308},
    url = {https://link.aps.org/doi/10.1103/PhysRevX.8.021046},
    doi = {10.1103/PhysRevX.8.021046},
    language = {en},
    number = {2},
    urldate = {2026-06-22},
    journal = {Physical Review X},
    author = {Harvey-Collard, Patrick and D’Anjou, Benjamin and Rudolph, Martin and Jacobson, N. Tobias and Dominguez, Jason and Ten Eyck, Gregory A. and Wendt, Joel R. and Pluym, Tammy and Lilly, Michael P. and Coish, William A. and Pioro-Ladrière, Michel and Carroll, Malcolm S.},
    month = may,
    year = {2018},
    pages = {021046},
}

@article{terrazos_theory_2021,
    title = {Theory of hole-spin qubits in strained germanium quantum dots},
    volume = {103},
    issn = {2469-9950, 2469-9969},
    url = {https://link.aps.org/doi/10.1103/PhysRevB.103.125201},
    doi = {10.1103/PhysRevB.103.125201},
    language = {en},
    number = {12},
    urldate = {2026-06-30},
    journal = {Physical Review B},
    author = {Terrazos, L. A. and Marcellina, E. and Wang, Zhanning and Coppersmith, S. N. and Friesen, Mark and Hamilton, A. R. and Hu, Xuedong and Koiller, Belita and Saraiva, A. L. and Culcer, Dimitrie and Capaz, Rodrigo B.},
    month = mar,
    year = {2021},
    pages = {125201},
}

@article{yu_optimising_2026,
    title = {Optimising germanium hole spin qubits with a room-temperature magnet},
    copyright = {2026 The Author(s)},
    issn = {2399-3650},
    url = {https://www.nature.com/articles/s42005-026-02634-3},
    doi = {10.1038/s42005-026-02634-3},
    language = {en},
    urldate = {2026-07-01},
    journal = {Communications Physics},
    author = {Yu, Cécile X. and van Straaten, Barnaby and Ivlev, Alexander S. and John, Valentin and Crielaard, Damien R. and Oosterhout, Stefan D. and Stehouwer, Lucas E. A. and Borsoi, Francesco and Scappucci, Giordano and Veldhorst, Menno},
    month = apr,
    year = {2026},
}

@misc{zeng_high-fidelity_2026,
    title = {High-{Fidelity} {Hole} {Spin} {Qubits} {Reveal} {Quadrupolar} {Nuclear}-{Bath} {Dynamics} in {Isotopically} {Purified} {Planar} {Germanium}},
    url = {http://arxiv.org/abs/2606.28695},
    doi = {10.48550/arXiv.2606.28695},
    urldate = {2026-07-03},
    publisher = {arXiv},
    author = {Zeng, Jian and Tan, Xiangjun and Wang, Hongzhang and Zhang, Yulei and Bian, Wendong and Wu, Lingting and Yang, Chenggang and Guo, Zhengshan and Li, Jiankun and Wang, Yongfeng and Lu, Jun and Luo, Jun-Wei and Pei, Tian},
    month = jun,
    year = {2026},
    note = {arXiv:2606.28695},
}

@article{zhang_universal_2025,
    title = {Universal control of four singlet–triplet qubits},
    volume = {20},
    copyright = {2024 The Author(s)},
    issn = {1748-3395},
    url = {https://www.nature.com/articles/s41565-024-01817-9},
    doi = {10.1038/s41565-024-01817-9},
    language = {en},
    number = {2},
    urldate = {2026-07-06},
    journal = {Nature Nanotechnology},
    author = {Zhang, Xin and Morozova, Elizaveta and Rimbach-Russ, Maximilian and Jirovec, Daniel and Hsiao, Tzu-Kan and Fariña, Pablo Cova and Wang, Chien-An and Oosterhout, Stefan D. and Sammak, Amir and Scappucci, Giordano and Veldhorst, Menno and Vandersypen, Lieven M. K.},
    month = feb,
    year = {2025},
    pages = {209--215},
}

@misc{nguyen_degenerate_2026,
    title = {A {Degenerate} {Singlet}-{Triplet} {Qubit} with {All}-{Electrical} {Orthogonal} {Control}},
    url = {http://arxiv.org/abs/2607.27067},
    doi = {10.48550/arXiv.2607.27067},
    urldate = {2026-07-30},
    publisher = {arXiv},
    author = {Nguyen, Phuong X. and Tsoukalas, Konstantinos and Ungerer, Jann H. and Santen, Julian and John, Valentin and Oosterhout, Stefan D. and Stehouwer, Lucas and Bosco, Stefano and Scappucci, Giordano and Veldhorst, Menno and Yacoby, Amir},
    month = jul,
    year = {2026},
    note = {arXiv:2607.27067},
}

@article{dijkema_cavity-mediated_2025,
    title = {Cavity-mediated {iSWAP} oscillations between distant spins},
    volume = {21},
    copyright = {2024 The Author(s)},
    issn = {1745-2481},
    url = {https://www.nature.com/articles/s41567-024-02694-8},
    doi = {10.1038/s41567-024-02694-8},
    language = {en},
    number = {1},
    urldate = {2026-08-02},
    journal = {Nature Physics},
    publisher = {Nature Publishing Group},
    author = {Dijkema, Jurgen and Xue, Xiao and Harvey-Collard, Patrick and Rimbach-Russ, Maximilian and de Snoo, Sander L. and Zheng, Guoji and Sammak, Amir and Scappucci, Giordano and Vandersypen, Lieven M. K.},
    month = jan,
    year = {2025},
    pages = {168--174},
}

@article{zwanenburg_single-qubit_2025,
    title = {Single-qubit gates beyond the rotating-wave approximation for strongly anharmonic low-frequency qubits},
    volume = {7},
    issn = {2643-1564},
    url = {https://link.aps.org/doi/10.1103/z62h-kcnh},
    doi = {10.1103/z62h-kcnh},
    language = {en},
    number = {4},
    urldate = {2026-08-02},
    journal = {Physical Review Research},
    author = {Zwanenburg, Martijn F. S. and Singh, Siddharth and Huang, Eugene Y. and Yilmaz, Figen and Stefanski, Taryn V. and Hu, Jinlun and Kumaravadivel, Piranavan and Andersen, Christian Kraglund},
    month = dec,
    year = {2025},
    pages = {043290},
}

@article{magesan_scalable_2011,
    title = {Scalable and {Robust} {Randomized} {Benchmarking} of {Quantum} {Processes}},
    volume = {106},
    copyright = {http://link.aps.org/licenses/aps-default-license},
    issn = {0031-9007, 1079-7114},
    url = {https://link.aps.org/doi/10.1103/PhysRevLett.106.180504},
    doi = {10.1103/PhysRevLett.106.180504},
    language = {en},
    number = {18},
    urldate = {2026-08-02},
    journal = {Physical Review Letters},
    author = {Magesan, Easwar and Gambetta, J. M. and Emerson, Joseph},
    month = may,
    year = {2011},
    pages = {180504},
}

@article{petersson2010,
    title = {Quantum {Coherence} in a {One}-{Electron} {Semiconductor} {Charge} {Qubit}},
    volume = {105},
    url = {https://link.aps.org/doi/10.1103/PhysRevLett.105.246804},
    doi = {10.1103/PhysRevLett.105.246804},
    number = {24},
    urldate = {2026-04-29},
    journal = {Physical Review Letters},
    publisher = {American Physical Society},
    author = {Petersson, K. D. and Petta, J. R. and Lu, H. and Gossard, A. C.},
    month = dec,
    year = {2010},
    pages = {246804},
}

@article{hayashi2003b,
    title = {Coherent {Manipulation} of {Electronic} {States} in a {Double} {Quantum} {Dot}},
    volume = {91},
    copyright = {http://link.aps.org/licenses/aps-default-license},
    issn = {0031-9007, 1079-7114},
    url = {https://link.aps.org/doi/10.1103/PhysRevLett.91.226804},
    doi = {10.1103/PhysRevLett.91.226804},
    language = {en},
    number = {22},
    urldate = {2026-08-03},
    journal = {Physical Review Letters},
    author = {Hayashi, T. and Fujisawa, T. and Cheong, H. D. and Jeong, Y. H. and Hirayama, Y.},
    month = nov,
    year = {2003},
    pages = {226804},
}

@article{vanriggelen-doelman2024,
    title = {Coherent spin qubit shuttling through germanium quantum dots},
    volume = {15},
    copyright = {2024 The Author(s)},
    issn = {2041-1723},
    url = {https://www.nature.com/articles/s41467-024-49358-y},
    doi = {10.1038/s41467-024-49358-y},
    language = {en},
    number = {1},
    urldate = {2026-08-05},
    journal = {Nature Communications},
    publisher = {Nature Publishing Group},
    author = {van Riggelen-Doelman, Floor and Wang, Chien-An and de Snoo, Sander L. and Lawrie, William I. L. and Hendrickx, Nico W. and Rimbach-Russ, Maximilian and Sammak, Amir and Scappucci, Giordano and Déprez, Corentin and Veldhorst, Menno},
    month = jul,
    year = {2024},
    pages = {5716},
}

@article{sommer_disentangling_2026,
    title = {Disentangling {Orbital} and {Confinement} {Contributions} to \textit{g} -{Factor} in {Ge}/{SiGe} {Hole} {Quantum} {Dots}},
    volume = {26},
    copyright = {https://creativecommons.org/licenses/by/4.0/},
    issn = {1530-6984, 1530-6992},
    url = {https://pubs.acs.org/doi/10.1021/acs.nanolett.6c00999},
    doi = {10.1021/acs.nanolett.6c00999},
    language = {en},
    number = {25},
    urldate = {2026-08-07},
    journal = {Nano Letters},
    author = {Sommer, L. and Seidler, I. and Schupp, F. J. and Paredes, S. and Hendrickx, N. W. and Massai, L. and Tsoukalas, K. and Orekhov, A. and Kelly, E. G. and Bedell, S. W. and Salis, G. and Mergenthaler, M. and Harvey-Collard, P. and Fuhrer, A. and Ihn, T.},
    month = jul,
    year = {2026},
    pages = {8117--8123},
}

@article{lawrie_simultaneous_2023,
    title = {Simultaneous single-qubit driving of semiconductor spin qubits at the fault-tolerant threshold},
    volume = {14},
    copyright = {2023 The Author(s)},
    issn = {2041-1723},
    url = {https://www.nature.com/articles/s41467-023-39334-3},
    doi = {10.1038/s41467-023-39334-3},
    language = {en},
    number = {1},
    urldate = {2026-08-10},
    journal = {Nature Communications},
    publisher = {Nature Publishing Group},
    author = {Lawrie, W. I. L. and Rimbach-Russ, M. and Riggelen, F. van and Hendrickx, N. W. and Snoo, S. L. de and Sammak, A. and Scappucci, G. and Helsen, J. and Veldhorst, M.},
    month = jun,
    year = {2023},
    pages = {3617},
}

@article{abraham_digitally_2026,
    title = {A digitally controlled silicon quantum processing unit},
    volume = {655},
    copyright = {2026 The Author(s)},
    issn = {1476-4687},
    url = {https://www.nature.com/articles/s41586-026-10754-7},
    doi = {10.1038/s41586-026-10754-7},
    language = {en},
    number = {8125},
    urldate = {2026-08-10},
    journal = {Nature},
    publisher = {Nature Publishing Group},
    author = {{Members of the HRL Quantum Team and Collaborators}},
    month = jul,
    year = {2026},
    pages = {1154--1159},
}

@article{carballido_compromise-free_2025,
    title = {Compromise-free scaling of qubit speed and coherence},
    volume = {16},
    copyright = {2025 The Author(s)},
    issn = {2041-1723},
    url = {https://www.nature.com/articles/s41467-025-62614-z},
    doi = {10.1038/s41467-025-62614-z},
    language = {en},
    number = {1},
    urldate = {2026-08-10},
    journal = {Nature Communications},
    author = {Carballido, Miguel J. and Svab, Simon and Eggli, Rafael S. and Patlatiuk, Taras and Chevalier Kwon, Pierre and Schuff, Jonas and Kaiser, Rahel M. and Camenzind, Leon C. and Li, Ang and Ares, Natalia and Bakkers, Erik P. A. M. and Bosco, Stefano and Egues, J. Carlos and Loss, Daniel and Zumbühl, Dominik M.},
    month = aug,
    year = {2025},
    note = {Publisher: Nature Publishing Group},
    pages = {7616},
}

@misc{smet_spin_2026,
    title = {Spin qubit operations by conveyor-mode shuttling},
    url = {http://arxiv.org/abs/2606.21452},
    doi = {10.48550/arXiv.2606.21452},
    urldate = {2026-08-11},
    publisher = {arXiv},
    author = {Smet, M. De and Matsumoto, Y. and Fernández-Fernández, D. and Tryputen, L. and Snoo, S. L. de and Michalak, D. J. and Eenink, H. G. J. and Platero, G. and Bosco, S. and Scappucci, G. and Vandersypen, L. M. K.},
    month = jun,
    year = {2026},
    note = {arXiv:2606.21452 [cond-mat.mes-hall]},
}

@article{tosato2026,
    title = {A crossbar chip for benchmarking semiconductor spin qubits},
    volume = {9},
    copyright = {2026 The Author(s)},
    issn = {2520-1131},
    url = {https://www.nature.com/articles/s41928-026-01569-5},
    doi = {10.1038/s41928-026-01569-5},
    language = {en},
    number = {3},
    urldate = {2026-08-13},
    journal = {Nature Electronics},
    publisher = {Nature Publishing Group},
    author = {Tosato, Alberto and Elsayed, Asser and Poggiali, Federico and Stehouwer, Lucas Erik Adriaan and Costa, Davide and Hudson, Karina Louise and Degli Esposti, Davide and Scappucci, Giordano},
    month = mar,
    year = {2026},
    pages = {324--333},
}

@article{PhysRevLett.95.076805,
  title = {Spin Relaxation and Decoherence of Holes in Quantum Dots},
  author = {Bulaev, Denis V. and Loss, Daniel},
  journal = {Phys. Rev. Lett.},
  volume = {95},
  issue = {7},
  pages = {076805},
  numpages = {4},
  year = {2005},
  month = {Aug},
  publisher = {American Physical Society},
  doi = {10.1103/PhysRevLett.95.076805},
  url = {https://link.aps.org/doi/10.1103/PhysRevLett.95.076805}
}

@article{PhysRevB.84.195314,
  title = {Strong spin-orbit interaction and helical hole states in {Ge}/{Si} nanowires},
  author = {Kloeffel, Christoph and Trif, Mircea and Loss, Daniel},
  journal = {Phys. Rev. B},
  volume = {84},
  issue = {19},
  pages = {195314},
  numpages = {8},
  year = {2011},
  month = {Nov},
  publisher = {American Physical Society},
  doi = {10.1103/PhysRevB.84.195314},
  url = {https://link.aps.org/doi/10.1103/PhysRevB.84.195314}
}

@article{PhysRevB.85.205308,
  title = {Extraction of large valence-band energy offsets and comparison to theoretical values for strained-{Si}/strained-{Ge} type-{II} heterostructures on relaxed SiGe substrates},
  author = {Teherani, James T. and Chern, Winston and Antoniadis, Dimitri A. and Hoyt, Judy L. and Ruiz, Liliana and Poweleit, Christian D. and Men\'endez, Jos\'e},
  journal = {Phys. Rev. B},
  volume = {85},
  issue = {20},
  pages = {205308},
  numpages = {10},
  year = {2012},
  month = {May},
  publisher = {American Physical Society},
  doi = {10.1103/PhysRevB.85.205308},
  url = {https://link.aps.org/doi/10.1103/PhysRevB.85.205308}
}

@article{guilloy2016germanium,
  author    = {Guilloy, Kevin and Pauc, Nicolas and Gassenq, Alban and Niquet, Yann-Michel and Escalante, Jose-Maria and Duchemin, Ivan and Tardif, Samuel and Osvaldo Dias, Guilherme and Rouchon, Denis and Widiez, Julie and Hartmann, Jean-Michel and Geiger, Richard and Zabel, Thomas and Sigg, Hans and Faist, J{\'e}r{\^o}me and Chelnokov, Alexei and Reboud, Vincent and Calvo, Vincent},
  title     = {Germanium under High Tensile Stress: Nonlinear Dependence of Direct Band Gap vs Strain},
  journal   = {ACS Photonics},
  volume    = {3},
  number    = {10},
  pages     = {1907--1911},
  year      = {2016},
  publisher = {American Chemical Society},
  doi       = {10.1021/acsphotonics.6b00429}
}

@book{voigt1928lehrbuch,
  author    = {Voigt, Woldemar},
  title     = {Lehrbuch der Kristallphysik},
  publisher = {B. G. Teubner},
  address   = {Leipzig},
  year      = {1928},
  pages     = {739}
}

@article{reuss1929berechnung,
  author    = {Reuss, Albert},
  title     = {Berechnung der Flie{\ss}grenze von Mischkristallen auf Grund der Plastizit{\"a}tsbedingung f{\"u}r Einkristalle},
  journal   = {Zeitschrift f{\"u}r Angewandte Mathematik und Mechanik},
  volume    = {9},
  number    = {1},
  pages     = {49--58},
  year      = {1929},
  doi       = {10.1002/zamm.19290090104}
}

@article{hill1952elastic,
  author    = {Hill, Rodney},
  title     = {The Elastic Constants of Polycrystalline Aggregates},
  journal   = {Proceedings of the Physical Society. Section A},
  volume    = {65},
  number    = {5},
  pages     = {349--354},
  year      = {1952},
  publisher = {IOP Publishing},
  doi       = {10.1088/0370-1298/65/5/307}
}

@book{madelung2004semiconductors,
  author    = {Madelung, Otfried},
  title     = {Semiconductors: Data Handbook},
  edition   = {3rd},
  publisher = {Springer},
  address   = {Berlin, Heidelberg},
  year      = {2004},
  isbn      = {978-3-540-40488-0},
  doi       = {10.1007/978-3-642-18865-7}
}

@misc{morozova_observation_2026,
    title = {Observation of magnetic quantum phase crossovers in a semiconductor spin ladder},
    url = {http://arxiv.org/abs/2608.17789},
    doi = {10.48550/arXiv.2608.17789},
    urldate = {2026-08-24},
    publisher = {arXiv},
    author = {Morozova, Elizaveta and Zhang, Xin and Bhattacharya, Utso and Fariña, Pablo Cova and Jirovec, Daniel and Nico-Katz, Alexander and Oosterhout, Stefan D. and Bose, Sougato and Scappucci, Giordano and Veldhorst, Menno and Demler, Eugene and Vandersypen, Lieven M. K.},
    month = aug,
    year = {2026},
    note = {arXiv:2608.17789 [cond-mat.mes-hall]},
}

@misc{massai_engineering_2026,
    title = {Engineering two-qubit gates via anisotropic exchange in germanium spin qubits},
    url = {http://arxiv.org/abs/2608.16716},
    doi = {10.48550/arXiv.2608.16716},
    urldate = {2026-08-24},
    publisher = {arXiv},
    author = {Massai, Leonardo and Hetényi, Bence and Kelly, Eoin G. and Seidler, Inga and Tsoukalas, Konstantinos and Aldeghi, Michele and Orekhov, Alexei and Sommer, Lisa and Pita-Vidal, Marta and Lüpke, Uwe von and Paredes, Stephan and Bedell, Stephen W. and Schupp, Felix J. and Mergenthaler, Matthias and Salis, Gian and Fuhrer, Andreas and Harvey-Collard, Patrick},
    month = aug,
    year = {2026},
    note = {arXiv:2608.16716 [cond-mat.mes-hall]},
}

@article{tahan_relaxation_2014,
    title = {Relaxation of excited spin, orbital, and valley qubit states in ideal silicon quantum dots},
    volume = {89},
    copyright = {http://link.aps.org/licenses/aps-default-license},
    issn = {1098-0121, 1550-235X},
    url = {https://link.aps.org/doi/10.1103/PhysRevB.89.075302},
    doi = {10.1103/PhysRevB.89.075302},
    language = {en},
    number = {7},
    urldate = {2026-08-24},
    journal = {Physical Review B},
    author = {Tahan, Charles and Joynt, Robert},
    month = feb,
    year = {2014},
    pages = {075302},
}

@misc{valvo2025,
    title = {Electrically {Tuneable} {Variability} in {Germanium} {Hole} {Spin} {Qubits}},
    url = {https://arxiv.org/abs/2512.12702v1},
    language = {en},
    urldate = {2026-08-25},
    journal = {arXiv.org},
    author = {Valvo, Edmondo and Jakob, Michele and Del Vecchio, Patrick and Rimbach-Russ, Maximilian and Bosco, Stefano},
    month = dec,
    year = {2025},
}

@article{mauro2025,
    title = {Strain engineering in {Ge/Ge-Si} spin-qubit heterostructures},
    volume = {23},
    shorttitle = {Strain engineering in {\textless}math xmlns="http},
    doi = {10.1103/PhysRevApplied.23.024057},
    number = {2},
    journal = {Physical Review Applied},
    author = {Mauro, Lorenzo and A. Rodríguez-Mena, Esteban and Martinez, Biel and Niquet, Yann-Michel
},
    year = {2025},
}

@article{froning2021,
    title = {Strong spin-orbit interaction and $g$-factor renormalization of hole spins in {Ge}/{Si} nanowire quantum dots},
    volume = {3},
    url = {https://link.aps.org/doi/10.1103/PhysRevResearch.3.013081},
    doi = {10.1103/PhysRevResearch.3.013081},
    number = {1},
    urldate = {2026-08-25},
    journal = {Physical Review Research},
    publisher = {American Physical Society},
    author = {Froning, F. N. M. and Rančić, M. J. and Hetényi, B. and Bosco, S. and Rehmann, M. K. and Li, A. and Bakkers, E. P. A. M. and Zwanenburg, F. A. and Loss, D. and Zumbühl, D. M. and Braakman, F. R.},
    month = jan,
    year = {2021},
    pages = {013081},
}

@article{geyer2024,
    title = {Anisotropic exchange interaction of two hole-spin qubits},
    volume = {20},
    copyright = {2024 The Author(s)},
    issn = {1745-2481},
    url = {https://www.nature.com/articles/s41567-024-02481-5},
    doi = {10.1038/s41567-024-02481-5},
    language = {en},
    number = {7},
    urldate = {2026-08-25},
    journal = {Nature Physics},
    publisher = {Nature Publishing Group},
    author = {Geyer, Simon and Hetényi, Bence and Bosco, Stefano and Camenzind, Leon C. and Eggli, Rafael S. and Fuhrer, Andreas and Loss, Daniel and Warburton, Richard J. and Zumbühl, Dominik M. and Kuhlmann, Andreas V.},
    month = jul,
    year = {2024},
    pages = {1152--1157},
}

@article{baum1985,
    title = {Broadband and adiabatic inversion of a two-level system by phase-modulated pulses},
    volume = {32},
    url = {https://link.aps.org/doi/10.1103/PhysRevA.32.3435},
    doi = {10.1103/PhysRevA.32.3435},
    number = {6},
    urldate = {2026-08-25},
    journal = {Physical Review A},
    publisher = {American Physical Society},
    author = {Baum, J. and Tycko, R. and Pines, A.},
    month = dec,
    year = {1985},
    pages = {3435--3447},
}

@article{makhlin_dephasing_2004,
    title = {Dephasing of {Solid}-{State} {Qubits} at {Optimal} {Points}},
    volume = {92},
    copyright = {http://link.aps.org/licenses/aps-default-license},
    issn = {0031-9007, 1079-7114},
    url = {https://link.aps.org/doi/10.1103/PhysRevLett.92.178301},
    doi = {10.1103/PhysRevLett.92.178301},
    language = {en},
    number = {17},
    urldate = {2026-08-25},
    journal = {Physical Review Letters},
    author = {Makhlin, Yuriy and Shnirman, Alexander},
    month = apr,
    year = {2004},
    pages = {178301},
}

@article{rodriguez-mena_sweet-spot_2026,
    title = {Sweet-spot protection of hole spins in sparse arrays via spin-dependent magnetotunneling},
    volume = {113},
    issn = {2469-9950, 2469-9969},
    url = {http://arxiv.org/abs/2510.25857},
    doi = {10.1103/qmdt-4plp},
    number = {20},
    urldate = {2026-08-27},
    journal = {Physical Review B},
    author = {Rodríguez-Mena, Esteban A. and Martínez, Biel and Kalo, Ahmad Fouad and Niquet, Yann-Michel and Abadillo-Uriel, José C.},
    month = may,
    year = {2026},
    note = {arXiv:2510.25857 [cond-mat.mes-hall]},
    pages = {205303},
}

@article{bosco2024,
    title = {High-{Fidelity} {Spin} {Qubit} {Shuttling} via {Large} {Spin}-{Orbit} {Interactions}},
    volume = {5},
    doi = {10.1103/PRXQuantum.5.020353},
    number = {2},
    journal = {PRX Quantum},
    author = {Bosco, Stefano},
    year = {2024},
}


\clearpage


\renewcommand{\thefigure}{S\arabic{figure}}
\renewcommand{\theHfigure}{S\arabic{figure}}
\renewcommand{\thetable}{S\arabic{table}}
\renewcommand{\theequation}{S\arabic{equation}}
\renewcommand{\theHequation}{S\arabic{equation}}
\renewcommand{\thepage}{S\arabic{page}}
\renewcommand{\thesubsection}{S\arabic{subsection}}
\renewcommand{\theHsubsection}{S\arabic{subsection}}
\setcounter{figure}{0}
\setcounter{table}{0}
\setcounter{equation}{0}
\setcounter{page}{1} 
\setcounter{subsection}{0}


\begin{center}

\end{center}


\newpage


\onecolumngrid

\section*{Supplementary Materials for: Coherent and ultra-low-power EDSR with a flopping-mode spin qubit in germanium}



\subsection{Experimental setup}\label{sec:Experimental setup}

The measurements are performed in a Bluefors LD25 dilution refrigerator (bottom-loader) with a base temperature of around \qty{10}{\milli\kelvin}, equipped with a three-axis superconducting vector magnet (American Magnetics) with maximum range $\pm$(1,1,6) \qty{}{\tesla} for (z,x,y) directions (as defined in \autoref{fig:fig1}a) operated in direct current mode. As shown in the measurement setup schematic in Supplementary~\autoref{fig:Supple_setup}, the dc gate voltages are applied with a 24-channel QDevil QDAC-II digital-to-analog converter (DAC) through twisted pair cables made of phosphor bronze. This is followed by a low-pass filtering stage at the mixing chamber plate using a QDevil filter box (225 MHz cut-off) and on-PCB low-pass filtering (50 kHz cut-off). We note that the PCB in this experiment is not light-tight (there is no sample lid), which offers a possible route to the improvement of $T_1$. RF pulsing and readout are performed using the Quantum Machines OPX1000, passing through a total attenuation of \qty{26}{\decibel} and combined with the dc voltages through on-PCB bias tees (\qty{0.1}{\second} time constant). The source and drain of the SHT are connected to two individual Basel Precision Instruments SP 983C IV converters with a gain of \qty{1e6}{\volt/\ampere} and \qty{30}{\kilo\hertz} bandwidth. The voltage signals are subtracted and amplified using a Stanford Research Systems SR560 differential amplifier with a gain of $10^2$ and \qty{100}{\kilo\hertz} bandwidth. The \qty{50}{\ohm} output port of the amplifier is fed to a readout channel of the OPX1000 for signal integration, while the \qty{600}{\ohm} output port is fed into a Keysight 34465A digital multimeter (DMM) for acquiring dc current maps. 

\begin{figure*}[h]
    \includegraphics[width=0.65\textwidth]{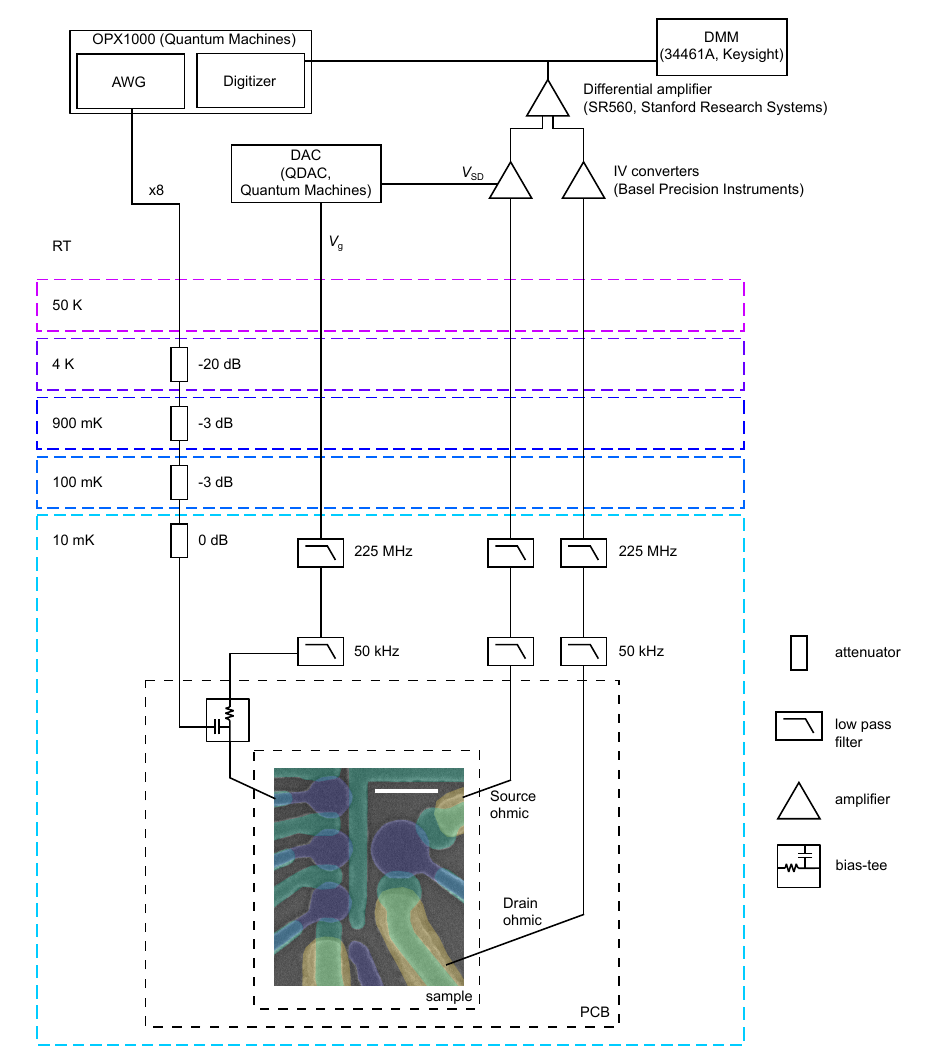}
	\caption{Measurement setup schematic.}
    \label{fig:Supple_setup}
\end{figure*}

\subsection{Initialization and readout}\label{sec:Readout and initialization}

The dc voltage setpoint is set to approximately the middle of the (0,1,1) charge region. Initialization into $\ket{\downarrow,\downarrow}_{\mathrm{QD2},\mathrm{QD3}}$ is performed by ramping with a \qty{3}{\micro\second} ramp (adiabatic with respect to the ST$^-$ anti-crossing) into the (0,0,2) charge region and waiting for \qty{15}{\micro \second} to ensure spin relaxation, before symmetrically ramping back to the (0,1,1) region. This is followed by a voltage pulse of approximately +\qty{8}{\milli\volt} on the virtual barrier $\overline{V}_{\rm B23}$, which ensures the suppression of the exchange interaction with the ancilla spin. We then ramp down the voltage on the virtual barrier $\overline{V}_{\rm B12}$ using a ramp of \qty{5000}{\nano\second}, by a value of approximately -\qty{15}{\milli\volt}, depending on the target $t_c$. This is followed by a ramp to the charge symmetry point between (1,0,1) and (0,1,1), the operating point of the FM qubit, using a \qty{500}{\nano\second} ramp that is adiabatic with respect to $t_c$. For all measurements except $T_1$ measurements:
Before performing gates we wait for a default duration of  \qty{3.5}{\micro\second}. This ensures stabilization of the voltage level. For $T_1$ measurements:
We perform an $X_\pi$ pulse immediately after ramping to the charge symmetry point. After performing the qubit operations we symmetrically ramp back to (0,1,1) and reverse the sequence of ramps on $\overline{V}_{\rm B12}$ and $\overline{V}_{\rm B23}$. This is followed by an adiabatic \qty{3}{\micro\second} ramp to the PSB region between (0,1,1) and (0,0,2), followed by a \qty{16}{\nano\second} ramp to the latched region in (0,1,2) for readout. We perform current-based SHT readout (see experimental setup section) with a \qty{50}{\micro\second} settling time, followed by a \qty{50}{\micro\second} integration time.

The readout fidelity is strongly dependent on magnetic field magnitude and orientation. This may be due to relaxation and mapping errors inherent to the latched PSB readout \cite{kelly_identifying_2025}. We do not observe any significant reduction in the visibility from detuning to the inter-dot unless the wait time is significant compared to the $T_1$ time.

\subsection{Detuning and gate calibration}\label{sec:Detuning and sweet-spot calibration}

\begin{figure*}[t]
    \includegraphics[width=0.65\textwidth]{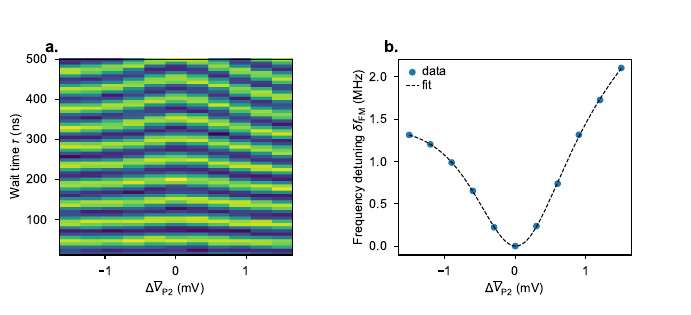}
	\caption{\textbf{a} Representative example of a Hahn echo measurement used to calibrate the detuning sweet-spot condition, with $\Delta\virttwo{}$ denoting the deviation from a previous calibration of the sweet-spot. \textbf{b} Qubit frequency detuning extracted from the data in \textbf{a} using a cosine fit. Black dashed line corresponds to a fit to the FM model Hamiltonian. The point of zero derivative of the model is used to extract the new sweet-spot condition (in this case closely coinciding with the previous calibration).  }
    \label{fig:figSweet}
\end{figure*}

The FM qubit operation point is sensitive to the inter-dot detuning which is subject to temporal drifts and jumps. We therefore rely on a modified spin-echo protocol to re-calibrate the detuning setpoint:
\begin{equation}
X_{\pi/2}~-~\Delta\virttwo{}(\tau)~-Y_\pi~-~\tau~-Z(\varphi)~-~X_{\pi/2}.
\end{equation}
Here, the rf-driven gates are performed at the FM qubit operating point ($\FMdetuning{}=0$). $\Delta \virttwo{}$ denotes the detuning from the sweet-spot, given by an adiabatic pulse on \virttwo{} of duration $\tau$. A virtual phase update $\varphi = 2\pi f_{\mathrm{virt}}\tau$ is applied before the final spin projection for easier fitting. When sweeping   $\Delta \virttwo{}$, a phase is picked up when the qubit is displaced from its initial detuning position. This results in a change in the frequency of the probability oscillations as demonstrated in a representative calibration measurement in \autoref{fig:figSweet}a. The oscillations are fitted for each detuning value, with the corresponding deviation from $f_{\mathrm{virt}}$ plotted in \autoref{fig:figSweet}b. We fit the data to Eq.~\eqref{eq:FM Hamiltonian} and use this to find the qubit frequency minimum for achieving sweet-spot operation. The zero-detuning point ($\FMdetuning{}=0$) and $t_c$ are calibrated using the same protocol, but taking a wider range of detunings ($|\FMdetuning{}|\gg t_c$), with an example shown in \autoref{fig:figS3}c.

\subsection{Magnetic field offset calibration}
The magnetic field shows hysteresis, which results in sub-mT offsets that change approximately linearly with applied magnetic field strength \cite{seidler2025, hendrickxSweetspot2024}. Due to the large out-of-plane $g$-tensor tilts (\qty{1.9}{\degree} and \qty{3.7}{\degree}) in our sample it is important to initially calibrate the offsets in all three axes of the vector magnet. The magnetic-field offsets are determined recursively for the axes
$i\in\{x,y,z\}$. The nominal field $B_i$ is swept while the remaining field components are held fixed, and the field that minimizes the qubit frequency,
$B^{+}_{\mathrm{min},i}$, is extracted from spectroscopy measurements. The signs of the two orthogonal field components are then reversed and the
measurement is repeated to obtain $B^{-}_{\mathrm{min},i}$. The offset along axis $i$ is estimated as
$B_{\mathrm{offset},i}=(B^{+}_{\mathrm{min},i}+B^{-}_{\mathrm{min},i})/2$. The procedure is performed sequentially for $z$, $y$, and $x$, and the entire sequence is repeated twice to refine the offset estimates. We find the offsets  $B_{offset, z} = \qty{-297}{\micro\tesla} $, $B_{offset, y} = \qty{-473}{\micro\tesla}$ and $B_{offset, x} = \qty{402}{\micro\tesla}$. Subsequent corrections are done using only the $z$ component as in other works.

\subsection{Flopping-mode qubit Hamiltonian}\label{sec:FM hamiltonian}
In this section we describe the comprehensive model that takes into account the $g$-tensors of QD$_1$ and QD$_2$, their gate-voltage dependence, and the spin-orbit interaction \cite{terrazos_theory_2021, hendrickxSweetspot2024, seidler2025, massai_engineering_2026}, to describe the dynamics of the FM qubit in our system. 

The Hamiltonian of an FM qubit which does not take into account the spin-flip tunneling term due to the spin-orbit interaction (SOI) is typically written as \eqref{eq: FM_hamiltonian_wo_soi} \cite{benito2019}.
\begin{equation}
H_0= \frac{\FMdetuning}{2}\tau_z+ t_\mathrm{0} \tau_x +\sum_{i=1,2}\frac{\mu_B\textbf{B}\cdot g_{i}(\virtone{}, \virttwo{})\pmb{\sigma}}{2}\frac{\tau_0+ (-1)^i \tau_z}{2} \ ,
\label{eq: FM_hamiltonian_wo_soi}
\end{equation}

Here, $\tau_i$ corresponds to the Pauli-$i$ operator in the charge basis $\lvert L \rangle = \lvert(1,0,1)\rangle$ and $\lvert R \rangle = \lvert(0,1,1)\rangle$, and $\pmb{\sigma} = (\sigma_x, \sigma_y, \sigma_z)$ is the Pauli spin operator.
$g_{1 (2)}(\virtone{}, \virttwo{})$ is the $g$-tensor of QD$_{1(2)}$ dependent on the gate-voltages \virtone{} and \virttwo{} in the lab frame, where we assume a linear dependence of $g$-tensors on the gate-voltages as shown in \eqref{eq: g_tensor_gate_dependence}.
\begin{equation}
g_{1 (2)}(\virtone{}, \virttwo{}) = g_{1 (2)}(\virtone{}^{1 (2),0}, \virttwo{}^{1 (2),0}) + \frac{\partial g_{1(2)}}{\partial \virtone} (\virtone - \virtone^{1 (2),0}) + \frac{\partial g_{1(2)}}{\partial \virttwo} (\virttwo - \virttwo^{1 (2),0})
\label{eq: g_tensor_gate_dependence}
\end{equation}
Here, $\virtone{}^{1 (2),0}, \virttwo{}^{1 (2),0}$ are the dc-voltages at which $g_{1(2)}$ were characterized by sweeping the magnetic field orientation \cite{hendrickxSweetspot2024}.
We show the extraction of $g_{1(2)}(\virtone{}^{1 (2),0}, \virttwo{}^{1 (2),0})$, $\partial g_{1(2)} / \partial \virtone$ and $\partial g_{1(2)} / \partial \virttwo$ in Supplementary~\autoref{sec:gtensor_SOI}. 

SOI results in a spin-flip tunneling term \eqref{eq: spin_flip_tunnneling} \cite{froning2021, geyer2024}.
\begin{equation}
H_\mathrm{SOI} = t_\mathrm{0} \tan(\theta_\mathrm{so}) \tau_y \mathbf{n_\mathrm{so}}\cdot \pmb{\sigma}
\label{eq: spin_flip_tunnneling}
\end{equation}
where $\mathbf{n_\mathrm{so}}$ is a spin-orbit vector whose norm is unity, $\theta_\mathrm{so}$ is the spin-orbit angle which describes the amount of the spin rotation around $\mathbf{n_\mathrm{so}}$ in case of a tunneling event. 
The effective tunnel coupling $t_\mathrm{c}$ is equivalent to $t_\mathrm{c} = \sqrt{t_0^2 + t_0^2 \tan^2 (\theta_\mathrm{so})}=t_0/\cos(\theta_\text{so})$

The total Hamiltonian of the FM qubit is written as \eqref{eq: total_FM_qubit_Hamiltonian}.

\begin{equation}
H_\mathrm{FM} = H_0 + H_\mathrm{SOI}
\label{eq: total_FM_qubit_Hamiltonian}
\end{equation}

In the spin-orbit frame, where the $g$-tensors are effectively transformed by $g_1^\mathrm{so} = g_1 R_\mathrm{so}(\theta_\mathrm{so})$ and $g_2^\mathrm{so} = g_2 R_\mathrm{so}(-\theta_\mathrm{so})$, the Hamiltonian reduces to Eq.~\eqref{eq:FM_hamiltonian_so_frame} \cite{geyer2024}. 
$R_\mathbf{so}(\theta)$ is a matrix representing the rotation around $\mathbf{n_\mathrm{so}}$ by an angle $\theta$.

\begin{equation}
H_\mathrm{FM}= \frac{\FMdetuning}{2}\tau_z+ t_\mathrm{c} \tau_x +\sum_{i=1,2}\frac{\mu_B\textbf{B}\cdot g_{i}^\mathrm{so}(\virtone{}, \virttwo{})\pmb{\sigma}}{2}\frac{\tau_0+ (-1)^i \tau_z}{2} \ ,
\label{eq:FM_hamiltonian_so_frame}
\end{equation}

We note that we have neglected contributions from magnetotunneling due to the low qubit frequencies considered here \cite{rodriguez-mena_sweet-spot_2026}. 

\subsection{$g$-tensor characterization and spin-orbit interaction measurement}\label{sec:gtensor_SOI}

\begin{figure}[h]
	\centering
    \includegraphics[width=0.5\columnwidth]{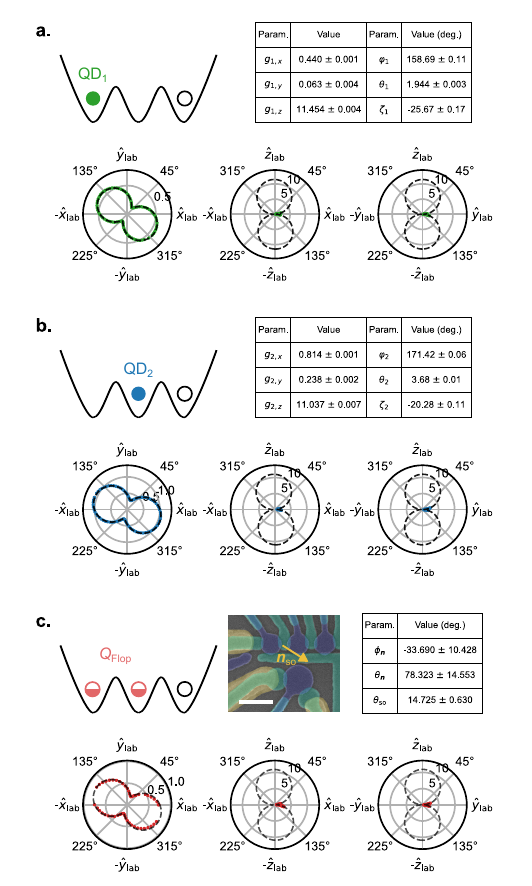}
	\caption{
    \textbf{a. (b.)} $g$-tensor characterization of the hole in QD$_1 (2)$. Bottom three panels correspond to the measured effective $g$-factor $g^* = E_\mathrm{z}/(\mu_\mathrm{B}|\mathbf{B}|)$ for different magnetic orientations in the $xy$-, $xz$- and $yz$-plane in the lab frame respectively. The black dashed line in each panel represents the fit to the $g$-tensor model. The corresponding set of parameters ($g_{1(2),x}, g_{1(2),y}, g_{1(2),z}, \varphi_{1(2)}, \theta_{1(2)}, \zeta_{1(2)}$) extracted from the fit are shown in the table in the top right panel. 
    \textbf{c.} Effective $g$-factor of the FM qubit for different magnetic field orientations in the $xy$-, $xz$- and $yz$-plane in the lab frame (bottom three panels). The black dashed line in each panel is the fit to the FM qubit model (Supplementary~\autoref{sec:FM hamiltonian}) which also takes into account the spin-orbit interaction. Yellow arrow in the device SEM (top middle panel) illustrates the direction of the fitted spin-orbit vector $\bm{n}_\mathrm{so}$. The fitted values representing the spin-orbit vector $(\phi_{\bm{n}}, \theta_{\bm{n}}, \theta_\mathrm{so})$ are presented in the table (top right panel, see text).
	}
    \label{fig:fig2}
\end{figure}

The Zeeman splitting $E_\mathrm{z}$ of a heavy-hole (HH) state is governed by an anisotropic $3 \times 3$ $g$-tensor $\mathbf{g}$, resulting in a field-direction–dependent splitting $E_\mathrm{z} = \mu_\mathrm{B} |\mathbf{B} \cdot\mathbf{g}|$.
In practice, this anisotropy is site-dependent, for instance due to strain gradient present in the material, and electrostatic confinement anisotropy \cite{terrazos_theory_2021, scappucciGermanium2021, valvo2025, seidler2025, mauro2025}.
We therefore characterize the $g$-tensor of QD$_1$ and QD$_2$ respectively, to systematically map the FM qubit operational sweet-line as a function of magnetic field direction shown in \autoref{fig:fig3}.

To characterize the $g$-tensor of QD$_{1 (2)}$, we prepare the $\lvert L (R), \downarrow \rangle$ state deep in (1,0,1) ((0,1,1)) charge configuration, by first initializing in (0,0,2)S, then adiabatically pulsing to a point in the charge stability diagram, $(\Delta\virtone{}, \Delta\virttwo{}) = (-5, +5) $~mV ($(+5, -5)$~mV) away from the charge symmetry point. The qubit frequency is measured as a function of \bfield{} direction using adiabatic rapid passage (ARP) spectroscopy with a chirped rf tone \cite{baum1985}, and fitted to $E_\mathrm{z}(\bfield) = \mu_\mathrm{B} |\bfield \cdot g_{1(2)}|$ to obtain $g_{1(2)}$ \cite{hendrickxSweetspot2024, crippa_electrical_2018}. Representative measurements of the effective $g$-factor $g^* = E_\mathrm{z}/(\mu_\mathrm{B}|\mathbf{B}|)$ of QD$_{1 (2)}$ as a function of the azimuthal ($\phi_\mathrm{lab}$) and the zenith ($\theta_\mathrm{lab}$) angles in the lab frame are shown in \autoref{fig:fig2}a(b). 
The $g$-tensor, ${g}_{1(2)}$, in its diagonalized form is given by $\mathrm{\bf{diag}}(g_{1(2),x}, g_{1(2),y}, g_{1(2),z})$, with the principal values $g_{1(2),y} < g_{1(2),x} < g_{1(2),z}$ defining the $g$-tensor frame. 
The laboratory frame $g$-tensor is obtained by applying the Euler rotations $R(\varphi_{1(2)})R(\theta_{1(2)})R(\zeta_{1(2)})$ to $\mathrm{\bf{diag}}(g_{1(2),x}, g_{1(2),y}, g_{1(2),z})$ with \textit{zyz} Euler rotation convention \cite{hendrickxSweetspot2024}.
The extracted parameters of ${g}_{1(2)}$ are summarized in the tables in \autoref{fig:fig2}a(b), while the corresponding fits are shown as black dashed lines. 
As can be inferred from the measured $E_\mathrm{z}$ along $\phi_\mathrm{lab}$ for QD$_1$ and QD$_2$ (left panel in \autoref{fig:fig2}a and b), the $g$-tensor is not uniform across the QDs despite their similar lithographic sizes and shapes (see \autoref{fig:fig1}a).
Especially, the $g$-factors along the principal axes in QD$_2$ are larger by about a factor of two compared to QD$_1$. 
We attribute the inhomogeneity in both the overall magnitude and tilt to strain induced by the ohmic contact (brown structure on the left side of QD$_1$ in \autoref{fig:fig1}a) \cite{mauro2025}, and by lattice mismatch in the material stack of the reverse-graded heterostructure employed in this work \cite{massai_spin_2026}.

To accurately capture the system Hamiltonian it is also necessary to take into account the finite gate-voltage tunability of the $g$-tensors. 
We therefore examine the dependence of each $g$-tensor on the plunger gate voltages $\virtone$ and $\virttwo$, reconstructing the full $g$-tensor derivatives $\partial g_{1(2)}/\partial \virtone$, and $\partial g_{1(2)}/\partial \virttwo$ \cite{hendrickxSweetspot2024}. In a typical Hahn-echo sequence $X_{\pi/2}-\tau-Y_{\pi}-\tau-X_{\pi/2}$, the qubit frequency is kept fixed throughout both free evolution segments represented by $\tau$. In order to evaluate $\partial \Delta E_\mathrm{z}/\partial \overline{V}_{\mathrm{P}i}$ at an arbitrary point in the charge stability diagram, one can instead detune the qubit by $(\Delta \virtone{}, \Delta \virttwo{})$ relative to the charge symmetry point during both $\tau$ segments. The qubit phase pick-up due to an incremental change in $\overline{V}_{\mathrm{P}i}$ close to the point  $(\Delta \virtone{}, \Delta \virttwo{})$ can be measured by slightly changing the amplitude of the first and second $\tau$ segment, by a value $\pm\delta V$ on $\overline{V}_{\mathrm{P}i}$, as shown in Supplementary Fig.~\eqref{fig:deriv_measurement}b. Moreover, to get a more accurate estimate, we sweep the value of $\delta V$ in a range $\pm\qty{1.75}{\milli\volt}$ and fit the result to a linear model, which gives the qubit frequency derivative.  To find $\partial \Delta E_\mathrm{z}^{1(2)}/\partial \virtone{}$, we pulse to  $(\Delta \virtone{}, \Delta \virttwo{})$ = $(\qty{+6}{\milli\volt},0$) ($(\qty{-6}{\milli\volt},0$)).  To evaluate $\partial \Delta E_\mathrm{z}^{1(2)}/\partial \virttwo$, we pulse to $(\Delta \virtone{}, \Delta \virttwo{})$ =  $(0, +\qty{6}{\milli\volt})$ ($(0, -\qty{6}{\milli\volt})$).


In this way, we measure $\partial \Delta E_z^{1(2)} / \partial \virtone$ and $\partial \Delta E_\mathrm{z}^{1(2)} / \partial \virttwo$ as a function of magnetic field direction, and fit to Eq.~\ref{eq:g_tensor_deriv_meas} to extract $\partial g_{1(2)}/\partial \virtone$.
The measured $\Delta g_{1(2)}^* = (\mu_\mathrm{B} |\bfield|)^{-1} \partial \Delta E_\mathrm{z}^{1(2)}/\partial \virtone $ are represented by green (blue) points in Supplementary Fig.~\ref{fig:deriv_measurement}, with the dashed lines representing $\Delta g_{1(2)}^*$ calculated using the fitted parameters (see Table~\ref{tab:gtensor_deriv_parameters} for the extracted parameters).

\begin{equation}
\begin{aligned}
\frac{\partial \Delta E_\mathrm{z}^{1(2)}}{\partial \overline{V}_j}(\bfield)
&=
\mu_\mathrm{B}
|\frac{\bfield \cdot (g_{1(2)}^* - g_{1(2)}^0)}
{\delta \overline{V}_j}|
\\
g_{1(2)}^*
&=
g_{1(2)}^0
+
\delta \overline{V}_j
\frac{\partial g_{1(2)}}{\partial \overline{V}_j}
\end{aligned}
\label{eq:g_tensor_deriv_meas}
\end{equation}
Here, $\overline{V}_j$ is either \virtone{} or \virttwo{}.

\begin{figure}[h]
	\centering
    \includegraphics[width=0.5\columnwidth]{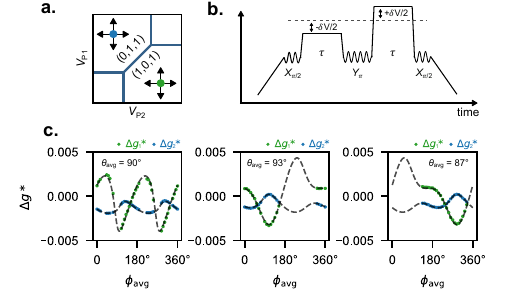}
	\caption{
    $g$-tensor derivative measurements. 
    \textbf{a.} Schematic of the $g$-tensor derivative measurement. The change of the Zeeman energies of each QD is measured by varying the gate-voltages in the direction denoted by the arrows at a fixed point in the charge stability diagram. 
    \textbf{b.} Schematic of the Hahn-echo like pulse sequence utilized for measuring the $g$-tensor derivative.
    \textbf{c.} $\Delta g_{1(2)}^*$ measured by varying \virtone{} at different magnetic field directions represented by $\phi_\mathrm{avg}$ for different values of $\theta_\mathrm{avg}$. 
	}
    \label{fig:deriv_measurement}
\end{figure}

\begin{table}
\centering
\caption{
Parameters representing the $g$-tensor derivatives
$\frac{\partial g_1}{\partial \virtone}$,
$\frac{\partial g_1}{\partial \virttwo}$,
$\frac{\partial g_2}{\partial \virtone}$, and
$\frac{\partial g_2}{\partial \virttwo}$.
Each parameter ($p'_\mathrm{i} = \partial p_i / \partial \overline{V}_j$)
is extracted by fitting the measured Zeeman splitting variation with
respect to the modulation of $\overline{V}_j$, to the model
Eq.~\ref{eq:g_tensor_deriv_meas}.
}
\label{tab:gtensor_deriv_parameters}

\renewcommand{\arraystretch}{1.25}
\setlength{\tabcolsep}{8pt}

\begin{tabular}{lcccc}
\toprule
 & $\displaystyle\frac{\partial g_1}{\partial \virtone}$
 & $\displaystyle\frac{\partial g_1}{\partial \virttwo}$
 & $\displaystyle\frac{\partial g_2}{\partial \virtone}$
 & $\displaystyle\frac{\partial g_2}{\partial \virttwo}$ \\
\midrule
$g'_\mathrm{i,x}  ~(\rm mV^{-1})$ 
& $(7.11 \pm 0.21) \times 10^{-4}$
& $(-1.05 \pm 0.02) \times 10^{-3}$
& $(-1.56 \pm 0.01) \times 10^{-3}$
& $(9.21 \pm 0.03) \times 10^{-4}$ \\

$g'_\mathrm{i,y} ~(\rm mV^{-1})$
& $(6.19 \pm 0.79) \times 10^{-4}$
& $(-8.27 \pm 0.87) \times 10^{-4}$
& $(-9.78 \pm 0.11) \times 10^{-4}$
& $(2.28 \pm 0.07) \times 10^{-4}$ \\

$g'_\mathrm{i,z} ~ (\rm mV^{-1})$
& $(6.51 \pm 0.07) \times 10^{-3}$
& $(-4.03 \pm 0.07) \times 10^{-3}$
& $(-1.36 \pm 0.03) \times 10^{-3}$
& $(7.28 \pm 0.02) \times 10^{-3}$ \\

$\varphi_i'~\rm(deg./mV)$ 
& $(4.01 \pm 0.02) \times 10^{-1}$
& $(-4.06 \pm 0.02) \times 10^{-1}$
& $(-4.83 \pm 0.04) \times 10^{-2}$
& $(-2.24 \pm 0.02) \times 10^{-2}$ \\

$\theta_i'~\rm(deg./mV)$ 
& $(5.11 \pm 0.08) \times 10^{-3}$
& $(-9.25 \pm 0.09) \times 10^{-3}$
& $(-3.05 \pm 0.04) \times 10^{-3}$
& $(-4.69 \pm 0.02) \times 10^{-3}$ \\

$\zeta_i'~\rm(deg./mV)$ 
& $(1.80 \pm 0.04) \times 10^{-1}$
& $(-2.28 \pm 0.05) \times 10^{-1}$
& $(-6.54 \pm 0.66) \times 10^{-3}$
& $(7.45 \pm 0.04) \times 10^{-2}$ \\
\bottomrule
\end{tabular}
\end{table}


Each of the $g$-tensor derivatives is represented by $(g_{1(2), x}', g_{1(2), y}', g_{1(2), z}', \phi_{1(2)}', \theta_{1(2)}', \zeta_{1(2)}')$ where $q'$ represents the derivative of the variable $q$ with respect to the gate voltage. 
Based on the extracted $\partial g_{1(2)}/\partial \virtone$ and $\partial g_{1(2)}/\partial \virttwo$, we reconstruct $g_{1(2)}(\virtone, \virttwo)$ for arbitrary $(\virtone, \virttwo)$ assuming its linear dependence on the gate-voltages, as anticipated in Eq.~\eqref{eq: g_tensor_gate_dependence}.

Having quantified the $g$-tensors and their dependence on gate voltages, we now turn to estimate the spin-orbit interaction (SOI) in the system. 
Finite SOI induces spin-flip tunneling between the two QDs (Eq.~\eqref{eq: spin_flip_tunnneling}) which also affects the FM qubit energy (Eq.~\eqref{eq: total_FM_qubit_Hamiltonian}).
We measure the FM qubit $g$-factor at the charge symmetry point ($\FMdetuning = 0$) for various \bfield{} directions, shown as red datapoints in the bottom three panels of Supplementary Fig.~\ref{fig:fig2}c. 
We estimate $g_{1(2)}(\virtone, \virttwo)$ at the charge symmetry point by linear extrapolation of the $g$-tensor values in QD$_{1(2)}$, using the measured $g$-tensor derivatives. We then fit the measured \qubitf{} to the FM qubit Hamiltonian Eq.~\eqref{eq: total_FM_qubit_Hamiltonian}, while putting the spin-orbit vector $\mathbf{n}_\mathrm{so}$ and the spin-orbit angle $\theta_\mathrm{so}$ as the fitting parameters. 
Here, $\mathbf{n}_\mathrm{so}$ is parameterized using the spherical coordinates $(\varphi_\mathbf{n}, \theta_\mathbf{n})$ as $\mathbf{n}_\mathrm{so} = (\cos(\varphi_\mathbf{n}) \sin(\theta_\mathbf{n}), \sin(\varphi_\mathbf{n}) \sin(\theta_\mathbf{n}), \cos(\theta_\mathbf{n}))$.
The black dashed lines in the bottom three panels of Supplementary Fig.~\ref{fig:fig2}c correspond to the model fit, with the fit parameters  shown in the table in the upper right panel of Supplementary Fig.~\ref{fig:fig2}c.
The spin-orbit vector $\mathbf{n}_\mathrm{so}$ is visualized as a yellow-arrow on the SEM image in the top middle panel of Supplementary Fig.~\ref{fig:fig2}.
We note that $\mathbf{n}_\mathrm{so}$ is expected to point in-plane (i.e. $\theta_\mathbf{n} = 90^\circ$) for a Rashba-type SOI, with the electric field along the out-of-plane direction. The fitted $\mathbf{n}_\mathrm{so}$ is tilting by about $12^\circ$ in the out-of-plane direction ($\theta_\mathbf{n} \sim 78.3^\circ$). 
This could possibly be explained by a small stray electric field in the in-plane direction.

\subsection{LSES and TSES calculation}\label{sec:LSES and TSES extraction}

Based on the Hamiltonian described in Supplementary~\ref{sec:FM hamiltonian}, and the parameters extracted in Supplementary~\ref{sec:gtensor_SOI}, the effective Zeeman splitting $E_\mathrm{z}(\virtone, \virttwo; \bfield)$ at an arbitrary point in the DQD stability diagram (\virtone{}, \virttwo{}) can be calculated for a given $\bfield$. 
By evaluating $E_\mathrm{z}(\virtone, \virttwo; \bfield)$ for different magnetic field directions, effective $g$-tensor $g(\virtone, \virttwo)$ is obtained, and utilized to build the FM qubit Larmor vector $\textbf{\textit{l}}(\virtone, \virttwo; \bfield)$, as reported in the main text.

The LSES of the FM qubit as a function of $\phi_\mathrm{avg}$ and $\theta_\mathrm{avg}$ with respect to $\virtone$ is presented in Supplementary Fig.~\ref{fig:Supple_LSES_TSES}a. It is numerically calculated, by first evaluating $g(\Delta\virtone = +\delta \virtone, \Delta\virttwo = 0)$, $g(\Delta\virtone = -\delta \virtone, \Delta\virttwo = 0)$ and $g_0 = g(\Delta\virtone = 0, \Delta\virttwo = 0)$. 
Then, we obtain $g_{0,\mathrm{P}1}' = \partial g_0 / \partial \virtone = (g(\Delta\virtone = +\delta \virtone, \Delta\virttwo = 0) - g(\Delta\virtone = -\delta \virtone, \Delta\virttwo = 0))/2\delta \virtone$.
Following the definition of LSES (Eq.~\ref{eq:LSES}), we obtain the LSES $\beta_{\parallel,\mathrm{P}1(2)}$ for arbitrary $\bfield$.
\begin{equation}
\beta_{\parallel,\mathrm{P}1(2)} = \frac{\mu_\mathrm{B}}{h|\bfield \cdot g_0|} (\bfield \cdot g_{0,\mathrm{P}1(2)}' ) \cdot (\bfield \cdot g_0)
\label{eq:LSES}
\end{equation}
The LSES with respect to $\virttwo$, $\beta_{\parallel,\rm P2}$ is presented in Supplementary Fig.~\ref{fig:Supple_LSES_TSES}d. We note that $\mu_\mathrm{B} \bfield \cdot g_{0,\mathrm{P}1(2)}'$ corresponds to the derivative of the Larmor vector $\frac{d\textbf{\textit{l}}}{d\overline{V}_{\mathrm{P}1(2)}}$ shown in Fig.~\ref{fig:fig3}a, with $\mu_\mathrm{B} \bfield \cdot g_0$ being $\textbf{\textit{l}}$ at the given configuration.

The TSES can be calculated using (Eq.~\ref{eq:TSES}), where the TSES with respect to $\virtone$ ($\beta_{\perp,\mathrm{P}1}$) and $\virttwo$ ($\beta_{\perp,\mathrm{P}2}$) are shown in Supplementary Fig.~\ref{fig:Supple_LSES_TSES}b and e respectively. 
\begin{equation}
\beta_{\perp,\mathrm{P}1(2)} = \frac{\mu_\mathrm{B}}{h|\bfield \cdot g_0|} |(\bfield \cdot g_{0,\mathrm{P}1(2)}' ) \times (\bfield \cdot g_0)|
\label{eq:TSES}
\end{equation}
The TSES normalized by the FM qubit frequency, $\beta_{\perp, \mathrm{P}1(2)}/f_\mathrm{FM}$ is  illustrated in Supplementary Fig.~\ref{fig:Supple_LSES_TSES}c and f respectively. 
We further show the calculated LSES, TSES and the normalized TSES of the Loss-DiVincenzo (LD) qubits in QD$_{1(2)}$ with respect to the corresponding virtual plunger gate voltages $\overline{V}_{\mathrm{P}1(2)}$  (Supplementary Fig.~\ref{fig:Supple_LSES_TSES}g-i(j-l)).
These results clearly demonstrate that the TSES of the FM qubit is two-orders-of-magnitude larger than that of the LD qubits in the same device, which allows ultra-low-power, and fast Rabi oscillation of the FM qubits.

\begin{figure*}[t]
	\centering
\includegraphics[width=\textwidth]{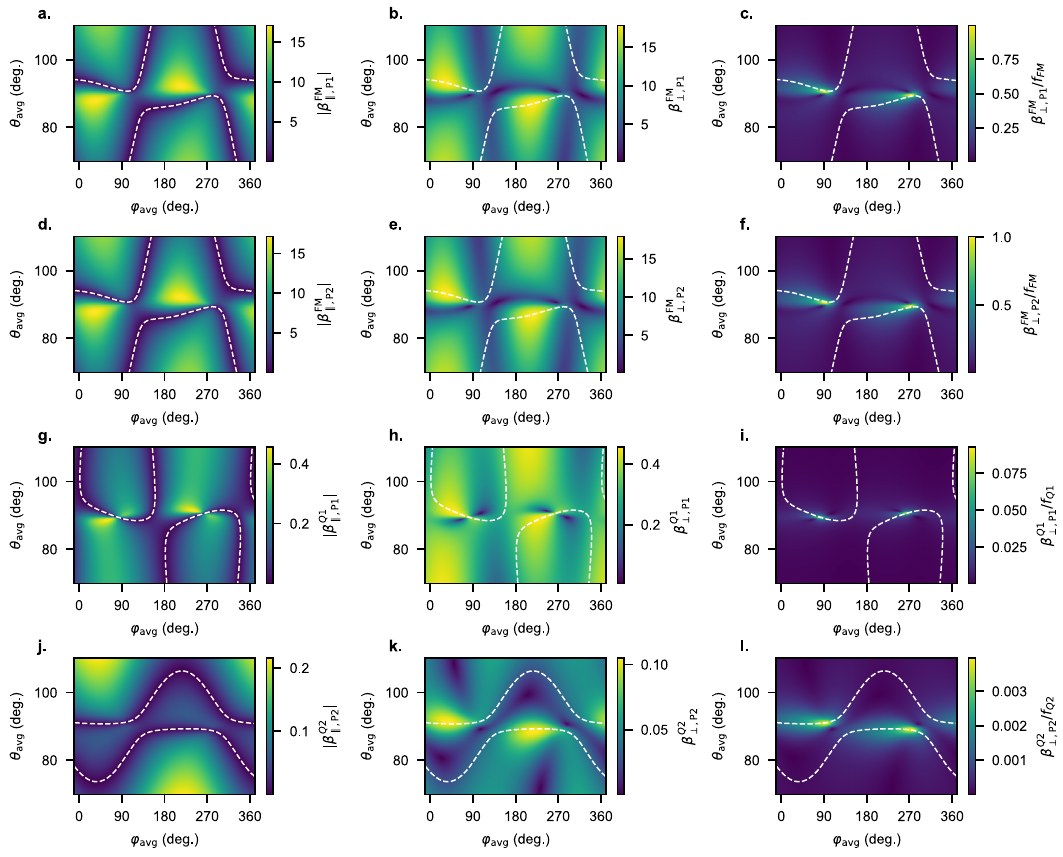}
	\caption{
    \textbf{a (d).} Calculated LSES of the FM qubit, $\beta^\mathrm{FM}_{\parallel,\mathrm{P}1(2)}$, as a function of the magnetic field angles $\varphi_\mathrm{avg}$ and $\theta_\mathrm{avg}$ with respect to \virtone{} (\virttwo{}) modulation.
    \textbf{b (e).} Calculated TSES of the FM qubit, $\beta^\mathrm{FM}_{\perp,\mathrm{P}1(2)}$ as a function of the magnetic field angle $\varphi_\mathrm{avg}$ and $\theta_\mathrm{avg}$ with respect to \virtone{} (\virttwo{}) modulation.  
    \textbf{c (f).} Calculated TSES normalized by FM qubit frequency $\beta^\mathrm{FM}_{\perp,\mathrm{P}1(2)}/f_\mathrm{FM}$ as a function of the magnetic field angle $\varphi_\mathrm{avg}$ and $\theta_\mathrm{avg}$ with respect to \virtone{} (\virttwo{}) modulation.
    \textbf{g (j).} Calculated LSES of the Loss-DiVincenzo (LD) qubit in QD$_1(2)$, $\beta^\mathrm{Q1(2)}_{\parallel,\mathrm{P}1(2)}$, as a function of the magnetic field angle $\varphi_\mathrm{avg}$ and $\theta_\mathrm{avg}$ with respect to \virtone{} (\virttwo{}) modulation.
    \textbf{h (k).} Calculated TSES of the LD qubit in QD$_1(2)$ with respect to the plunger gate voltage $P1(2)$, $\beta^\mathrm{Q1(2)}_{\perp,\mathrm{P}1(2)}$ as a function of the magnetic field angle $\varphi_\mathrm{avg}$ and $\theta_\mathrm{avg}$ with respect to \virtone{} (\virttwo{}) modulation.  
    \textbf{i (l).} Calculated TSES normalized by LD qubit frequency $\beta^\mathrm{Q1(2)}_{\perp,\mathrm{P}1(2)}/f_\mathrm{Q1(2)}$ as a function of the magnetic field angle $\varphi_\mathrm{avg}$ and $\theta_\mathrm{avg}$ with respect to \virtone{} (\virttwo{}) modulation.
    White dashed curves in each plot denote the sweet-line where the corresponding LSES becomes 0.
    }
    \label{fig:Supple_LSES_TSES}
\end{figure*}

\subsection{FM qubit coherence time extraction}\label{sec:FM qubit coherence}
Qubit dephasing times \dephasing{} and \echo{} are obtained by fitting the probability decay to the model
\begin{equation}\label{eq:decay fit}
    P(t) = A\exp\left[-\left(\frac{t}{T_2}\right)^\beta\right] \cos(2\pi f t + \varphi) + C
\end{equation}
where $\beta$ is left as a free parameter. Additional corrections to this formula can apply due to the quadratic coupling to the noise at sweet-spots, beyond the scope of this work \cite{makhlin_dephasing_2004}.

Presumably due to the quasi-static fluctuations of the Rabi frequency, we find that the Rabi quality factor is not a representative number for gate fidelity. We report an example Rabi decay trace in \autoref{fig:fig6}c for \bfield{} direction \bthree{} and the same parameters as for the RB experiment in \autoref{fig:figRB}b of the main text. The quasi-static nature of the Rabi decay can be seen by performing a rotary echo experiment, where the phase of the resonant drive is flipped by $\pi$ at half time ($t_{\mathrm{burst}}/2$). This significantly prolongs the driven coherence time $T^{\mathrm{rot}}_{2\rho}$ (\autoref{fig:fig6}d). The rotary echo data is fitted using exponential decay (with $\beta=1$). In \autoref{tab:coherence} we report the coherence times and gate parameters  for the two configurations in \autoref{fig:figRB} of the main text.


\begin{table}[t]
\centering
\caption{Qubit coherence metrics measured for the same configurations as the RB results in \autoref{fig:figRB}.}
\label{tab:coherence}

\renewcommand{\arraystretch}{1.25}
\setlength{\tabcolsep}{10pt}

\begin{tabular}{lcc}
\toprule
 & \bone{} & \bthree{} \\
\midrule
$\bmag{}$
& \qty{5}{\milli\tesla}
& \qty{5}{\milli\tesla} \\

$t_c/h$
& \qty{13}{\giga\hertz}
& \qty{13}{\giga\hertz} \\

$\qubitf{}$
& \qty{58.8}{\mega\hertz}
& \qty{33.7}{\mega\hertz} \\

$\Adrive{}$
& \qty{460}{\micro\volt}
& \qty{550}{\micro\volt} \\

$f_\mathrm{Rabi}$
& \qty{5.7}{\mega\hertz}
& \qty{3.5}{\mega\hertz} \\

$\dephasing{}$
& \qty{1.44\pm 0.01}{\micro\second}, $\beta = \qty{1.83\pm 0.05}{}$
& \qty{1.4\pm 0.02}{\micro\second}, $\beta = \qty{1.64\pm 0.04}{}$ \\

$\echo{}$
& \qty{11.28\pm 0.07}{\micro\second}, $\beta = \qty{2.1\pm 0.05}{}$
& \qty{11.5\pm 0.06}{\micro\second}, $\beta = \qty{2.42\pm 0.05}{}$ \\

$T^{\mathrm{\phi, CPMG}}_2$
& \qty{82\pm 6}{\micro\second} (24 decoupling pulses), $\beta = \qty{1.1\pm 0.1}{}$
& \qty{130\pm 3}{\micro\second} (32 decoupling pulses), $\beta = \qty{3.2\pm 0.2}{}$ \\

$T_1$
& \qty{53\pm 2}{\micro\second}
& \qty{226\pm 6}{\micro\second} \\

$T^{\mathrm{rot}}_{2\rho}$
& \qty{11.8\pm 0.2}{\micro\second}
& \qty{21.1\pm 0.5}{\micro\second} \\

$\mathcal{F}_{\rm 1qb}$
& 99.76(1)\%
& 99.74(1)\% \\

\bottomrule
\end{tabular}
\end{table}


\subsection{Flopping-mode qubit noise spectrum}\label{sec:Flopping-mode qubit noise spectrum}

\begin{figure*}[h]
	\centering
    \includegraphics[width=\textwidth]{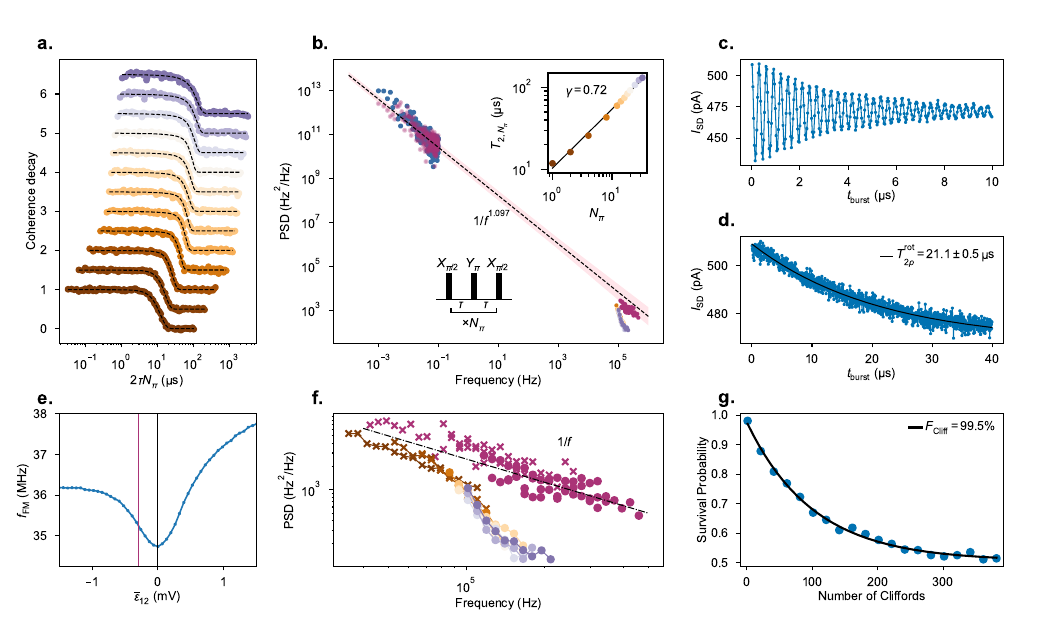} 
	\caption{Qubit characterization for the \bfield{} orientation \bthree{}, $\bmag=\qty{5}{\milli\tesla}$, $t_c/h = \qty{13}{\giga\hertz}$. \textbf{a.} CPMG decay traces for different number of $Y_{\pi}$ pulses: $N_\pi=1,2,4,8,12,14,16,18,20,24,28,32$. The x-axis corresponds to the total duration of the sequence as defined in the inset in \textbf{b.}, where $\tau$ is the varied inter-pulse separation. An offset is added for easier visual separation. Dashed lines correspond to model fits (see text). \textbf{b.} Frequency dependence of the power spectral density. PSD extracted from the CPMG traces in \textbf{a.} shown on the bottom right, excluding points with $N_\pi<8$. Magenta points on the bottom right are extracted from a CPMG measurement at the off-detuned point shown in \textbf{e.} Blue datapoints on the top left are obtained by averaging three PSDs extracted from time-series measurements of the FM qubit frequency at $\FMdetuning{}=0$, measured using Ramsey oscillations. Black dashed line corresponds to a linear fit to this data in log-log space, with the shaded blue region denoting the $1\sigma$ uncertainty of the fit.  Magenta datapoints on the top left are the analogous measurement at the off-detuned point.  \textbf{c.} Rabi oscillations at $\FMdetuning{}=0$ with a drive amplitude of \qty{550}{\micro\volt}. \textbf{d.} Rotary Echo measurement for the same parameters as \textbf{c.} fitted with an exponential decay model. \textbf{e.} FM qubit frequency as a function of detuning \FMdetuning{}, extracted from Ramsey oscillations. Black solid line denotes the parking position for sweet-spot operation ($\FMdetuning{}=0$). Magenta line denotes the value of \FMdetuning{} where the off-detuning PSD is extracted. \textbf{f.} Zoom-in on CPMG PSD data shown in \textbf{b.} Additionally the PSD extracted from $N_\pi<8$ is shown as crosses. \textbf{g.} Single-qubit randomized benchmarking decay, fitted to obtain an average Clifford gate fidelity of \qty{99.5}{\%} and average physical gate fidelity of \qty{99.7}{\%}.}
    \label{fig:fig6}
\end{figure*}

We characterize the FM qubit performance at the detuning sweet-spot for the field direction $\hat{\bm{b}}_3$, fixing $B=$~\qty{5}{\milli\tesla} and  $t_c/h=$~\qty{13}{\giga\hertz}. The corresponding chevron plot is shown in \autoref{fig:fig1}g of the main text. To examine the noise affecting the qubit energy splitting, we perform CPMG spectroscopy~\cite{rojas-arias2025}, varying the number of $Y_\pi$ pulses from $N_{\pi}=1$ to $N_{\pi}=32$, and sweeping the inter-pulse separation $\tau$ as defined in the schematic in \autoref{fig:fig6}b. The coherence decay traces, shown in \autoref{fig:fig6}a, are fitted with a decay model to find the CPMG dephasing time $T_{2, N_\pi}$ and normalized as described in Supp. Inf. \autoref{sec:CPMG Noise spectroscopy}. The extracted values of $T_{2, N_\pi}$ as a function of $N_\pi$ are shown in the top right inset of \autoref{fig:fig6}b, with a linear log-log fit giving the slope $\gamma = 0.72$ (\autoref{fig:fig6}b inset). If we assume a colored noise model $S(f) = S_0/f^\alpha$ for the PSD, we find $\alpha = \frac{\gamma}{1-\gamma} = 2.58$, which is not consistent with a simple charge-noise picture.

To further study this trend we plot the noise power spectral density (PSD) extracted from the CPMG data (see Supplementary \autoref{sec:CPMG Noise spectroscopy}), shown on the bottom right in \autoref{fig:fig6}b with the same color coding. Moreover, we also extract the PSD on the flank of the spectrum (magenta datapoints, bottom right in \autoref{fig:fig6}b), at the detuning $\FMdetuning{}/h= \qty{-15}{\giga\hertz}\approx -t_c/h$, where the LSES is measured to be the steepest. This condition is explicitly marked with a magenta line in \autoref{fig:fig6}e, where the qubit frequency is shown as a function of \FMdetuning{} (zoom-in of \autoref{fig:fig4} main text). The CPMG data displayed in \autoref{fig:fig6}b only includes datapoints with  $N_\pi\geq8$. In \autoref{fig:fig6}f a zoom-in on the data is shown, where we additionally include points obtained from $N_\pi<8$, marked with crosses. This allows for a comparison between the datasets over a wider frequency range, although we note that the reconstruction is less accurate for lower number of decoupling pulses \cite{rojas-arias2025}. The sweet-spot data reveals a localized feature in the PSD which could possibly be explained by a two-level fluctuator (TLS). We note that beatings likely induced by a TLS were observed in the Ramsey decay of the LD qubit in QD$_2$ (see Supplementary \autoref{sec:LD qubit performance}).  The $^{73}$Ge isotope is expected to give a peak in the PSD at a much lower frequency of 7.4~kHz for the chosen value of \bmag{} \cite{hendrickxSweetspot2024, stehouwer2025}. However, we cannot rule out other effects of hyperfine interactions \cite{zeng_high-fidelity_2026, stehouwer2025}. A comparison of the data on and off the sweet-spot reveals that detuning sweet-spot operation effectively suppresses noise above 100 kHz, lowering the PSD by more than an order of magnitude. Finally, we fit the CPMG PSD on the flank with a linear function in log-log space and fixing a noise exponent of $\alpha = 1$ (dash-dotted line in \autoref{fig:fig6}f). With this we extrapolate a charge noise level of $S_{1/f}(\qty{1}{\hertz})=\qty{2.0e-12}{e\volt^2\per\hertz}$, similar to values reported on the same material \cite{hendrickxSweetspot2024}, where we have used the slope of the red line in \autoref{fig:fig6}e for converting Hz to eV. We note that this value shows a discrepancy with the Ramsey measurement discussed later, potentially indicating another dominating noise source for lower frequencies, or a deviation from the $1/f$ model. 

The low frequency range of the PSD on the sweet-spot and on the flank is probed using Ramsey spectroscopy (see Supplementary \autoref{sec:Ramsey Noise spectroscopy}) \cite{rojas-arias2025, bluhm_dephasing_2011}, where the deviations from the virtual frequency are tracked for subsequent Ramsey measurements. This allows us to track qubit frequency fluctuations at a sampling rate of $f_s =\qty{0.2}{\hertz}$. The PSD is obtained from the Fourier transform of the data. The blue and magenta points shown on the top left in \autoref{fig:fig6}b are obtained by averaging three PSDs obtained from the data shown in Supplementary \autoref{fig:figS1}. The PSD on the flank is fitted to obtain a noise exponent $\alpha=1.097$. The extrapolation to high frequency values is shown with dashed line with an error bar shown as the shaded pink region in \autoref{fig:fig6}b. If we try to extract a charge noise amplitude from this data, we obtain $S_{1/f^{1.097}}(\qty{1}{\hertz})=\qty{1.4e-11}{e\volt^2\per\hertz}$, which is about an order of magnitude larger than previously reported values for charge noise in the same material \cite{hendrickxSweetspot2024}. We therefore expect other mechanisms to dominate the PSD in the low-frequency regime, such as magnetic noise, which would be consistent with the weak frequency dependence of the \dephasing{} (\autoref{fig:fig4}b) in this magnetic field configuration. 

Another source of noise in the FM qubit is the fluctuation of the Rabi frequency. This effect is expected to be more pronounced in this type of qubit, since the Rabi frequency is strongly dependent on detuning~\cite{benito2019}. The decay of the Rabi oscillation is shown in \autoref{fig:fig6}c. To verify that the noise is quasistatic we perform a rotary echo experiment (\autoref{fig:fig6}d), where the phase of the drive is flipped by $\pi$ in the middle of the sequence. The resulting coherence is improved by about a factor of five. We thus conclude that Rabi decay is mostly quasi-static.

\onecolumngrid

\subsection{CPMG noise spectroscopy}\label{sec:CPMG Noise spectroscopy}
\noindent The CPMG sequence follows the pulse protocol: 
\begin{equation}
X_{\pi/2}~-~\left(\tau~-~Y_\pi~-~\tau\right)^{N_{\pi}}~-~X_{\pi/2}.
\end{equation}
The measurement takes between 5 and 16 minutes to complete for $N_{\pi}=1$, and $N_{\pi}=32$ respectively. Before every measurement, we calibrate the amplitude of the detuning pulse to find the position of the detuning sweet-spot. We perform single-shot readout, averaging the outcomes of 4000 shots. The CPMG data at finite detunings $|\FMdetuning|>0$ (magenta points \autoref{fig:fig6}b,f) is obtained by inserting additional detuning pulses during each $\tau$ segment:
\begin{equation}
X_{\pi/2}~-~\left(\Delta \overline{V}_{\rm P2}(\tau)~-~Y_\pi~-~\Delta \overline{V}_{\rm P2}(\tau)\right)^{N_{\pi}}~-~X_{\pi/2}.
\end{equation}
and using an odd number of $Y_\pi$ pulses $N_\pi = 1, 3, 5, 9, 17, 33$. The PSD is extracted from the CPMG traces following the method described in Ref.~\cite{rojas-arias2025}, with the difference that here we also consider the effect of the $T_1$ relaxation as outlined in Ref.~\cite{bylander2011}.  Each coherence decay trace is fitted using the model 
\begin{equation}
    f(T) = A \exp\left[-\left(\frac{T}{T_{2, N_\pi}}\right)^\beta\right]\exp\left[-\frac{1}{2}\frac{T}{T_1}\right] + B,
\end{equation}

\noindent where $T=2N_\pi\tau$ is the total duration of the sequence, $T_{2, N_\pi}$ is the coherence time, $\beta$ is a free parameter capturing the noise color, $T_1$ relaxation is separately extracted and fixed, and $A$ and $B$ are fitted constants. Next, the datapoints are normalized between 0 and 1 and the effect of $T_1$ is divided out:  
\begin{equation}
    \tilde{f}(T) = \frac{f(T) - B} {A\exp(- \frac{1}{2}\frac{T}{T_1})}.
\end{equation}
We retain only the data points on the steep part of the decay curve, lying between 0.15 and 0.85. 
Finally, the PSD in units of $\qty{}{\hertz}^2/\qty{}{\hertz}$ is obtained using the formula
\begin{equation}
PSD\left(\frac{N_\pi}{2T} \right) = \frac{-\ln\tilde{f}(T)
}{2\pi^2T}.
\end{equation}

\subsection{Ramsey noise spectroscopy}\label{sec:Ramsey Noise spectroscopy}
\begin{figure*}
    \includegraphics[width=\textwidth]{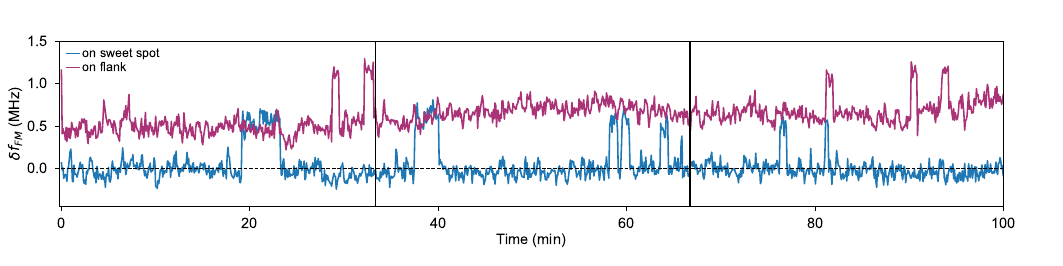}
	\caption{Time series of the extracted FM qubit frequency fluctuation ($\delta f_{\mathrm{FM}}$) on the flank (magenta) and on the sweet-spot of the qubit frequency spectrum shown in \autoref{fig:fig6}e. Black vertical lines indicate pauses in the acquisition, followed by a re-calibration of the sweet-spot detuning. Dashed line at $\delta f_{FM} = 0$ indicates zero deviation from the virtual frequency. The three time traces on the sweet-spot are taken consecutively, followed by the three traces on the flank consecutively. We overlay the time series for the two detuning setpoints for easier comparison of the noise amplitudes. }
    \label{fig:figS1}
\end{figure*}

The low frequency PSD on the sweet-spot is probed using Ramsey spectroscopy with the sequence
\begin{equation}
X_{\pi/2}~-~\tau~-~Z( 2\pi f_{\mathrm{virt}}\tau)~-~X_{\pi/2}, 
\end{equation}
where $\tau$ is a varied wait time, $Z(\varphi)$ is a virtual gate and $f_{\mathrm{virt}} = \qty{15.625}{\mega\hertz}$ is the virtual frequency. The Ramsey spectroscopy on the flank is obtained with the pulse sequence:
\begin{equation}
X_{\pi/2}~-~\Delta\virttwo{}(\tau)~-~Z(2\pi f_{\mathrm{virt}}\tau)~-~X_{\pi/2}, 
\end{equation} 
where $\Delta\virttwo{}(\tau)$ denotes a detuning pulse, adiabatic with respect to $t_c$, with a duration $\tau$. To acquire a single Ramsey trace we execute a single shot for each duration $\tau$ before repeating the cycle for a total of 50 shots. The outcomes are then averaged and saved in a buffer. Another Ramsey measurement is then directly repeated. All of the data acquisition is done on the OPX1000 with no software loops. We therefore rely on the deterministic timing of the hardware to ensure equal spacing between the measurements. The full duration of the experiment is approximately 33 minutes. 

The decay of each Ramsey oscillation is fitted using Eq.~\ref{eq:decay fit}, obtaining a time series of the qubit frequency detuning $\delta f_{FM}(\tau) = \qubitf{} - f_{\mathrm{virt}}$  (\autoref{fig:figS1}). The experiment is repeated multiple times with an interleaved calibration of the sweet-spot position, denoted with vertical black lines in \autoref{fig:figS1}. The PSD is calculated individually for each experiment and the results are then averaged together to obtain the PSD shown in \autoref{fig:fig6}. The PSD is obtained from the real fast Fourier transform (rFFT) of the data as $\text{PSD} = \frac{2|\text{rFFT}|^2}{Nfs}$. We overlay the results for on and off the flank for easier comparison of the noise amplitude. We note the presence of a two-level fluctuator which leads to frequency jumps of $\sim 15\%$ of the Rabi frequency present in both datasets. Another notable feature is the background drift which is present only in the on-flank dataset. We attribute this to the suppression of low-frequency noise in the tunnel coupling at the sweet-spot.

\subsection{LD qubit performance}\label{sec:LD qubit performance}

\begin{figure*}[t]
\includegraphics[width=0.85\textwidth]{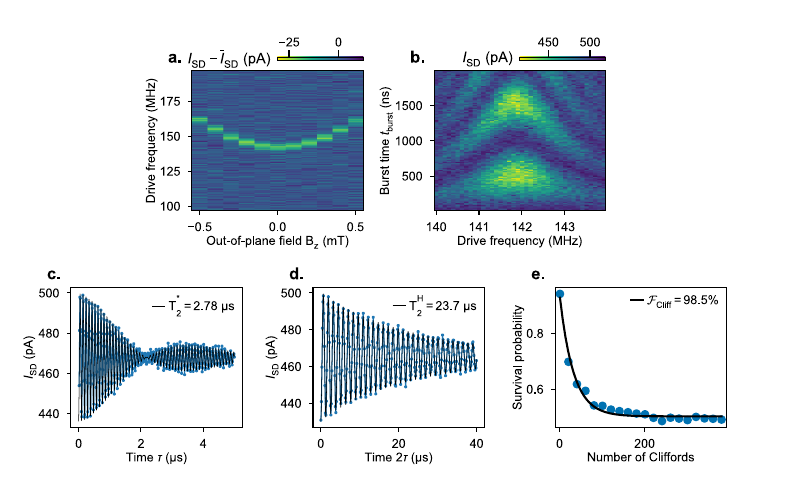}
	\caption{\textbf{a.} ARP spectroscopy as a function of out-of-plane magnetic field strength. \textbf{b.} Chevron oscillation for the qubit under $P_2$, driven with an amplitude of \qty{7.5}{\milli\volt} \textbf{c.(d.)},  Ramsey (Hahn echo) measurement at the hyperfine sweet-spot with $B_z=0$ and $\bmag{} = \qty{15}{\milli\tesla}$. \textbf{e.} Randomized benchmarking performed on the LD qubit at the hyperfine sweet-spot.}
    \label{fig:figS4}
\end{figure*}

We compare the performance of the FM qubit to an LD qubit defined in QD$_2$. The qubit is operated at the center of the (0,1,1) charge region with a closed inter-dot tunnel coupling between (1,0,1) and (0,1,1). We orient \bfield{} into the equatorial plane of $g_2$ by performing state spectroscopy as a function of the out-of plane \bfield{} component $\bfield{}_z$ (\autoref{fig:figS4}a) for an in-plane angle $\phi_{\mathrm{QD2}}=\qty{41.4}{\degree}$ in the $g_2$ reference frame. Operating close to the equatorial plane is expected to yield improved fidelities for $g$-TMR driven single-qubit gates \cite{hendrickxSweetspot2024, john_robust_2025}. At the frequency minimum, with $\bmag{} =\qty{15}{\milli\tesla}$ and a drive amplitude $\Adrive{} = \qty{7.5}{\milli\volt}$, we obtain the chevron plot in \autoref{fig:figS4}b. We find a Rabi frequency of \qty{0.95}{\mega\hertz} and a qubit frequency of \qty{141.9}{\mega\hertz}, resulting in $\eta_{\rm d} = \qty{0.13}{\mega\hertz/\milli\volt}$ and $\tilde{\eta}_{\rm d} = \qty{0.0009}{\per\milli\volt}$.  The corresponding decay traces for Ramsey and Hahn echo experiments are shown in \autoref{fig:figS4}c,d. We observe a beating in the Ramsey decay but not in the echo decay, suggesting the presence of quasi-static random telegraph noise from a two-level fluctuator. We fit the Ramsey decay to the model $Ae^{-\left(t/\dephasing\right)^\alpha}\cos(2\pi ft)\cos(\pi\delta f_{\mathrm{TLS}}t) + B$, finding $\dephasing=2.8\pm0.05~\qty{}{\micro\second} $. For the echo decay, we use the same model as in the main text, finding $\echo=23\pm\qty{0.5}{\micro\second}$. We then perform randomized benchmarking (\autoref{fig:figS4}e), obtaining a Clifford gate fidelity $98.5\pm0.1\%$ and an average primitive gate fidelity of $99.1\pm0.1\%$. This gives a factor 3 higher infidelity compared to the FM qubit.

\subsection{Lever arm and $t_c$ estimation}\label{sec:Lever arm extraction}
\begin{figure*}[th]
    \centering
    \includegraphics[width=\textwidth]{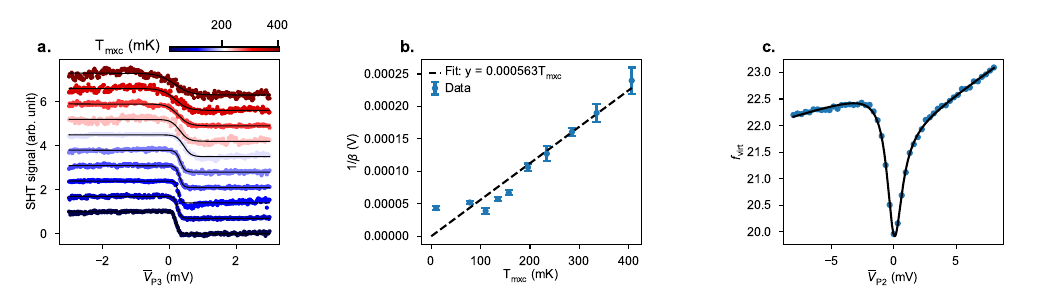}
	\caption{\textbf{a.} Normalized sensor current as a function of virtual plunger voltage \virtthree{} revealing the (0,1,2)-(0,1,1) reservoir transition. The measurement is repeated for different mixing chamber temperatures $T_{\mathrm{mxc}}$ and fitted with a Fermi-Dirac distribution (black lines). An offset is added for visual separation of the data. \textbf{b.} Extracted values of $\beta^{-1}=k_BT_\mathrm{mxc}/\alpha_{33}$ as a function of $T_{\mathrm{mxc}}$. Dashed line corresponds to a linear fit which is constrained to pass through zero. \textbf{c.} Typical measurement used for calibrating $t_c$ and the condition $\FMdetuning{}=0$. }
    \label{fig:figS3}
\end{figure*}

We first extract the virtual gate lever arm of \virtthree{} using temperature-dependent broadening of the reservoir transition \cite{sommer_disentangling_2026} between the charge sectors (0,1,1) and (0,1,2), probed by varying \virtthree{} in the range $\pm\qty{3}{\milli\volt}$ (\autoref{fig:figS3}a). We set the temperature using a heater on the mixing chamber plate and monitor the temperature with a sensor also placed on the mixing chamber plate. For each temperature, the SHT current as a function of \virtthree{} is fitted to the Fermi-Dirac distribution: 
\begin{equation*}
f(V) = \frac{A}{e^{\beta(V-V_0)} + 1},
\end{equation*}
with $\beta = \frac{\alpha}{k_B T_{\mathrm{mxc}}}$, where $T_{\mathrm{mxc}}$ is used as an estimate for the electron temperature and $\alpha$ is the lever arm. In \autoref{fig:figS3}b we show the extracted value of $\beta^{-1}$ as a function of $T_{\mathrm{mxc}}$. We overlay a least squares linear fit to the last five points, with an intercept constrained to zero. From the slope of the fit we find a lever arm of \virtthree{} of $\alpha_{33} = \qty{0.153}{e\volt/\volt}$, where we use the notation $\alpha_{ij}$ for gate $\overline{V}_{i}$ acting on QD$_j$. The slope of the inter-dot transition between (0,0,2) and (0,1,1) in virtual gate space (\autoref{fig:fig1}e) is given by the expression:
\begin{equation}
s_{23} = \frac{\alpha_{33} }{\alpha_{22}} 
\end{equation}
 We extract $s_{23} = 1.04$ from the measured charge stability diagram and use this to find $\alpha_{22} = \qty{0.147}{e\volt/\volt}$. Next, we estimate the lever arm $\alpha_{11}$. As a consequence of the gate $\overline{V}_\mathrm{B12}$ being pulsed to open the tunnel coupling, the lever arm of gate $P_2$ with respect to QD$_1$  changes from the uncoupled value, resulting in a relative lever arm ratio $\alpha_{21}/\alpha_{11} = 0.4$, while $\alpha_{12}=0$ still holds. We extract the lever arm of virtual gate \virtone{} as: 

\begin{equation}
   \alpha_{11} = \frac{\alpha_{22} }{s_{12} + \alpha_{21}/\alpha_{11}} = 0.10~\mathrm{eV/V}
\end{equation}

Next, we extract the detuning lever arm of the combined gate $\overline{\varepsilon}_{12} = \virttwo{} - \virtone{}$ as:

\begin{equation}
    \alpha_{\varepsilon} = \alpha_{22} + \alpha_{11}\left(1 - \frac{\alpha_{21}}{\alpha_{11}}\right) = 0.21 \mathrm{eV/V}
\end{equation}

The tunnel coupling $t_c$ is extracted by sweeping across the inter-dot transition between (0,1,1) and (1,0,1) with the gate \virttwo{} using the echo sequence described in the detuning calibration \autoref{sec:Detuning and sweet-spot calibration} and fitting the resulting qubit spectrum to the model:

\begin{equation}\label{eq:FM Hamiltonian}
    \qubitf{} = \frac{\mu_B B}{h} \frac{\sqrt{\frac{\varepsilon_{12}^2 + 2t_c^2}{2}(g_1^2 + g_2^2) + \frac{\varepsilon_{12}}{2}\sqrt{\varepsilon_{12}^2 + 4t_c^2}(g_2^2 - g_1^2) + 2g_1 g_2 t_c^2 \cos(\tilde{\theta}_{12})}}{\sqrt{\varepsilon_{12}^2 + 4t_c^2}}.
\end{equation}

Here, $\varepsilon_{12}$ is the energy detuning, $\tilde{\theta}_{12}$ is a free parameter capturing quantization axis misalignment due to $g$-tensor variation and spin-flip tunneling coming from SOI \cite{seidler2025},  and $g_1$ and $g_2$ are free parameters capturing the $g-$factors in QD$_1$ and QD$_2$. We also allow for linear variation of the $g_1$ and $g_2$ with $\varepsilon_{12}$ \cite{wang_operating_2024}. We use the standard convention where $2t_c$ is the separation between the bonding and anti-bonding states. An example of the model fit resulting in $t_c/h=\qty{13}{\giga\hertz}$ is shown in \autoref{fig:figS3}c.

\subsection{Virtual gate matrix}\label{sec:vmatrix}

We note that the following virtual gate matrix leads to independent chemical potentials of QD$_1$ and QD$_2$ only in the case before $\overline{V}_{\rm B12}$ is pulsed (see Supplementary~\autoref{sec:Readout and initialization}).  Pulsing of $\overline{V}_{\rm B12}$ leads to modified relative lever arms, which we consider in the lever arm extraction (\autoref{sec:Lever arm extraction}).

\begin{equation*}
\renewcommand{\arraystretch}{1.2}
\footnotesize
\left(\begin{array}{c}
V_{\mathrm{P4}} \\
V_{\mathrm{P3}} \\
V_{\mathrm{P2}} \\
V_{\mathrm{P1}} \\
V_{\mathrm{B45}} \\
V_{\mathrm{B34}} \\
V_{\mathrm{B23}} \\
V_{\mathrm{B12}} \\
V_{\mathrm{RB}} \\
V_{\mathrm{PS}}
\end{array}\right)
=
\left(\begin{array}{cccccccccc}
1 & 0 & 0 & 0 & 0 & 0 & 0 & 0 & 0 & 0 \\
0 & 1 & -0.1584 & -0.0075 & 0 & -0.948 & -0.4812 & 0.0522 & 0 & 0 \\
0 & -0.2146 & 1 & -0.1892 & 0 & 0.2034 & -0.8567 & -0.4171 & 0 & 0 \\
0 & -0.0029 & -0.327 & 1 & 0 & 0.0028 & 0.1211 & -1.5553 & 0 & 0 \\
0 & 0 & 0 & 0 & 1 & 0 & 0 & 0 & 0 & 0 \\
0 & 0 & 0 & 0 & 0 & 1 & 0 & 0 & 0 & 0 \\
0 & 0 & 0 & 0 & 0 & 0 & 1 & 0 & 0 & 0 \\
0 & 0 & 0 & 0 & 0 & 0 & 0 & 1 & 0 & 0 \\
0 & 0 & 0 & 0 & 0 & 0 & 0 & 0 & 1 & 0 \\
-0.0143 & -0.0258 & -0.0345 & -0.02 & -0.017 & -0.0122 & 0.0106 & 0.0269 & -0.0054 & 1
\end{array}\right)
\left(\begin{array}{c}
\overline{V}_{\mathrm{P4}} \\
\overline{V}_{\mathrm{P3}} \\
\overline{V}_{\mathrm{P2}} \\
\overline{V}_{\mathrm{P1}} \\
\overline{V}_{\mathrm{B45}} \\
\overline{V}_{\mathrm{B34}} \\
\overline{V}_{\mathrm{B23}} \\
\overline{V}_{\mathrm{B12}} \\
\overline{V}_{\mathrm{RB}} \\
\overline{V}_{\mathrm{PS}}
\end{array}\right)
\end{equation*}

\subsection{Randomized benchmarking}\label{sec:Randomized benchmarking}

We perform single qubit RB using the Clifford gate decomposition shown in \autoref{tab:cliffords}. For each of the 101 random Clifford sequences we vary the number of Cliffords in steps of 20 as $\{1, 21, 41, 61, ... , 381\}$ (20 lengths), followed by a recovery Clifford. The averaging is done with 500 shots in the inner loop, followed by varying the length of the sequence, and finally sweeping over 101 random Clifford sequences in the outer loop. For each of the 101 Clifford sequences we prepend an idling pulse and an $X_{\pi}$ pulse, which are then used to normalize the probability scale. The experiment is performed using the OPX1000 from Quantum Machines, which enables on-the-fly generation of random sequences and determination of their inversion gate using a stored Cayley lookup table. The total duration of the experiment is approximately 20 minutes. We fit the averaged decay curve using the model $Ap^n~+~B$ where $n$ is the number of Cliffords and $p$ is the decay rate. The Clifford gate fidelity is obtained as $\mathcal{F}_{\rm Cliff} = (1 + p) / 2$. The error in the fidelity is computed from the covariance of the fit. The average single qubit gate fidelity is obtained as $1-(1-\mathcal{F}_{\rm Cliff})/r$, where $r=1.875$ is the average number of gates per Clifford. 

\begin{table}[ht]
\centering
\begin{tabular}{c|l}
\hline
\textbf{Clifford} & \textbf{Pulse sequence} \\
\hline
$C_0$  & $I$ \\
$C_1$  & $X_{180}$ \\
$C_2$  & $Y_{180}$ \\
$C_3$  & $Y_{180}\,X_{180}$ \\
$C_4$  & $X_{90}\,Y_{90}$ \\
$C_5$  & $X_{90}\,Y_{-90}$ \\
$C_6$  & $X_{-90}\,Y_{90}$ \\
$C_7$  & $X_{-90}\,Y_{-90}$ \\
$C_8$  & $Y_{90}\,X_{90}$ \\
$C_9$  & $Y_{90}\,X_{-90}$ \\
$C_{10}$ & $Y_{-90}\,X_{90}$ \\
$C_{11}$ & $Y_{-90}\,X_{-90}$ \\
$C_{12}$ & $X_{90}$ \\
$C_{13}$ & $X_{-90}$ \\
$C_{14}$ & $Y_{90}$ \\
$C_{15}$ & $Y_{-90}$ \\
$C_{16}$ & $X_{-90}\,Y_{90}\,X_{90}$ \\
$C_{17}$ & $X_{-90}\,Y_{-90}\,X_{90}$ \\
$C_{18}$ & $X_{180}\,Y_{90}$ \\
$C_{19}$ & $X_{180}\,Y_{-90}$ \\
$C_{20}$ & $Y_{180}\,X_{90}$ \\
$C_{21}$ & $Y_{180}\,X_{-90}$ \\
$C_{22}$ & $X_{90}\,Y_{90}\,X_{90}$ \\
$C_{23}$ & $X_{-90}\,Y_{90}\,X_{-90}$ \\
\hline
\end{tabular}
\caption{The 24 single-qubit Clifford operations and their corresponding pulse decompositions.}
\label{tab:cliffords}
\end{table}

\subsection{$T_1$ relaxation theory}\label{sec:T1 theory}
 \subsubsection{Flopping mode}
The flopping mode Hamiltonian is 
\begin{equation}
H= \frac{\FMdetuning}{2}\tau_z+ {t_c}\tau_x +\sum_{i=1,2}\frac{\mu_B\textbf{B}\cdot g_{i}\pmb{\sigma}}{2}\frac{\tau_0+ (-1)^i \tau_z}{2} \ ,
\label{eq: bare_FM_hamiltonian}
\end{equation}
in the position basis of each dot and with the gate-dependent $g$-tensors $g_{i}(\bar{V}_{P1}, \bar{V}_{P2})$ defined in Eq.~\eqref{eq: g_tensor_gate_dependence}, written in shorthand.
\begin{equation}
    g_i=R_{avg}^T\cdot R_i \cdot g_\text{diag}^i \cdot R_i^T \cdot R_{avg}\cdot R_{so,i} \ \ , \ \ \ g_\text{diag}^i=\text{diag}(g_1^i,g_2^i,g_3^i) \ .
\end{equation}
where the principal axis g-tensor $g^{i}_{\rm diag}$ and its $zyz$-rotation in the lab frame $R_{i}$ can be found in Fig.\ref{fig:fig2}. A further rotation $R_{avg}$ is applied to rotate into the averaged lab frame. Finally, an antisymmetric part given by SOI rotation $g_i=g_s^i R_{so,i}$ is applied to rotate the averaged lab frame into SO frame.

For convenience we introduce the Larmor vector of the $i$th qubit
\begin{equation}
   \textbf{l}_{i}=\mu_B \textbf{ B}\cdot g_{i} \ ,
\end{equation}
whose amplitude corresponds to the measured Larmor frequency of the $i$th dot  $ E_{z,i}=|\textbf{l}_{i}|$\ .
We remark that $E_{z,i}$ is independent of the spin-flip tunneling contribution $R_{so,i}$, but depends on the principal axes ($R_i$) and principal values ($g^i_{1,2,3}$) of $g_s^i$. For this reason, the individual $g$-tensor characterization only allows us to extract information on $g_s^i$.



\subsubsection{Effective Theory}

To develop an effective FM qubit model  we first diagonalize the $B=0$ part of the Hamiltonian by a unitary transformation
\begin{equation}
\label{eq:orbital-eig}
\tilde{U}=e^{-i \theta \tau_y} \ , \ \  \tan(2\theta)=2t_c/\FMdetuning \ ,
\end{equation}
with identity $e^{-i\theta(\hat{n}\cdot\sigma)} = \sigma_{0}\cos{\theta} - i(\hat{n}\cdot \sigma)\sin{\theta}$, leading to
\begin{align}
\tilde{H}&=\tilde{U}^{\dagger}H \tilde{U} =\frac{\Omega}{2}\tau_z +\left(
\begin{array}{cc}
 \sin ^2(\theta ) & \sin(2\theta)/2 \\
 \sin(2\theta)/2 & \cos ^2(\theta ) \\
\end{array}
\right)\frac{ \textbf{l}_1\cdot\pmb{\sigma}}{2}+ \left(
\begin{array}{cc}
 \cos ^2(\theta ) & -\sin(2\theta)/2 \\
 -\sin(2\theta)/2 & \sin ^2(\theta ) \\
\end{array}
\right)\frac{ \textbf{l}_2\cdot\pmb{\sigma}}{2} \ , 
\end{align}
with orbital energy 
\begin{equation}
    \hbar\Omega=\sqrt{4t_c^2+\FMdetuning^2} \ .
\end{equation}

The FM qubit is encoded in the ground state projection of $\tilde{H}$, ignoring the bonding states energy offset $\frac{\hbar \Omega}{2}$:
\begin{equation}
\label{eq:FM-hamiltonian}
    H_\text{FM}=\frac{ \textbf{l}\cdot\pmb{\sigma}}{2} \ , 
\end{equation}
where we defined the FM Zeeman frequency vector $\textbf{l}$   that interpolates between the Larmor vectors $\textbf{l}_{1,2}=\mu_B \textbf{ B}\cdot g_{1,2}$ of left and right QDs according to the detuning-dependent hybridization parameter $\eta\in[0,1]$ :
\begin{align}
   \textbf{l}&=  \mu_B\textbf{B}\cdot[\cos ^2(\theta )  g_1 +\sin ^2(\theta ) g_2 ] =\mu_B \textbf{B}\cdot[(1-\eta)  g_1 +\eta  g_2 ] = (1-\eta)  \textbf{l}_1 +\eta  \textbf{l}_2\ , \ \ \\
 \eta&=\sin^2\theta =\frac{1}{2}-\frac{\FMdetuning }{2 \sqrt{4t_c^2+\FMdetuning ^2}}\ .
\end{align}
The parameter  $\eta=0$ when the wavefunction is fully localized in QD$_1$ ($\FMdetuning\gg t_c$) and  $\eta=1$ when the wavefunction is localized in QD$_2$ ($\FMdetuning\ll -t_c$); in FM regime $\FMdetuning=0$,  $\eta=1/2$.
We emphasize that in our experiment the tunneling $t_c$ is much larger than all Zeeman energies, such that higher order corrections to $H_\text{FM}$ that lead to the emergence of sweet-spots in FM qubits are negligible here. The lowest order corrections are estimated below.

Importantly, even if the vector $\textbf{l}$ is a linear interpolation of the two Larmor vectors $\textbf{l}_1$ and $\textbf{l}_2$, the  measured Zeeman energy of the FM qubit, i.e. the absolute value of $\textbf{l}$, can present a non monotonic behavior against $\FMdetuning$ when the two Larmor vectors are misaligned. This effect can be understood by explicitly observing that 
\begin{align}
    \Delta E =|\textbf{l}|&=\sqrt{[(1-\eta)E_{z,1}+\eta E_{z,2}]^2-2 \eta (1-\eta) (E_{z,1} E_{z,2}-\textbf{l}_1\cdot \textbf{l}_2)}  \ ,
\end{align}
where $E_{z,i}=|\textbf{l}_i|$ is the Zeeman energy of the $i$-th dot. 
There are two terms determining $\Delta E$. The first term $(1-\eta)E_{z,1}+\eta E_{z,2}$ is the detuning-dependent interpolation of the Zeeman energies of each dot, that is a monotonic function of $\FMdetuning$. The second term  $2 \eta (1-\eta) (E_{z,1} E_{z,2}-\textbf{l}_1\cdot \textbf{l}_2)$ appears at finite hybridization parameters ($\eta\neq 0$ and $\eta \neq 1$) and captures the tilt between Larmor vectors, vanishing when $\textbf{l}_1\parallel \textbf{l}_2$.
Introducing the relative angle between Larmor vectors
\begin{equation}
\label{eq:costheta12}
    \cos(\theta_{12})=\frac{\textbf{l}_1\cdot \textbf{l}_2}{E_{z,1} E_{z,2}},
\end{equation} and
\begin{equation}
\label{eq:sintheta12}
\sin(\theta_{12})=\frac{|\textbf{l}_1\times \textbf{l}_2|}{E_{z,1} E_{z,2}}
\end{equation} 
and in the FM regime $\FMdetuning=0$ (i.e. $\eta=1/2$)
\begin{align}
    \Delta E_\text{FM} =\sqrt{\left(\frac{E_{z,1}+E_{z,2}}{2}\right)^2-E_{z,1} E_{z,2}\sin^2\left(\frac{\theta_{12}}{2}\right)}  \ ,
\end{align}

Note that the angle $\theta_{12}$ depends on both symmetric measured $g$ tensors ($g_s^i$) and on the accumulated spin-flip tunneling rotation $R_{so,i}$ between dot 1 and 2 
\begin{equation}
    \cos(\theta_{12})= \frac{\mu_B^2    }{\hbar^2E_{z,1} E_{z,2}}  (\textbf{B} g_1 g_2^T\textbf{B})=\frac{\mu_B^2     }{\hbar^2E_{z,1} E_{z,2}}(\textbf{B} g_s^1 R_{so} g_s^2\textbf{B}) \ ,
\end{equation} 
where $R_{so}=R_{so,1}R_{so,2}^T$. 

Explicitly, the transformation that diagonalizes the $H_\text{FM}$ in Eq.~\eqref{eq:FM-hamiltonian} is 
\begin{equation}
\label{eq:utrafo}
    u=e^{-i\psi \sigma_z}e^{-i\phi \sigma_y} \ , \ \text{with} \ \   \cos(2\psi)= l_y/ l_x \ , \ \  \tan(2\phi)= l_{z}/\Delta E \ .
\end{equation}

\begin{figure}
    \centering
    \includegraphics[width=\linewidth]{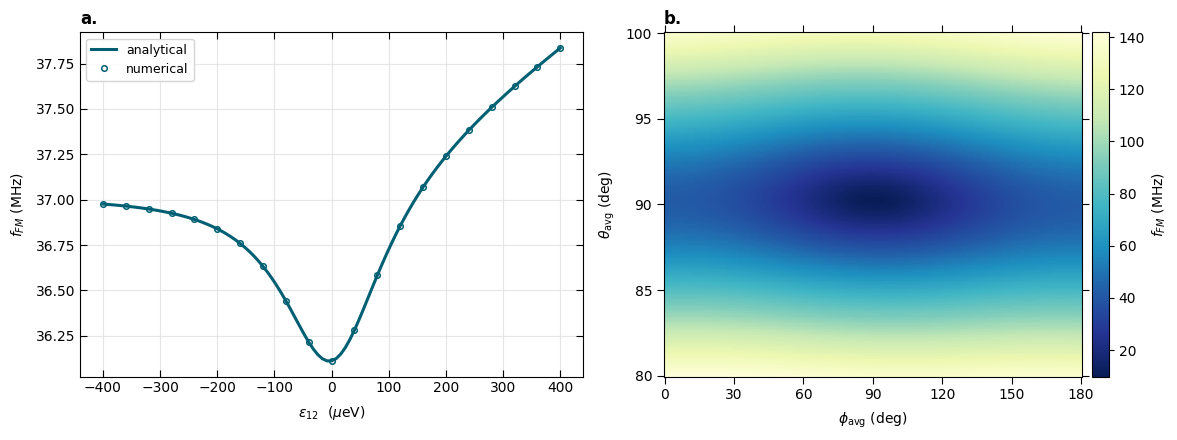}
    \caption{$\textbf{a.}$ FM qubit frequency as function of detuning $\epsilon_{12}$. We use experimentally extracted tunnel coupling $t_c = 52\mu eV$, $g_1$ and $g_2$ and the ZYZ rotation angle in Fig. \ref{fig:fig2}. The numerical line is obtained by numerically diagonalizing the 4 by 4 Hamiltonian in Eq.~\eqref{eq: bare_FM_hamiltonian} and the analytical equation is obtained by plugging in the Eq.~\eqref{eq:FM-hamiltonian}.
    $\textbf{b.}$ The FM qubit frequency as function of in plane $\phi_{avg}$ and out of plane $\theta_{avg}$ magnetic field angles.}
    \label{fig:S6_FM_qubit_frequency}
\end{figure}

\subsubsection{Correction to the FM Hamiltonian}
We now analyze corrections to the FM Hamiltonian $H_\text{FM}$ in Eq.~\eqref{eq:FM-hamiltonian} arising from the coupling between the bonding and antibonding orbital states and show that they scale at least as $B^3/\Omega^2$ and can therefore be safely neglected in the low-field limit where we are operating the qubit. By defining
\begin{align}
 \bar{\textbf{l}}&= \eta  \textbf{l}_1+(1-\eta)  \textbf{l}_2 =\textbf{l}|_{\theta\to\theta+\pi/2}\ , \\
 \Delta{\textbf{l}}&= \frac{\textbf{l}_1-\textbf{l}_2}{2}= \frac{\mu_B\textbf{B}\cdot[g_1 -g_2 ]}{2 } 
\end{align}
the DQD Hamiltonian reduces to 
\begin{align}
\label{eq.htilde}
\tilde{H}&=\frac{1}{2}\left(
\begin{array}{cc}\Omega\sigma_0+\bar{\textbf{l}}\cdot\pmb{\sigma} & \sin(2\theta)\Delta \textbf{l}\cdot \pmb{\sigma}\\
 \sin(2\theta)\Delta \textbf{l}\cdot\pmb{\sigma} &-\Omega\sigma_0 +{\textbf{l}}\cdot\pmb{\sigma} \\
\end{array}
\right)\ .
\end{align}
The notation $\textbf{l}|_{\theta\to\theta+\pi/2}$ indicates that $\bar {\textbf{l}}$ is connected to the Larmor vector of the FM $\textbf{l}$ by substituting $\theta$ with $\theta+\pi/2$, or equivalently by changing the sign of the detuning, i.e. $\FMdetuning\to -\FMdetuning$, which leads to $\eta\to (1-\eta)$ and $(1-\eta)\to \eta$. 

The correction to $H_\text{FM}$ can be derived by a Schrieffer-Wolff (SW) transformation.
To second order and for low magnetic fields, the generator $S$ of the SW transformation is 
\begin{align}
\label{eq:S-gen}
S =\frac{\sin (2 \theta )}{2\Omega} \left(
\begin{array}{cc}
 0 &  {\Delta\textbf{l}}\cdot\pmb{\sigma}\\
 -{\Delta\textbf{l}}\cdot\pmb{\sigma}& 0 \\
\end{array}
\right) \ ,
\end{align}
resulting in the effective Hamiltonian
\begin{equation}
    H_\text{eff} \approx \text{diag}\left(\frac{\bar{\textbf{l}}\cdot\pmb{\sigma}}{2}+ \frac{\Omega}{2}\left(1+\frac{\sin^2(2\theta) |{\Delta \textbf{l}}|^2}{\textcolor{black}{2}\Omega^2}\right)\sigma_0,\frac{\textbf{l}\cdot\pmb{\sigma}}{2}- \frac{\Omega}{2}\left(1+\frac{\sin^2(2\theta) |{\Delta \textbf{l}}|^2}{\textcolor{black}{2}\Omega^2}\right)\sigma_0\right),
    \label{eq:heff4by4}
\end{equation}
the bonding block gives the effective FM subspace
\begin{equation}
\label{eq:heff}
H_\text{eff}=\frac{\textbf{l}\cdot\pmb{\sigma}}{2}- \frac{\Omega}{2}\left(1+\frac{\sin^2(2\theta) |{\Delta \textbf{l}}|^2}{\textcolor{black}{2}\Omega^2}\right)\sigma_0
\end{equation}
with 
\begin{equation}
   |{\Delta \textbf{l}}|^2=\Delta {l}_x^2+\Delta {l}_y^2+\Delta {l}_z^2=\frac{(E_{z,1}-E_{z,2})^2}{4}+E_{z,1}E_{z,2}\sin^2\left(\frac{\theta_{12}}{2}\right)\propto B^2 \ .
\end{equation}
This result shows that the lowest order corrections to the FM Hamiltonian~\eqref{eq:FM-hamiltonian} are spin independent. Spin dependent terms emerge in higher orders and are thus at least $\propto B^3/\Omega^2$. 

The effective Hamiltonian can be fully diagonalized with additional rotation Eq.~\eqref{eq:utrafo}
\begin{equation}
\label{eq:ul}
    U_{\textbf{l}} = \text{diag} (\bar{u}, u),
\end{equation} the transformation in antibonding states is defined as $$\bar u = e^{-i\bar\psi\sigma_z}e^{-i\bar\phi\sigma_y},\qquad
\tan(2\bar\psi)=\bar l_y/\bar l_x,\qquad \cos(2\bar\phi)=\bar l_z/|\bar{l}|$$.

Applying the unitary transformation ~\eqref{eq:ul} on block diagonalized effective Hamiltonian~\eqref{eq:heff4by4}, we get
\begin{equation}
    \tilde{H}_\text{eff} = U_{\textbf{l}}^\dagger H_{\text{eff}}U_{\textbf{l}} = \text{diag}\left(\frac{|\bar{\textbf{l}}|\sigma_{z}}{2}+ \frac{\Omega}{2}\left(1+\frac{\sin^2(2\theta) |{\Delta \textbf{l}}|^2}{\textcolor{black}{2}\Omega^2}\right)\sigma_0,\frac{|\textbf{l}|\sigma_{z}}{2}- \frac{\Omega}{2}\left(1+\frac{\sin^2(2\theta) |{\Delta \textbf{l}}|^2}{\textcolor{black}{2}\Omega^2}\right)\sigma_0\right).
    \label{eq:heff4by4diag}
\end{equation}

\subsection{Decoherence}
To characterize decoherence in the FM qubit, we study the effect on the FM qubit of fluctuations of tunneling, detuning, and $g$-tensors, $\delta_t$, $\delta_\epsilon$, and $\delta g_i$ respectively. These are characterized by the Hamiltonian 
\begin{align}
    H_N &= \frac{\delta_\epsilon}{2}\tau_z+ {\delta_t}\tau_x + \sum_{i=1,2}\frac{ \pmb{\delta}_l^i \cdot \pmb{\sigma}}{2}\frac{\tau_0+ (-1)^i \tau_z}{2} \ ,
    \end{align}
    where $\pmb{\delta}_l^i =\mu_B\textbf{B}\cdot \delta g_{i}$. In the orbital eigenstates (see Eq.~\eqref{eq:orbital-eig}) reads
    \begin{align}
    \tilde{H}_N &=\tilde{U}^\dagger H_{N}\tilde{U} =\frac{ \delta_\epsilon \cos(2\theta)+2\delta_t \sin(2\theta) }{2}\tau_z- \frac{\delta_\epsilon \sin(2\theta)-2\delta_t \cos(2\theta)}{2}\tau_x+ \frac{1}{2}\left(
\begin{array}{cc}
 \bar{\pmb{\delta}_l}\cdot\pmb{\sigma} & \sin(2\theta)\Delta\pmb{\delta}_l \cdot \pmb{\sigma}\\
 \sin(2\theta)\Delta\pmb{\delta}_l\cdot\pmb{\sigma} &\pmb{\delta}_l\cdot\pmb{\sigma} \\
\end{array}
\right) \ , 
\end{align}
where  in analogy to above we introduce the quantities
\begin{align} 
\label{eq:g_tensor_noise}
     \pmb{\delta}_l&= (1-\eta)   \pmb{\delta}_l^1+\eta \pmb{\delta}_l^2 \ , \\
    \bar{\pmb{\delta}}_l&= \eta\pmb{\delta}_l^1+ (1-\eta)  \pmb{\delta}_l^2 \ , \\
    \Delta\pmb{\delta}_l&=\frac{\pmb{\delta}_l^1-\pmb{\delta}_l^2}{2} \ ,
\end{align}
and, explicitly, 
\begin{equation}
\sin(2\theta)= \frac{2t_c}{\hbar\Omega}=2\sqrt{\eta(1-\eta) } \ , \ \ \cos(2\theta)=\frac{\FMdetuning}{\hbar\Omega} = 1-2\eta\ .
\end{equation}

Including perturbatively the hybridization between antibonding and bonding states via the second order SW generator $S$ in Eq.~\eqref{eq:S-gen}, we find, keeping only the terms linear in $B$,

\textcolor{black}{\begin{subequations}
\label{eq:Hnoise}    
\begin{align}
 H_{\text{eff},N}&=\frac{\delta_\epsilon}{2}\cos(2\theta)\tau_z-\frac{\delta_\epsilon}{2}\sin(2\theta)\tau_x+\frac{\delta_\epsilon}{2} \sin(2\theta)\frac{  {{\Delta \textbf{l}}\cdot\pmb{\sigma}}}{\Omega}\left(
\begin{array}{cc}
-\sin(2\theta) & -\cos(2\theta)\\
 -\cos(2\theta)  &\sin(2\theta)\\
\end{array}
\right)\\
 &+{\delta_t}\sin(2\theta)\tau_z+{\delta_t}\cos(2\theta)\tau_x+ \delta_t\sin(2\theta)\frac{ {{\Delta \textbf{l}}\cdot\pmb{\sigma}}}{\Omega}\left(
\begin{array}{cc}
\cos(2\theta) & -\sin(2\theta)\\
 -\sin(2\theta)  &-\cos(2\theta)\\
\end{array}
\right) \\
&+ \frac{1}{2}\left(
\begin{array}{cc}
 {\bar{\pmb{\delta}_l}}\cdot\pmb{\sigma} & \sin(2\theta){\Delta\pmb{\delta}_l} \cdot \pmb{\sigma}\\
 \sin(2\theta){\Delta\pmb{\delta}_l} \cdot \pmb{\sigma} &{\pmb{\delta}_l}\cdot\pmb{\sigma} \\
\end{array}
\right)  \ + \mathcal{O}(B^2) \ .
\end{align}
\end{subequations}}

\subsubsection{First-order decoherence process}
From the noise Hamiltonian in Eq.~\eqref{eq:Hnoise}, we can directly extract the matrix elements resulting in dephasing and relaxation for the FM qubit.
By diagonalizing the qubit energy by the transformation in Eq.~\eqref{eq:utrafo}, which rotates the vectors defined in lab into the qubit frame
\textcolor{black}{\begin{equation}
    \Delta\textbf{l}\to \tilde{\Delta \textbf{l}}=R_y(-2\phi)R_z(-2\psi)\Delta \textbf{l} \ , \ \  \pmb{\delta}_l\to \tilde{\pmb{\delta}_l}=R_y(-2\phi)R_z(-2\psi)\pmb{\delta}_l
\end{equation}}
we find
\begin{equation}
    H_\text{FM,N}=\frac{\tilde{\pmb{\delta}_N}\cdot\pmb{\sigma}}{2} \ \ ,\ \ \ \tilde{\pmb{\delta}_N}= \tilde{\pmb{\delta}_l}+\frac{\tilde{\Delta \textbf{l}}}{\Omega} \left[\sin(2\theta)\textcolor{black}{^2} \delta\epsilon-2\cos(2\theta)\textcolor{black}{\sin(2\theta)}\delta t \right]\ ,
\end{equation}
including fluctuations of $g$-tensors that directly couple to the spin and of detuning that couple to the spin via spin-orbit interaction.
Dephasing is characterized by the longitudinal part of $\tilde{\pmb{\delta}_N}$ ($\tilde{\pmb{\delta}_N})_z$,
\begin{equation}
\begin{aligned}(\tilde{\pmb{\delta}_N})_z&=\pmb{\delta}_{N}\cdot\frac{\textbf{l}}{|\textbf{l}|} = \frac{(1-\eta)^2\pmb{\delta}_l^1\cdot\textbf{l}_1+\eta^2\pmb{\delta}_l^2\cdot\textbf{l}_2+\eta(1-\eta)(\pmb{\delta}_l^1\cdot\textbf{l}_2+\pmb{\delta}_l^2\cdot\textbf{l}_1)}{\Delta E} \\&+2\sqrt{(1-\eta)\eta}\frac{[(1-\eta)E_{z,1}+\eta E_{z,2}](E_{z,1}-E_{z,2})+2E_{z,1}E_{z,2}(1-2\eta)\sin^2(\theta_{12}/2)}{\Omega \Delta E }\left[\sqrt{\eta(1-\eta) } \delta\epsilon+(1-2\eta)\delta t \right] \ ,
\end{aligned}
\end{equation}
and relaxation by its transversal part ($\tilde{\pmb{\delta}_N})_\perp$, 
that can be written as 
\begin{align}
(\tilde{\pmb{\delta}_N})_\perp^2&= \left| \pmb{\delta}_{N}\times\frac{\textbf{l}}{|\textbf{l}|} \right| ^2 =(\tilde{\pmb{\delta}_N})_x^2+(\tilde{\pmb{\delta}_N})_y^2=|\tilde{\pmb{\delta}_N}|^2-(\tilde{\pmb{\delta}_N})_z^2 \ . \\
|\tilde{\pmb{\delta}_N}|^2&\approx\sum_{n=(x,y,z)}(\pmb{\delta}_l)_{n}^2+\frac{4(1-\eta)\eta}{\Omega^2}\left[(E_{z,1}-E_{z,2})^2+4E_{z,1}E_{z,2}\sin^2\left(\frac{\theta_{12}}{2}\right)\right] \left[\eta(1-\eta) \delta\epsilon^2+(1-2\eta)^2\delta t^2 \right] \ .
\end{align}
We neglect cross terms coupling fluctuations of different parameters.
At zero-detuning $\eta=1/2$ and $\Omega=2t_c/h$, the FM qubit decoherence is determined by

\begin{equation}
(\tilde{\pmb{\delta}_N})_z= \frac{(\pmb{\delta}_l^1+\pmb{\delta}_l^2)\cdot(\textbf{l}_1+\textbf{l}_2)}{4\Delta E_\text{FM}} +\frac{E_{z,1}^2-E_{z,2}^2}{8 t_c \Delta E_\text{FM} } \delta\epsilon  \ , 
\label{eq:T2_sweetspot_1st_order}
\end{equation}
\begin{equation}
(\tilde{\pmb{\delta}_N})_\perp=\frac{|(\pmb{\delta}_l^1+\pmb{\delta}_l^2)\times(\textbf{l}_1+\textbf{l}_2)|}{4\Delta E_\text{FM}} +\frac{E_{z,1}E_{z,2}\sin\left(\theta_{12}\right)}{4t_c \Delta E_\text{FM} }\delta\epsilon \ .
\label{eq:T1_sweetspot_1st_order}
\end{equation}

\subsubsection{Second-order relaxation process}\label{sec:second-order relaxation process}

The second-order process involves the off diagonal block matrix elements. In the qubit frame, the modulus of transition matrix off diagonal block elements are (with $\tilde{M}_{\epsilon} = U_{\textbf{l}}^\dagger M_{\epsilon}U_{\textbf{l}}$, $\tilde{M}_{t} = U_{\textbf{l}}^\dagger M_{t}U_{\textbf{l}} $ and $\tilde{M}_{\pmb{\delta}}$):
\begin{equation}
\begin{aligned}
    &|\tilde{M}_{t,14}| = \left|\frac{|\epsilon_{12}|}{ \Omega}\sin(\frac{\gamma}{2})-\frac{4 t_{c}^2}{ \Omega^3}(|\Delta \textbf{l}_{\perp}|\cos(\frac{\gamma}{2})-\Delta \textbf{l}_{\parallel}\sin(\frac{\gamma}{2}))\right|,\\
    &|\tilde{M}_{t,23}| = \left|\frac{|\epsilon_{12}|}{ \Omega}\sin(\frac{\gamma}{2})+\frac{4 t_{c}^2}{\Omega^3}(|\Delta \textbf{l}_{\perp}|\cos(\frac{\gamma}{2})-\Delta \textbf{l}_{\parallel}\sin(\frac{\gamma}{2}))\right|\\
    &|\tilde{M}_{t,13}| = \left|\Big(\frac{|\epsilon_{12}|}{ \Omega}+\frac{4 t_{c}^2 \Delta \textbf{l}_{\parallel}}{ \Omega^3}\Big)\cos{(\frac{\gamma}{2})+\frac{4t_{c}^2 |\Delta l_{\perp}|}{ \Omega^3}\sin(\frac{\gamma}{2})} \right|,\\
    &|\tilde{M}_{t,24}| = \left|\Big(\frac{|\epsilon_{12}|}{ \Omega}-\frac{4 t_{c}^2 \Delta \textbf{l}_{\parallel}}{ \Omega^3}\Big)\cos{(\frac{\gamma}{2})-\frac{4t_{c}^2 |\Delta l_{\perp}|}{\Omega^3}\sin(\frac{\gamma}{2})} \right|
    \label{eq:Mtilde_t}
\end{aligned}
\end{equation}

\begin{equation}
\begin{aligned}
    &\left|\tilde{M}_{\epsilon,14}\right| = \left|\frac{t_{c}}{ \Omega}\sin(\frac{\gamma}{2})+\frac{t_{c}|\epsilon_{12}|}{ \Omega^3}(|\Delta \textbf{l}_{\perp}|\cos(\frac{\gamma}{2})-\Delta \textbf{l}_{\parallel}\sin(\frac{\gamma}{2}))\right|,\\
    &|\tilde{M}_{\epsilon,23}| = \left|\frac{t_{c}}{ \Omega}\sin(\frac{\gamma}{2})-\frac{t_{c}|\epsilon_{12}|}{\Omega^3}(|\Delta \textbf{l}_{\perp}|\cos(\frac{\gamma}{2})-\Delta \textbf{l}_{\parallel}\sin(\frac{\gamma}{2}))\right|,\\
    &|\tilde{M}_{\epsilon,13}| = \left|\Big(\frac{t_{c}}{ \Omega}-\frac{t_{c}|\epsilon_{12}| \Delta \textbf{l}_{\parallel}}{ \Omega^3}\Big)\cos(\frac{\gamma}{2})-\frac{t_{c}|\epsilon_{12}| |\Delta \textbf{l}_{\perp}|}{ \Omega^3}\sin(\frac{\gamma}{2})\right|,\\
    &|\tilde{M}_{\epsilon,24}| = \left|\Big(\frac{t_{c}}{ \Omega}+\frac{t_{c}|\epsilon_{12}| \Delta \textbf{l}_{\parallel}}{ \Omega^3}\Big)\cos(\frac{\gamma}{2})+\frac{t_{c}|\epsilon_{12}| |\Delta \textbf{l}_{\perp}|}{ \Omega^3}\sin(\frac{\gamma}{2})\right|,
    \label{eq:Mtilde_e}
\end{aligned}
\end{equation}where we define the transverse B field gradient $\Delta l_{\perp}=|\Delta l \times\hat{l}| = \frac{E_{z,1}E_{z,2} \sin(\theta_{12})}{2 \Delta E}$, longitudinal B-field gradient $\Delta l_{\parallel} = \frac{E_{z,1}^2-E_{z,2}^2}{4 \Delta E}+ \frac{\epsilon_{12}}{ \Omega} \frac{|\Delta \textbf{l}|^2}{\Delta E}$ and the angle $\gamma$ between the FM Zeeman vector in the bonding and anti-bonding states $\cos(\gamma) = \hat{\textbf{l}}\cdot\hat{\bar{\textbf{l}}}$.


At zero detuning ($\hat{\textbf{l}} = \hat{\bar{\textbf{l}}}$, $\gamma = 0$), we observe that the second-order channel survives only in the tunneling channel, mediated by spin-preserving inter-orbital tunneling
\begin{equation}
    |\tilde{M}_{t,13}|=|\tilde{M}_{t,24}|= |\frac{E_{z,1}^2-E_{z,2}^2}{8 t_{c} \Delta E_{FM}}|,
\end{equation} and spin flip inter orbital tunneling
\begin{equation}
|\tilde{M}_{t,14}|=|\tilde{M}_{t,23}|\approx \frac{|\Delta l_{\perp}|}{2t _{c}}=\frac{1}{4 t_{c}}\frac{E_{z,1}E_{z,2} \sin(\theta_{12})}{ \Delta E_{FM}}. 
\label{eq:Mtildet_2nd_order}
\end{equation}

\subsection{Relaxation}
\subsubsection{General model}

The complete Hamiltonian of the FM qubit including an environment reads
\begin{equation}
\tilde{H}=\tilde{H}_{\text{eff}}+\tilde{H}_\text{eff,N}+H_\text{bath}\ , 
\end{equation}
with $\tilde{H}_{\text{eff}}$ and $H_\text{eff,N}$ given in Eqs.~\eqref{eq:heff4by4diag} and~\eqref{eq:Hnoise} with an additional rotation $\tilde{H}_{\text{eff,N}} = \tilde{U}_{\textbf{l}}^\dagger H_{\text{eff,N}}\tilde{U}_{\textbf{l}}$ into the qubit frame, respectively, and $H_\text{bath}$ representing a general environment Hamiltonian. 
The bath-system Hamiltonian in interaction picture is
\begin{equation}
    \tilde{H}_{\text{eff, N}} = \sum_{\nu\in{\epsilon, t}}\tilde{M}_{\nu}\delta_{\nu} \ ,  \ \ \ \tilde{M}_{\nu} = \sum_{n,m}\tilde{M}_{\nu,nm}e^{i\omega_{nm}t}\ket{n}\bra{m},
    \label{eq:heff_n_M_delta}
\end{equation}$\tilde{M}_{\nu}$ is the bath-system coupling matrix for the $\nu^\mathrm{th}$ channel and $\delta_{\nu}$ is the fluctuation of noise of the $\nu^\mathrm{th}$ channel caused by the bath in the unit of energy. The detailed matrix elements are shown in Eq. ~\eqref{eq:Mtilde_t} and ~\eqref{eq:Mtilde_e}.
Consider the general multilevel system, the equation of motion of a qubit system operator $C$ under the framework of the quantum Langevin equation is
\begin{align}
&\dot{C}= -\frac{i}{\hbar}[C, \tilde{H}]-\sum_{\nu}\sum_{nm}{\Gamma_{\nu}^{+}(\omega_{nm})[C, \tilde{M}_{\nu,nm}]\tilde{M}_{\nu, nm}^\dagger -\Gamma_{\nu}^{-}(\omega_{nm})\tilde{M}_{\nu, n m}^\dagger[C, \tilde{M}_{\nu, n m}]},
\label{eq:master_equation}
\end{align} the asymmetric noise spectral density is
\begin{equation}
    \Gamma_{\nu}^{+}(\omega_{nm}) = \frac{1}{\hbar^2}\int_{0}^{\infty}d \tau e^{i\omega_{nm}\tau}\braket{\delta_{\nu}(\tau)\delta_{\nu}(0)},
    \label{eq:half_spectral_density_plus}
\end{equation}

\begin{equation}
    \Gamma_{\nu}^{-}(\omega_{n,m}) = \frac{1}{\hbar^2}\int_{0}^{\infty}d \tau e^{i\omega_{nm}\tau}\braket{\delta_{\nu}(0)\delta_{\nu}(\tau)}.
    \label{eq:half_spectral_density_minus}
\end{equation}

Relaxation processes in our system occur not only because of direct spin-flip transitions in the bonding orbital ground state, but also occur because of higher-order processes involving fast orbital transitions between bonding and antibonding states.
To describe this effect, we consider the rate equation for the populations $P_\mu = \ket{\mu}\bra{\mu}$ of each flopping mode state $\mu$. Since $P_{\mu}$ is a projection operator, the equation ~\eqref{eq:master_equation} becomes
\begin{equation}
\begin{aligned}
    &\dot{P}_{\mu}= \sum_{\nu}\sum_{n,m}{\Gamma_{\nu}^{+}(\omega_{n,m})[P_{\mu} ,\tilde{M}_{\nu,nm}]\tilde{M}_{\nu, n m}^\dagger -\Gamma_{\nu}^{-}(\omega_{n, m})\tilde{M}_{\nu, n m}^\dagger[P_{\mu}, \tilde{M}_{\nu, n m}]}\\
    &=\sum_{\nu}\sum_{n}|\tilde{M}_{\nu, \mu n}|^2\{[\Gamma_{\nu}^{+}(\omega_{n,\mu})+\Gamma_{\nu}^{-}(\omega_{\mu,n})]P_{n}-[\Gamma_{\nu}^{+}(\omega_{\mu,n})+\Gamma_{\nu}^{-}(\omega_{n,\mu})]P_{\mu}\},
\end{aligned} 
\end{equation}
where in the second line we have used $\tilde{M}_{\nu,nm}^\dagger[P_{\mu}, \tilde{M}_{\nu,nm}] = [\tilde{M}_{\nu,nm}, P_{\mu}],\tilde{M}_{\nu,nm}^\dagger = |\tilde{M}_{\nu,nm}|^2(\delta_{\mu n}P_{m} - \delta_{\mu m} P_{\mu})$. Then we can write a set of rate equations describing the occupation rate of each energy level $P_i$

\begin{equation}
\label{eq:final_master_equation}
    \dot{P}_\mu=\sum_j(W_{\mu j}P_j-W_{ j\mu }P_\mu)  = -W_\mu P_\mu+ \sum_{j\neq \mu}W_{\mu j}P_j \ ,
\end{equation}
where $W_{\mu j}=W_{j\to \mu}$ are the transition rates from state $j$ to state $\mu$ with energies $E_{j}$ and $E_\mu$, respectively, and we introduce the total rate $W_\mu=\sum_{j\neq \mu} W_{j\mu}$. 
The energy level transition rates are given by

\textcolor{black}{
\begin{align}
\label{eq:rel-rates-general}
  W_{nm}=  W_{m\to n}&=\sum_{\nu}|\tilde{M}_{\nu,nm}|^2[\Gamma_{\nu}^{+}(\omega_{mn})+\Gamma_{\nu}^{-}(\omega_{nm})] = \sum_{\nu}|\tilde{M}_{\nu,nm}|^2\frac{S_{\nu}(\omega_{nm})}{\hbar^2},
\end{align}} which connects to the symmetric power spectral density defined as
\begin{equation}
    S_{\nu}(\omega_{nm}) = \int_{- \infty}^\infty d\tau e^{- i \omega_{nm} \tau} \braket{\delta_{\nu}(\tau)\delta_{\nu}(0)}.
    \label{eq:psd}
\end{equation}

To estimate the relaxation rates of the FM qubit encoded in the spin state of the bonding state, we perform adiabatic elimination of the spin states in the antibonding states. This procedure assumes that the excited orbital states (antibonding) quickly decay to the bonding state and their population \textcolor{black}{transition rate} is constantly zero; we thus set $\dot{P}_{3,4} = 0$. We also note that the direct spin transition between the spin antibonding states is much slower than the orbital transition rate, thus for simplicity we neglect the corrections coming from $W_{34}$ and $W_{4,3}$. Solving for $P_{3}$ and $P_{4}$ and inserting back into the equation \ref{eq:final_master_equation}, we find
\begin{align}
    \dot{P}_{1} &= -P_{1}\left(W_{1}-\frac{W_{13}W_{31}}{W_{3}}-\frac{W_{14}W_{41}}{W_{4}}\right)+P_{2}\left(W_{12}+\frac{W_{13}W_{32}}{W_{3}}+\frac{W_{14}W_{42}}{W_{4}}\right) \ , \\
    \dot{P}_{2} &= P_{1}\left(W_{21}+\frac{W_{23}W_{31}}{W_{3}}+\frac{W_{24}W_{41}}{W_{4}}\right) -P_{2}\left(W_{2}-\frac{W_{23}W_{32}}{W_{3}}-\frac{W_{24}W_{42}}{W_{4}}\right) \ .
\end{align}

From this system of equations, we find the relaxation rate $1/T_1$ as the sum of its off-diagonal elements:
\begin{equation}
\frac{1}{T_1} = \frac{1}{T_1^{(1)}}+ \frac{1}{T_1^{(2)}} \ , 
\end{equation} 
where we distinguish between first and second order processes
\begin{equation}
\begin{aligned}
\label{eq:T1_first_second_order}
\frac{1}{T_1^{(1)}}&= W_{12}+W_{21} \ , \\
\frac{1}{T_1^{(2)}} &=  \frac{W_{13}W_{32}+W_{23}W_{31}}{W_{3}}+\frac{W_{14}W_{42}+W_{24}W_{41}}{W_{4}} \ .
\end{aligned} 
\end{equation}

We now discuss explicitly different possible sources of relaxation.

\subsubsection{Relaxation by a bosonic bath}

We model the environment by considering a bosonic bath of $n$ orthogonal modes with wavevector $\textbf{q}$ and frequency $\omega_{n,\textbf{q}}$, i.e., $H_\text{bath}=\sum_{n,\textbf{q}}\hbar\omega_{n,\textbf{q}}a_{\textbf{q},n}^\dagger a_{\textbf{q},n}$, that couples to tunneling, detuning, and $g$-tensor as
\begin{align}
\delta_\epsilon=\sum_{n,\textbf{q}} \delta\epsilon_{n,\textbf{q}}^* a_{n,\textbf{q}}+ \delta\epsilon_{n,\textbf{q}}a^\dagger_{n,\textbf{q}} \ , \ \ \ \delta_t= \sum_{n,\textbf{q}} \delta t_{n,\textbf{q}}^* a_{n,\textbf{q}}+  \delta t_{n,\textbf{q}} a^\dagger_{n,\textbf{q}}\ , \ \ \ \pmb{\delta}_l= \sum_{n,\textbf{q}} \delta \textbf{l}_{n,\textbf{q}}^* a_{n,\textbf{q}}+  \delta \textbf{l}_{n,\textbf{q}} a^\dagger_{n,\textbf{q}} \ .
\label{eq:bosonic_system_coupling}
\end{align} 
We postpone the discussion about the explicit form of the couplings $\delta\epsilon_{n,\textbf{q}}, \ \delta t_{n,\textbf{q}}, \ \delta \textbf{l}_{n,\textbf{q}}$ to the next section and provide here general results for arbitrary bosonic baths.

For convenience, we group all the prefactors of $a_{n,\textbf{q}}$ in $H_\text{eff,N}$ ~\eqref{eq:Hnoise} in definition ~\eqref{eq:heff_n_M_delta}. The system-environment coupling Hamiltonian is
\begin{equation}
    \tilde{H}_\text{eff,N} = \tilde{M}_{\epsilon}\delta_{\epsilon}+\tilde{M}_{t}\delta_{t},
\end{equation} with $\tilde{M}_{\epsilon} = U_{\textbf{l}}^\dagger M_{\epsilon}U_{\textbf{l}}$ and $\tilde{M}_{t} = U_{\textbf{l}}^\dagger M_{t}U_{\textbf{l}}$  where
\begin{align}
    \label{eq:Me_analytical}
    M_{\epsilon} = \frac{1}{2}\cos(2\theta)\tau_{z} -\frac{1}{2}\sin(2\theta) \tau_{x}+\frac{\Delta \textbf{l}\cdot\sigma}{2\Omega}\sin(2\theta)\left(
\begin{array}{cc}
-\sin(2\theta) & -\cos(2\theta)\\
 -\cos(2\theta)  &\sin(2\theta)\\
\end{array}
\right)
\end{align}

\begin{align}
        \label{eq:Mt_analytical}
        M_{t} = \sin(2\theta)\tau_{z} +\cos(2\theta) \tau_{x}+\frac{\Delta \textbf{l}\cdot\sigma}{\Omega}\sin(2\theta)\left(
\begin{array}{cc}
\cos(2\theta) & -\sin(2\theta)\\
 -\sin(2\theta)  &-\cos(2\theta)\\
\end{array}
\right)
\end{align}
Further, the g-tensor can be modulated by the gate voltage, as shown in Tab. \ref{tab:gtensor_deriv_parameters}. We connect the bosonic-bath-induced gate fluctuation to the local g-tensor fluctuation by  $$\pmb{\delta}_{l}^{i}=\mu_{B}\textbf{B}\cdot (\frac{\partial \textbf{g}_{i}}{\partial \bar{V}_{P2}}\frac{\alpha_{21}+\alpha_{11}}{\alpha_{11}\alpha_{22}}-\frac{\partial \textbf{g}_{i}}{\partial \bar{V}_{P1}}\frac{1}{\alpha_{11}}) \delta_\epsilon,$$ resulting in additional transition components to the detuning channel
$\tilde{M}_{\pmb{\delta_{l}}} = U^{\dagger}_{\textbf{l}}M_{\pmb{\delta_{l}}}U_{\textbf{l}}$:
\begin{equation}
    M_{\pmb{\delta_{l}}} =  \frac{1}{2}\left(
\begin{array}{cc}
 {\bar{\pmb{\delta}_l}}\cdot\pmb{\sigma} & \sin(2\theta){\Delta\pmb{\delta}_l} \cdot \pmb{\sigma}\\
 \sin(2\theta){\Delta\pmb{\delta}_l} \cdot \pmb{\sigma} &{\pmb{\delta}_l}\cdot\pmb{\sigma} \\
\end{array}
\right)  \ .
\end{equation}

We will show that the gate modulated g-tensor noise has a small contribution compared to the detuning and tunneling noise channels in the FM and small-detuning regime. In the large-detuning regime, where we return to the LD qubit limit, relaxation is dominated by the gate-modulated g-tensor noise channel that couples to the Johnson-Nyquist noise source. 

Assuming that the bath is in a thermal state with population $N_{n,\textbf{q}}=(e^{\hbar\omega_{n,\textbf{q}}/k_BT}-1)^{-1}$ the relaxation rates in Eq.~\eqref{eq:rel-rates-general} equation reduces to the Fermi Golden rule (using $\delta(\hbar\omega_i)\to\frac{1}{\hbar}\delta(\omega_{i})$)
\begin{align}
\label{eq-rate-bosons}
     W_{ij}=W_{j\to i}&=\frac{2\pi}{\hbar}\sum_{\nu,n,\textbf{q}}|\delta_{n,\textbf{q}}|^2|\tilde{M}_{\nu, ij}|^2 \Big[N_{n,\textbf{q}} \delta(E_{i}-E_j-\hbar\omega_{n,\textbf{q}}) + (1+N_{n,\textbf{q}}) \delta(E_{i}-E_j+\hbar\omega_{n,\textbf{q}}) \Big] \ = \frac{1}{\hbar^2}\sum_{\nu} S_{\nu}(\omega_{ij})|\tilde{M}_{\nu, ij}|^2 ,  \\
     W_{ji}=W_{i\to j}&=W_{j\to i}e^{(E_i-E_j)/k_BT} \ ,
\end{align}
with the power spectral density ~\eqref{eq:psd} under the rotating wave approximation (dropping all $\delta(2\omega_{ij}\pm\omega_{n,\mathbf{q}})$ terms) is
\begin{equation}
    S_{\delta}(\omega_{i,j}) = 2 \pi\sum_{n,\mathbf{q}}|\delta_{n, \mathbf{q}}|^2[N_{n,\mathbf{q}}\delta(\omega_{ij}-\omega_{n,\mathbf{q}})+(N_{n,\mathbf{q}}+1)\delta(\omega_{ij}+\omega_{n,\mathbf{q}})]
\end{equation}
leading to
\begin{equation}
\begin{aligned}
\Gamma_{1}^{(1)} = \frac{1}{T_1^{(1)}}&=  \frac{2\pi}{\hbar}\coth\left(\frac{\Delta E}{2k_BT}\right)\sum_{\nu, n,\textbf{q}} |\delta_{n,\mathbf{q}}|^2|M_{\nu, ij}|^2  \delta(\Delta E-\hbar\omega_{n,\textbf{q}})\ , \\
\Gamma_{1}^{(2)} = \frac{1}{T_1^{(2)}} & \approx 2e^{-\frac{\Omega}{k_BT}}\left(\frac{1}{W_{13}^{-1}+W_{23}^{-1}}+\frac{1}{W_{14}^{-1}+W_{24}^{-1}} \right)\ .
\label{eq:relaxation_T1_T2}
\end{aligned}
\end{equation}
In the last approximation we considered that in our experiment thermal energy $k_\mathrm{B}T$ is higher than the Zeeman splitting, but not larger than the orbital energy $\Omega$. This equation shows that second-order decay rates are exponentially suppressed as a function of the orbital energy.

\subsubsection{Phonons}
We now estimate explicitly the couplings $\delta\epsilon_{n,\textbf{q}}, \ \delta t_{n,\textbf{q}}$ caused by phonons.
A phonon bath couples to the charge degree of freedom by 
\begin{equation}
H_\text{ph-charge}= i \sum_{n,\textbf{q}} C_{n,\textbf{q}} \Big(\Phi_n(\textbf{q}) e^{i\textbf{q}\cdot \textbf{r}}a_{n,\textbf{q}}-\Phi_n^*(\textbf{q})e^{-i\textbf{q}\cdot \textbf{r}}a^\dagger_{n,\textbf{q}}\Big) \ ,
\end{equation}
where the potential for a 3D phonon with wavevector $\textbf{q}$ and polarization $\textbf{c}_n$ coupled to pure HH states (confined along z) is \cite{PhysRevLett.95.076805}
\begin{equation}
\Phi_n(\textbf{q}) =  \left(a+\frac{b}{2}\right) \textbf{q}\cdot \textbf{c}_n -\frac{3}{2} b q_z (\textbf{c}_n)_z \ ,
\end{equation}
with normalization
\begin{equation}
    C_{n,\textbf{q}}=\sqrt{\frac{\hbar}{2\rho V \omega_{n,\textbf{q}}}} \ .
\end{equation}
Here, $V$ is the volume of the sample, $\rho$ is the density, and $a$, $b$ are deformation potentials. We neglect anisotropies in the deformation potential.

There are 3 bulk phonon modes, one longitudinal with velocity $v_l=\sqrt{(\lambda+2\mu)/\rho}$ and $\textbf{c}_l =\textbf{q}/q=(\sin \Theta  \sin \Phi ,\sin \Theta  \cos \Phi ,\cos \Theta )$ and two transversal with velocity $v_t=\sqrt{\mu/\rho}$  and $\textbf{c}_{t1}\perp \textbf{q}$ (e.g. $(-\cos\Phi,\sin\Phi,0)$) and $\textbf{c}_{t2}=\textbf{c}_{t1}\times \textbf{q}=(\cos \Theta  \sin \Phi ,\cos \Theta  \cos \Phi ,-\sin \Theta )$ \cite{PhysRevB.84.195314}.

Using spherical coordinates for $\textbf{q}$
\begin{align}
\Phi_l(\textbf{q}) &=  q \left[\left(a+\frac{b}{2}\right) -\frac{3}{2} b  \cos^2(\Theta)\right] \ , \\
\Phi_{t1}(\textbf{q}) &=0 \ , \\
\Phi_{t2}(\textbf{q})&=\Phi_{t}(\textbf{q}) = \frac{3b}{4} q \sin(2\Theta)  \ ,
\end{align}
such that the coupling is only from two phonon modes.

In a DQD the coupling to detuning and tunneling is 
\begin{equation}
H_\text{ph-DQD}= \frac{1}{2}\tau_z \sum_{n,\textbf{q}} \big(\delta\epsilon_{n,\textbf{q}} a_{\textbf{q},n}+\delta\epsilon_{n,\textbf{q}}^* a^\dagger_{\textbf{q},n}\Big) + \tau_x \sum_{n,\textbf{q}}  \big(\delta t_{n,\textbf{q}} a_{\textbf{q},n}+\delta t_{n,\textbf{q}}^* a^\dagger_{\textbf{q},n}\Big)
\end{equation}
with
\begin{align}
\delta\epsilon_{n,\textbf{q}}&=\Phi_n(\textbf{q})C_{n,\textbf{q}}  \left[\langle \psi_L | e^{i \textbf{q}\cdot \textbf{r}}|\psi_L\rangle - \langle \psi_R | e^{i \textbf{q}\cdot \textbf{r}}|\psi_R\rangle \right] \ , \\
\delta t_{n,\textbf{q}}&= \frac{\Phi_n(\textbf{q})}{2}C_{n,\textbf{q}} \Big[\langle \psi_L | e^{i \textbf{q}\cdot \textbf{r}}|\psi_R\rangle + \langle \psi_R | e^{i \textbf{q}\cdot \textbf{r}}|\psi_L\rangle \Big] \ .
\end{align}
We neglect here the direct coupling of the phonons with the spin degree of freedom through the electric tunability of the $g$-tensor.

Assuming for simplicity that the two QDs are identical isotropic  Gaussians with width $l$, shifted by $\pm d/2$, having an overlap $s=e^{-d^2/4l^2}\ll1$,  we find that
\begin{subequations}
\label{eq:delta t and episol}
\begin{align}
\delta\epsilon_{n,\textbf{q}}&\approx \textcolor{black}{2}C_{n,\textbf{q}}  {\Phi_n(\textbf{q})}e^{-\frac{q^2 l^2}{4}} \sin \left(\frac{qd}{2}  \sin \Theta \sin \Phi\right)\ , \\
\delta t_{n,\textbf{q}}&\approx  C_{n,\textbf{q}}  \Phi_n(\textbf{q}) s e^{-\frac{q^2 l^2}{4}  }\ \ .
\end{align}
\end{subequations}

Combining Eqs.~\eqref{eq:delta t and episol} and~\eqref{eq-rate-bosons}, we find that the transition rates can be written as
\begin{equation}
     W_{ij}=|\tilde{M}_{\epsilon,ij}|^2 \sum_n(\gamma_\epsilon)_{ij}^n + |\tilde{M}_{ t,ij}|^2 \sum_n(\gamma_t)_{ij}^n \ ,
     \label{eq:Wij}
\end{equation}
with
\begin{align}
    (\gamma_\epsilon)_{ij}^n&=\frac{2\pi}{\hbar}\sum_{\textbf{q}}|\delta \epsilon_{n,\textbf{q}}|^2 \Big[N_{n,\textbf{q}} \delta(E_{i}-E_j-\hbar\omega_{n,\textbf{q}}) + (1+N_{n,\textbf{q}}) \delta(E_{i}-E_j+\hbar\omega_{n,\textbf{q}}) \Big] \\
    (\gamma_t)_{ij}^n&=\frac{2\pi}{\hbar}\sum_{\textbf{q}}|\delta t_{n, \textbf{q}}|^2 \Big[N_{n,\textbf{q}} \delta(E_{i}-E_j-\hbar\omega_{n,\textbf{q}}) + (1+N_{n,\textbf{q}}) \delta(E_{i}-E_j+\hbar\omega_{n,\textbf{q}}) \Big] \ . 
\end{align}
Using $\sum_{q}\rightarrow\frac{V}{(2\pi)^3} \int q^2 \sin(\theta_{q})dq d\theta_{q}d\phi_{q}$, $\omega_{n,\textbf{q}} = v_n q$, $N(\omega_{n,\textbf{q}})=N_{n,\textbf{q}}$ $\hbar\omega_{ij}=E_i-E_j$, and $q_{ij}^n = \omega_{ij}/ v_n$ we find
\begin{equation}
\begin{aligned}
    (\gamma_\epsilon)_{ij}^n&= \frac{(q_{ij}^n)^2 e^{-\frac{(q_{ij}^n)^2 l^2}{2}}}{2\rho\pi^2\hbar \omega_{ij} v_n} \Big[N(\omega_{ij})\Theta(\omega_{ij}) + [1+N(\omega_{ji})]\Theta(\omega_{ji}) \Big]\int   d\theta_{q}d\phi_{q} |\Phi_n(q_{ij}^n,\theta_q)|^2 \sin\theta_{q} \sin^2 \left(\frac{q_{ij}^nd}{2}  \sin \theta_q \sin \phi_q\right) \\
    &\approx \frac{(q_{ij}^n)^{2}e^{-\frac{(q_{ij}^n)^2 l^2}{2}}}{2\rho\pi^2\hbar \omega_{ij} v_n} \Big[N(\omega_{ij})\Theta(\omega_{ij}) + [1+N(\omega_{ji})]\Theta(\omega_{ji}) \Big]I_\epsilon^{n}, n \in \{l, t2\} 
    \label{eq:phonon_level_transition_rate_gammae}
\end{aligned}
\end{equation}
\begin{equation}
\begin{aligned}
    (\gamma_t)_{ij}^n&= s^2\frac{(q_{ij}^n)^2e^{-\frac{(q_{ij}^n)^2 l^2}{2}}}{8\rho\pi^2\hbar \omega_{ij} v_n} \Big[N(\omega_{ij})\Theta(\omega_{ij}) + [1+N(\omega_{ji})]\Theta(\omega_{ji}) \Big]\int   d\theta_{q}d\phi_{q} |\Phi_n(q_{ij}^n,\theta_q)|^2 \sin\theta_{q} \\
    &\approx s^2\frac{(q_{ij}^n)^2e^{-\frac{(q_{ij}^n)^2 l^2}{2}}}{8\rho\pi^2\hbar \omega_{ij} v_n} \Big[N(\omega_{ij})\Theta(\omega_{ij}) + [1+N(\omega_{ji})]\Theta(\omega_{ji}) \Big]I_t^{n}, n \in \{l, t2\}   , \label{eq:phonon_level_transition_rate_gammat}
\end{aligned} 
\end{equation}
where the step function $\Theta(\omega)$ is defined as
\begin{equation}
    \Theta(x)=\begin{cases}1, & x>0\\ 0, & x<0\end{cases}.
\end{equation}

Here, each integration has an analytical expression:
\begin{equation}
\begin{aligned}
\label{eq:phonon_detune_integrand}
 I_\epsilon^{l} &=\frac{\pi  x_l^2}{d^2} \left(a^2 \left(2-2 J_0\left(x_l\right)\right)+a b \left(\frac{6 J_1\left(x_l\right)}{x_l}-2 J_0\left(x_l\right)\right)+b^2 \left(-\frac{J_0\left(x_l\right)}{2}+\frac{3 J_1\left(x_l\right)}{x_l}-\frac{27 J_2\left(x_l\right)}{2 x_l^2}+\frac{2}{5}\right)\right)\\
I_\epsilon^{t2} &=\frac{\pi x_t^2}{d^2}  b^2 \left(\frac{3}{5}-\frac{9 J_1\left(x_t\right)}{2 x_t}+\frac{27 J_2\left(x_t\right)}{2 x_t^2}\right),
\end{aligned}
\end{equation}
and
\begin{equation}
\begin{aligned}
\label{eq:phonon_tunneling_integrand}
 I_t^{l} &=\frac{4 \pi  \left(5 a^2+b^2\right) x_l^2}{5 d^2}\\
I_t^{t2} &=\frac{6 \pi  b^2 x_t^2}{5 d^2},
\end{aligned}
\end{equation} where we have defined $x_l = \frac{d\omega_{ij}}{v_l}$ and $x_t = \frac{d \omega_{ij}}{v_{t}}$. $J_{n}$ is the $n$-th Bessel function.
\

\begin{figure}[htbp]
    \centering
    \includegraphics[width=1\linewidth]{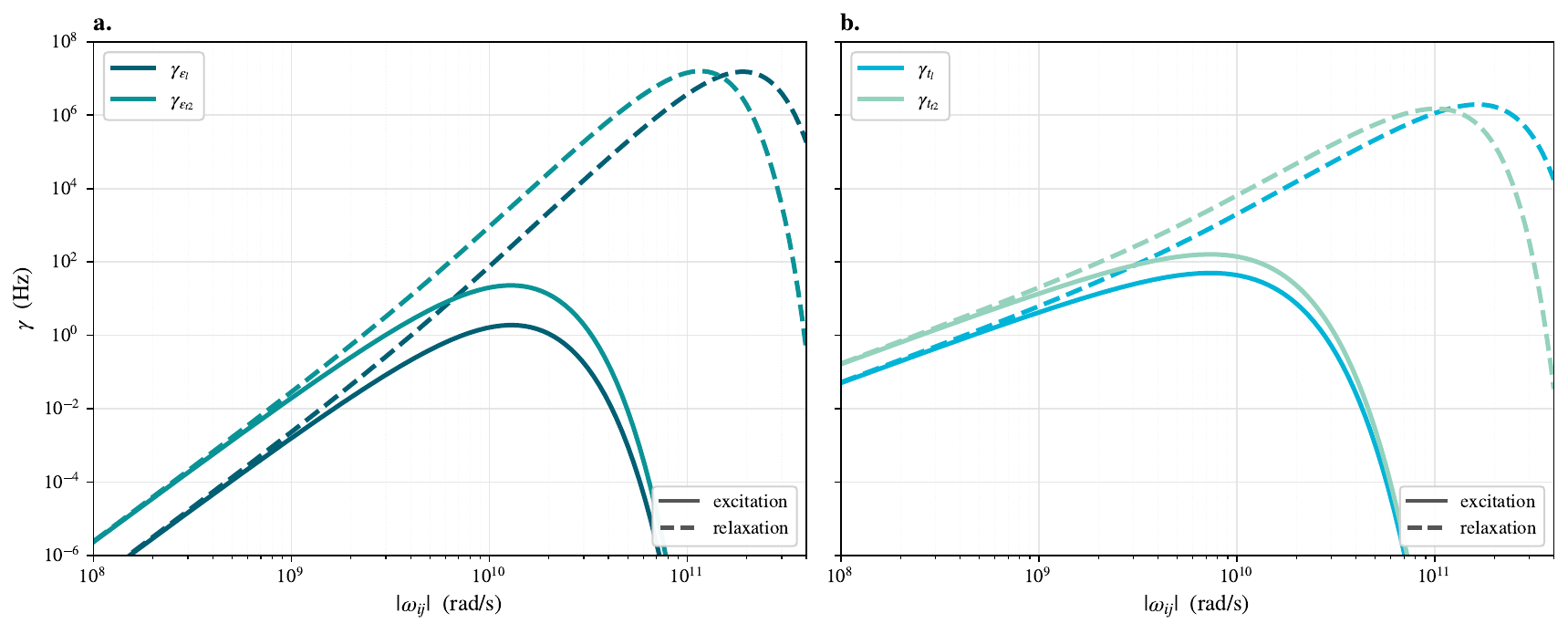}
    \caption{The excitation and relaxation rate of energy levels for  $\epsilon$ channel and $t$ channel are presented in \textbf{a} using Eq. ~\eqref{eq:phonon_level_transition_rate_gammae} and in \textbf{b} using Eq.~\eqref{eq:phonon_level_transition_rate_gammat} respectively.}
    \label{fig:phonon_rate}
\end{figure}

We introduce the parameter that we used for estimating the energy level transition rate induced by phonon bath in Eq. ~\eqref{eq:phonon_level_transition_rate_gammae} and ~\eqref{eq:phonon_level_transition_rate_gammat}. The hole-phonon coupling is governed by the valence-band absolute and shear deformation potentials, with parameters $a_{v} = 2.1$ eV \cite{PhysRevB.85.205308} and $b = -2.32$ eV \cite{guilloy2016germanium}, respectively. The Lamé constants, estimated from the elastic constants of germanium with the Voigt‐Reuss‐Hill Approximation \cite{voigt1928lehrbuch,reuss1929berechnung,hill1952elastic} ($c_{11} = 126$ GPa, $c_{12} = 44$ GPa, and $c_{44} = 67.7$ GPa), yield $\lambda = 34.4$GPa and $\mu = 55.4$GPa. The longitudinal and transverse sound velocities are then calculated from the Lamé constants, $c_l =\sqrt{(\lambda + 2\mu)/\rho} =5220 $ m/s and $c_t = \sqrt{\mu/\rho} = 3230$ m/s, using a Ge mass density of $\rho = 5330$ kg/m$^3$ \cite{madelung2004semiconductors}. The geometry of the double quantum dot is modeled with an inter-dot distance of $d= 100$ nm and a characteristic dot wavelength of $l_{dot} = 55$ nm. The effective temperature of the phonon thermal bath is assumed to be $T_{ph} = 20$ mK. Furthermore, the spin-orbit rotation and the effective magnetic field orientation are defined by the angles $\theta_{so} \approx 14.7^\circ$, $\theta_n \approx 78.3^\circ$, and $\phi_n \approx -33.7^\circ$.

The results of the energy level transition rates induced by phonon are plotted in Fig.\ref{fig:phonon_rate}. Around the FM qubit frequency ($\omega_{ij}<1$GHz), we observe that for both channels, the excitation and relaxation are comparable, indicating that the system is immersed in a phonon bath. At the higher frequency, the relaxation is larger than the excitation by several orders of magnitude, indicating the consequences of spontaneous emission and detailed balance.

\subsubsection{Johnson noise}

We consider the relaxation induced by thermal photon, the transition matrix $\tilde{M}$ in Eq.~\eqref{eq:Wij} is identical to the phonon case but with different level transition rate given by
\begin{align}
    &(\gamma_{\nu})_{ij} =\frac{S_{\nu }(\omega_{ij})}{\hbar^2}= \frac{|\alpha_{\nu}|^2}{ \hbar^2}2\pi\Big[|\delta(\omega_{ij})|^2 N(\omega_{ij})\Theta(\omega_{ij})+|\delta(\omega_{ji})|^2[N(\omega_{ji})+1]\Theta(\omega_{ji})\Big].
    \label{eq:johnson_level_transition_rate}
\end{align} 
Here we consider the general bosonic bath coupled to the system via the fluctuation of the top gates, with the same voltage fluctuation amplitude $\delta$ for different channels $\nu \in\{t, \epsilon\}$. This describes the fluctuation of the gates induced shifting in detuning and tunneling ~\eqref{eq:bosonic_system_coupling}. The lever arms $\alpha_{\nu}$ are extracted from the experiment.

We consider the Johnson-Nyquist noise, with asymmetric power spectral density
\begin{equation}
    S_{jn, \nu}(\omega_{ij}) =  2|\alpha_{\nu}|^2 R_{J}\Big[  \hbar \omega_{ij} N(\omega_{ij}) \Theta(\omega_{ij})+\hbar \omega_{ji} (N(\omega_{ji})+1)\Theta(\omega_{ji})\Big]
    \label{eq:johnson_nyquist} \ .
\end{equation} This is equivalent to the standard symmetric Johnson-Nyquist PSD for $\omega_{ij} = -\omega_{ji}>0$ 
\begin{equation}
    S_{VV}(\omega_{ij}) = S_{\nu}(\omega_{ij}) + S_{\nu}(\omega_{ji}) =  2|\alpha_{\nu}|^2 R_{J}\Big[  \hbar \omega_{ij} N(\omega_{ij}) +\hbar \omega_{i j} (N(\omega_{ij})+1)\Big] = 2  |\alpha_{\nu}|^2R\hbar \omega_{ij}\coth{\frac{\hbar \omega_{ij}}{2 k_{B}T}}, 
\label{eq:johnson_nyquist_symmetric}
\end{equation}

In the high temperature limit, this results in  white noise with
\begin{equation}
    S_{VV}(\omega_{ij}) = 4  \alpha_{\nu}^2R_{J} k_{B}T  \ .
\end{equation}

The Johnson-Nyquist noise-equivalent resistance is extracted from the experimental PSD by measuring thermal photons and converting the signal into resistance, which gives $R_{J} \approx 250 \Omega$ at temperature $T = 300$ mK.

\begin{figure}[htbp]
    \centering
    \includegraphics[width=1\linewidth]{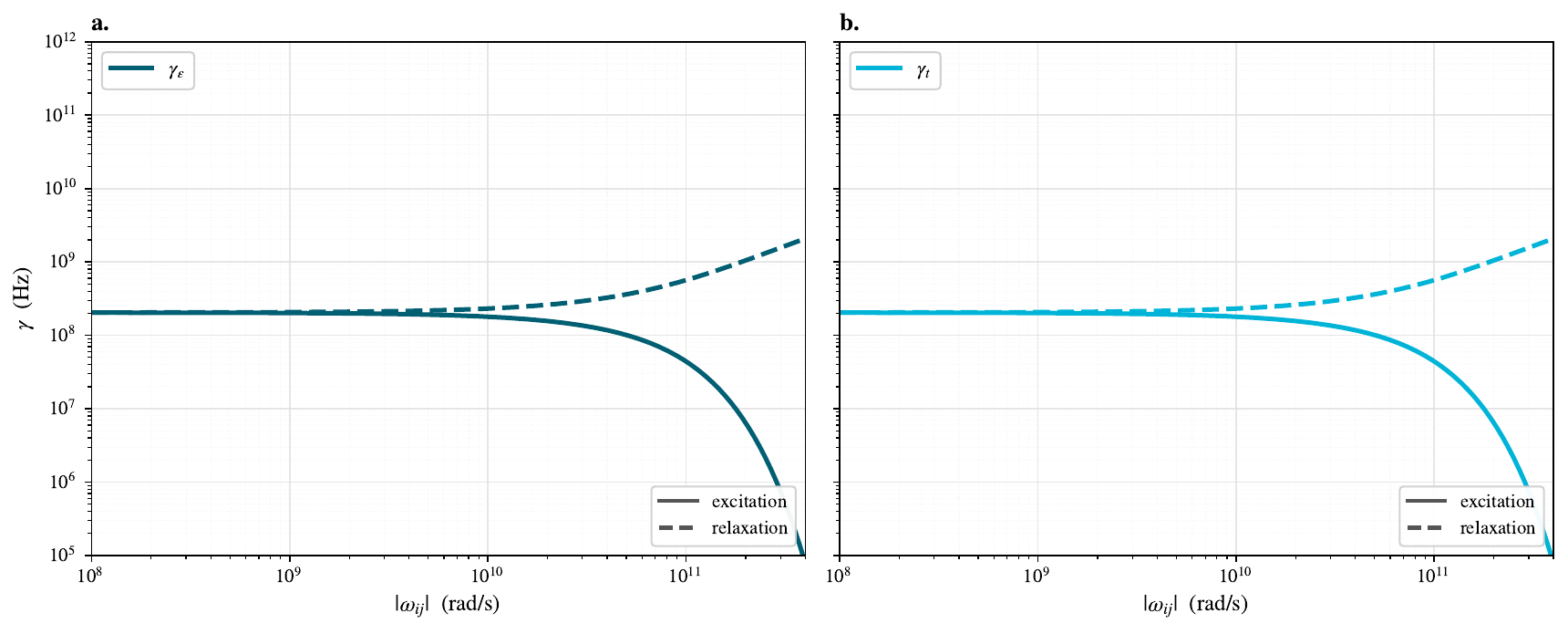}
    \caption{The excitation and relaxation rate of energy levels for  $\epsilon$ channel and $t$ channel are presented in \textbf{a} and in \textbf{b} using Eq.~\eqref{eq:johnson_level_transition_rate}. We assumed a equal lever arm for detuning and tunnling channel ($\alpha_{t} = \alpha_{\epsilon}$).}
    \label{fig:photon_rate}
\end{figure}


\subsubsection{1/f noise}
The  power spectral density of $1/f$ charge noise is 
\begin{equation}
\label{eq:1f_PSD}
    S_{1/f}(\omega_{nm})= S_{1/f}(-\omega_{nm}) = \int_{- \infty}^\infty d\tau e^{- i \omega_{nm} \tau}\braket{\delta(\tau)\delta(0)} = \frac{A}{|\omega_{nm}/2\pi|^{\alpha}} ,
\end{equation} where $\delta$ is defined as the classical fluctuation, and $A$ is the amplitude of gate fluctuations at 1 Hz in units of $V^2$.


The transition rate equation ~\eqref{eq:rel-rates-general} holds generally for any system-environment coupling; therefore, the transition rate associated with $1/f$ charge noise is

\begin{equation}
    W_{nm} = |M_{\epsilon,nm}|^2\gamma_\epsilon(\omega_{nm})+|M_{t,nm}|^2\gamma_t(\omega_{nm}), \qquad \gamma_{\nu}(\omega_{nm}) = \begin{cases}
\frac{|\alpha_{\epsilon}|^2 S_{1/f}(\omega_{nm})}{\hbar^2}\\
\frac{|\alpha_{t}|^2 S_{1/f}(\omega_{nm})}{\hbar^2}
    \end{cases}.
    \label{eq:1f_level_transition_rate}
\end{equation}
The exponent extracted at zero detuning is $\alpha = 1$, and the PSD after conversion to energy is experimentally determined to be $\alpha_{\epsilon}^2S_{1/f^{1.3}} = 2 \times 10^{-12} \text{ eV}^2/Hz$. We neglected the tunneling related 1/f noise channel as it was not measured in the experiment. We emphasize that the high-frequency PSD cannot be measured in this experimental setup. It is
hence only possible to extrapolate the PSD to the qubit frequency using
Eq.~\eqref{eq:1f_level_transition_rate}. This extrapolation may overestimate the PSD and hence
the $1/f$ charge-noise-induced relaxation rate.

\subsubsection{Comparison of different relaxation mechanisms}
\label{subsubsection:Comparison_of_different_relaxation_mechanisms}

\begin{figure}[htbp]
    \centering
    \includegraphics[width=\linewidth]{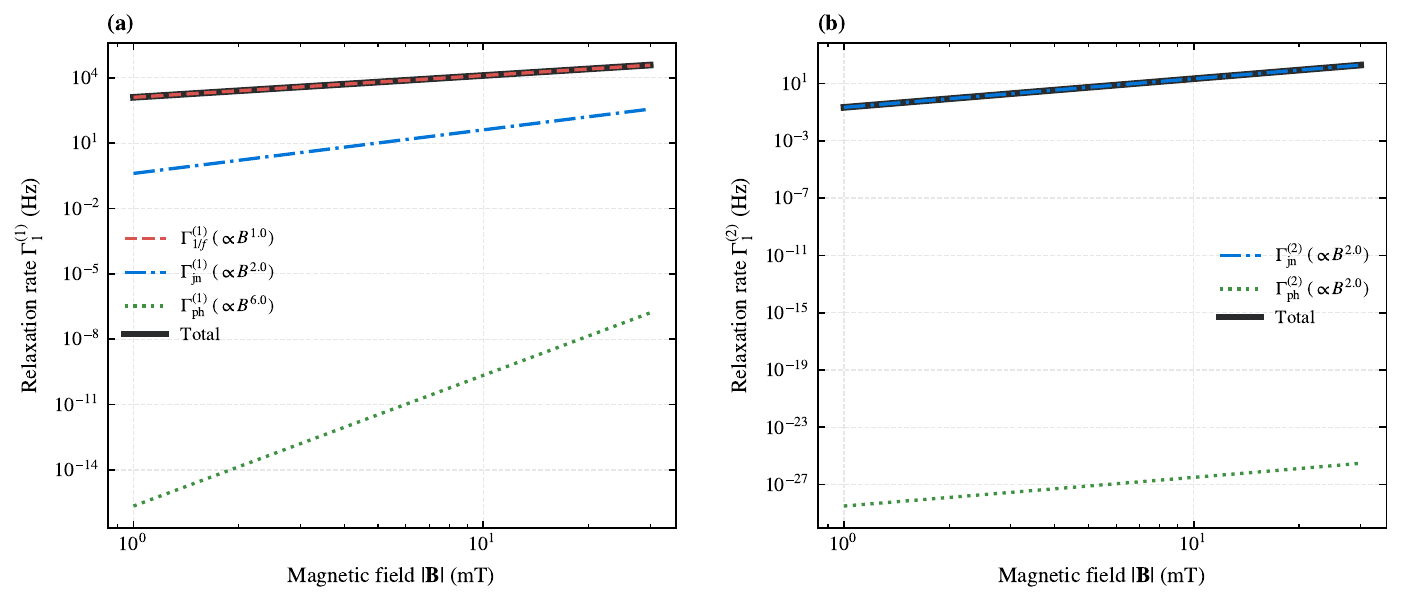}
    \caption{The first order relaxation rate $\textbf{(a)}$ and the second order relaxation $\textbf{(b)}$ from all major relaxation source including $\Gamma_{1/f}^{(1)}$, $\Gamma_{jn}^{(1)}$, and $\Gamma_{ph}^{(1)}$ as function of magnitude of $\textbf{B}$ field are calculated via Eq.~\eqref{eq:relaxation_T1_T2} with the level transition rate for each source obtained from Eq.~\eqref{eq:1f_level_transition_rate},~\eqref{eq:johnson_level_transition_rate}, ~\eqref{eq:phonon_level_transition_rate_gammae} and \eqref{eq:phonon_level_transition_rate_gammae} respectively.}
    \label{fig:S8}
\end{figure}

\begin{figure}[htbp]
    \centering
    \includegraphics[width=\linewidth]{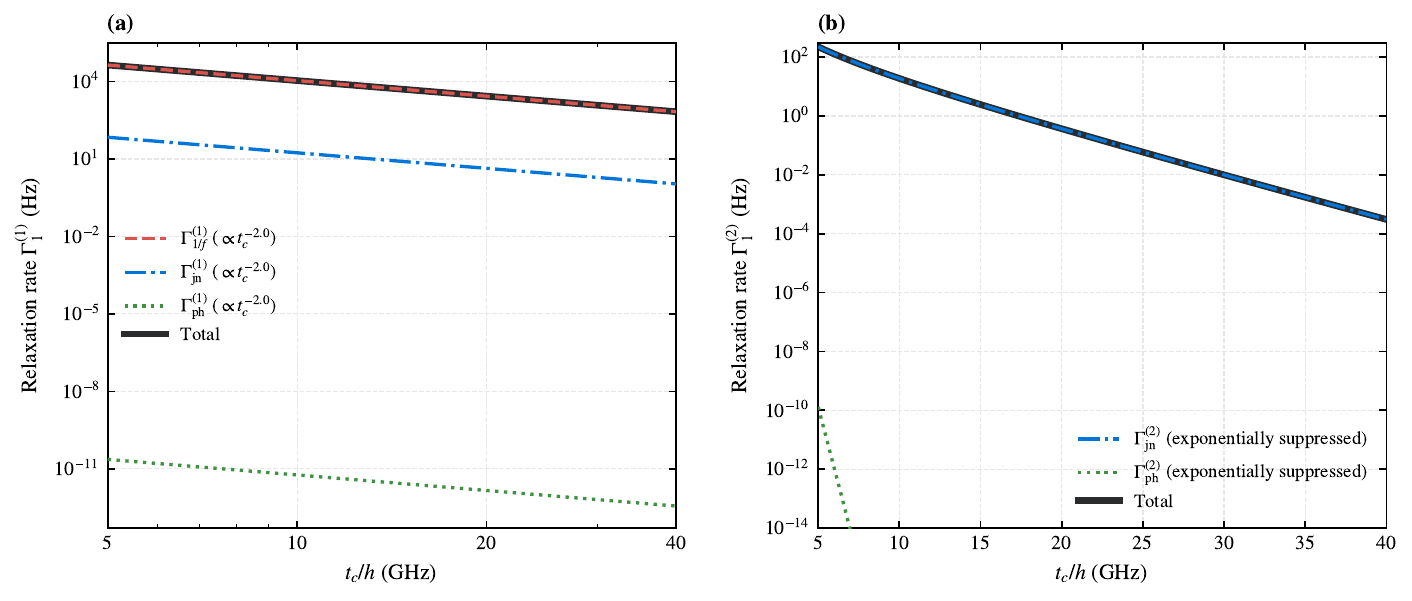}
    \caption{The first order relaxation rate $\textbf{(a)}$ and the second order relaxation $\textbf{(b).}$ as function of $t_{c}$ from all major relaxation source including $\Gamma_{1/f}$, $\Gamma_{jn}$, and $\Gamma_{ph}$ are calculated via Eq.~\eqref{eq:relaxation_T1_T2} with the level transition rate for each source obtained from Eq.~\eqref{eq:1f_level_transition_rate},~\eqref{eq:johnson_level_transition_rate} and~\eqref{eq:phonon_level_transition_rate_gammae} and ~\eqref{eq:phonon_level_transition_rate_gammat} respectively.}
    \label{fig:S9}
\end{figure}

We discuss the magnetic-field dependence of the first- and second-order relaxation processes for different noise sources at the FM operation point $\epsilon_{12} = 0$ at low qubit frequencies, and the results are summarized in Fig. ~\ref{fig:S8}.

First, for all noise sources, the second-order relaxation process is proportional to $B^2$. This is because the second-order process involves only spin-flip transitions between bonding and antibonding states. According to Fermi's Golden Rule in Eq.~\eqref{eq-rate-bosons}, the transition frequency is well approximated as $\hbar\omega_{ij} = 2 t_{c} \pm \Delta E_\text{FM} \approx 2t_{c}$. The magnetic-field dependence arises because the  $B$-field dependent Larmor-vector gradient between the two QDs mediates the spin-flip transition, providing the finite matrix elements in Eq.~\eqref{eq:Mtilde_e} and Eq.~\eqref{eq:Mtilde_t}. Note that at the zero-detuning point, the detuning relaxation channel $\tilde{M}_{\epsilon}$ in Eq.~\eqref{eq:Mtilde_e} only permits first-order processes, whereas the tunneling relaxation channel $\tilde{M}_{t}$ in Eq.~\eqref{eq:Mtilde_t} only permits second-order relaxation processes.

For the first order relaxation process, different noise sources have different $B$ field dependence. We first consider two relaxation channels, each consisting of two phonon modes. By combining Eq.~\eqref{eq:phonon_level_transition_rate_gammae},~\eqref{eq:phonon_level_transition_rate_gammat},~\eqref{eq:phonon_detune_integrand} and ~\eqref{eq:phonon_tunneling_integrand}, we find that the detuning phonon channel $\gamma_{\epsilon}$ shows a $\Gamma^{(1)}_{ph, \epsilon}\propto B^6$ dependence, while the tunneling phonon channel $\gamma_{t}$ shows a $\Gamma^{(1)}_{ph, t} \propto B^4$ relation, with a factor of $B^2$ contributed by the matrix elements $|\tilde{M}|^2$ from both channels.
\begin{align}
&(\gamma_{\epsilon})_{ij}^{l} \approx \frac{d^2  k_{B} T_{ph} \left(35 a^2+14 a b+5 b^2\right) \omega _{\text{ij}}^4}{210 \pi  \rho  \hbar ^2 v_l^7}\\
&(\gamma_{\epsilon})_{ij}^{t} \approx \frac{3 b^2 d^2  k_{B} T_{ph} \omega _{\text{ij}}^4}{70 \pi  \rho  \hbar ^2 v_t^7}\\
&(\gamma_{t})_{ij}^{l} \approx \frac{ s^2  k_{B}T_{ph} \left(5 a^2+b^2\right) \omega _{\text{ij}}^2}{10 \pi  \rho  \hbar ^2 v_l^5}\\
&(\gamma_{t})_{ij}^{t} \approx \frac{3 b^2  s^2 k_{B} T_{ph} \omega _{\text{ij}}^2}{20 \pi  \rho  \hbar ^2 v_t^5}
\end{align}

The Johnson-Nyquist noise is a white noise at high temperature limits ($E_{z}\ll k_{B}T$).  The transition rate Eq.~\eqref{eq:johnson_level_transition_rate} at high temperatures for detuning and tunneling channels is
\begin{align}
    &\gamma_{\epsilon} = \frac{2 \alpha_{\epsilon} ^2 k_{B} R T}{\hbar ^2},\\
    &\gamma_{t} = \frac{2 \alpha_{t} ^2 k_{B} R T}{\hbar ^2}.
\end{align} This indicates that the energy transition rate at qubit subspace is frequency independent. 
For the orbital energy, the Johnson-Nyquist noise is not a white noise ($ \Omega \gg k_{B}T$), but it is also not B-field dependent. Considering the $B^2$ contribution from the matrix elements, we find that both first- and second-order relaxation processes induced by Johnson–Nyquist noise show a $\Gamma^{(1)}_{jn}\propto B^2$ dependence.

The $1/f$ charge noise PSD in Eq.~\eqref{eq:1f_PSD} and Eq.~\eqref{eq:1f_level_transition_rate} shows $S_{1/f}\propto \frac{1}{B^{\alpha}}$, where $\alpha$ is fitted to be 1. Therefore, the first-order relaxation process shows a $\Gamma_{1/f}^{(1)} \propto B$ dependence.

In the experiment, a clear relation of $\Gamma_{1} = \Gamma_{1}^{(1)}+\Gamma_{1}^{(2)} \propto B^2$ is shown in Fig. \ref{fig:fig5} ($\textbf{d}$). Based on this B-field dependence, we conclude that the likely dominant noise source is either second-order phonons or a combination of first- and second-order Johnson–Nyquist noise. However, as shown in Fig.~\ref{fig:S8}, at low temperature ($T_{ph}=20$~mK) and low magnetic
field ($B=5$~mT), the phonon contribution is many orders of magnitude lower than the
Johnson--Nyquist noise for both first- and second-order processes. Therefore, Johnson--Nyquist
noise is more likely the main relaxation process than phonons.

\begin{figure}[htp]
    \centering
    \includegraphics[width=1\linewidth]{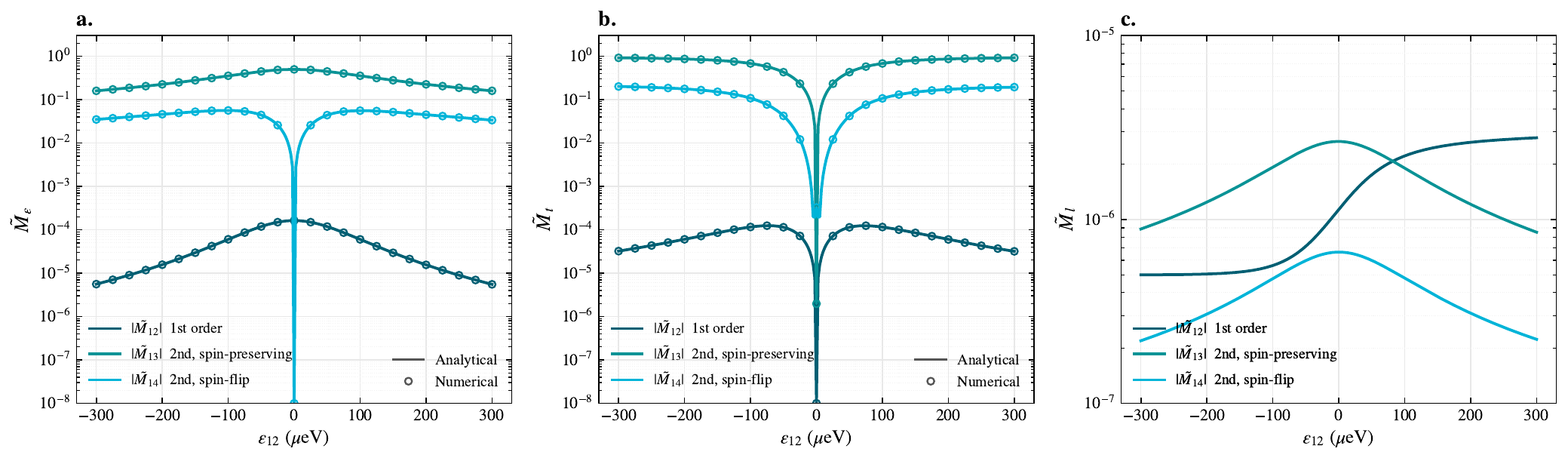}
    \caption{The detuning channel \textbf{a}, tunneling channel \textbf{b}, and g-tensor modulation channel \textbf{c} are plotted. In \textbf{a} and \textbf{b}, we provide the comparison of analytical expression in Eq. ~\eqref{eq:Mtilde_t} and ~\eqref{eq:Mtilde_e} with the numerical expression $\tilde{M}_{t} = U_{q}^\dagger M_{t} U_{q}$ and $\tilde{M}_{\epsilon} = U_{q}^\dagger M_{\epsilon}U_{q}$. In plot \textbf{c}, we only numerically calculate the transition matrix element as their analytical expression is lengthy and the contribution from g-tensor modulation channel at FM regime compared to the other two channel is negligible.}
    \label{fig:Mepsilon_Mt_Mg_modulation}
\end{figure}

\begin{figure}[htp]
    \centering
    \includegraphics[width=1\linewidth]{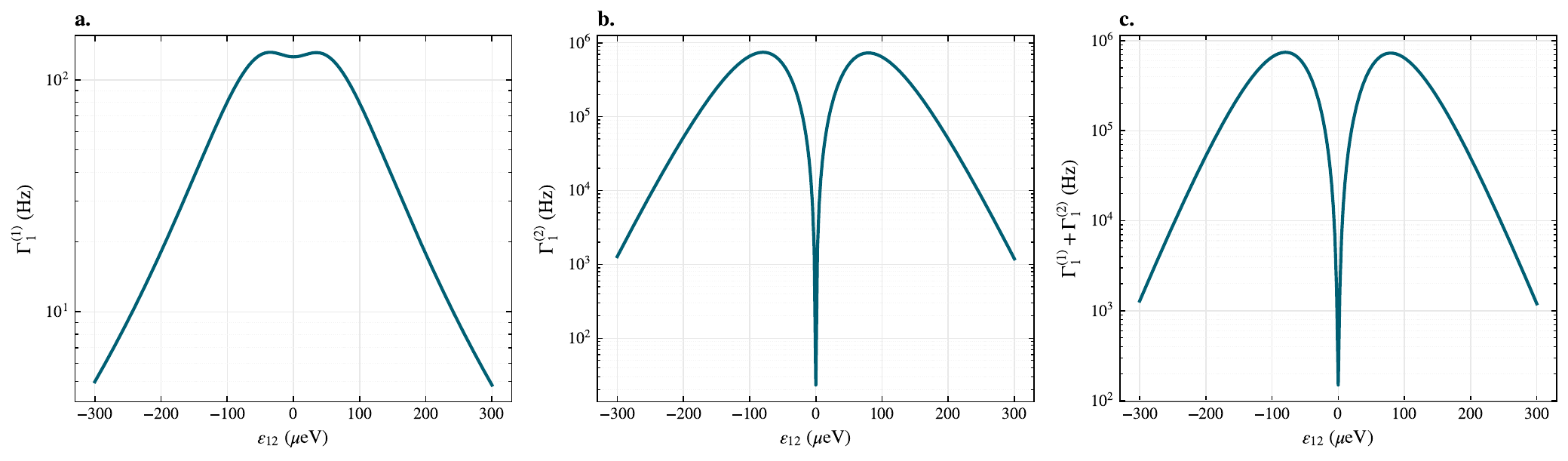}
    \caption{The relaxation rate for the first order process in \textbf{a}, the second order process in \textbf{b} and total process in \textbf{c}.}
\label{fig:S17}
\end{figure}

To investigate whether the first-order or second-order relaxation process is dominant, we introduce additional evidence from Fig. \ref{fig:fig5} ($\textbf{e}$): as the tunnel coupling increases, the B-field dependence becomes $\Gamma_{1} \propto B$, indicating that at high $t_{c}$, $T_{1}$ is limited by $1/f$ charge noise.

To understand this, we examine the relaxation rates from different noise sources as a function of the tunnel coupling $t_{c}$. For all first-order processes, the PSD is independent of $t_{c}$, and the matrix element is proportional to $t_{c}^{-1}$, as indicated by Eq.~\eqref{eq:Mt_analytical}, Eq.~\eqref{eq:Me_analytical} and Eq.~\eqref{eq:T1_sweetspot_1st_order}. Therefore, all the first order relaxation processes give $\Gamma^{(1)} \propto t_{c}^{-2}$. For a bosonic bath, the bosonic PSD has a polynomial dependence on $t_{c}$, similar to its $B$-field dependence. However, due to detailed balance, the second-order relaxation process is exponentially suppressed by the energy gap to higher orbitals, $\Gamma_{1}\propto e^{-2t_{c}/k_{B}T}$, as indicated by Eq.~\eqref{eq:relaxation_T1_T2}. Classical $1/f$ charge noise is likely negligible at $t_{c}/h = 11$ GHz. Nonetheless, if we naively consider a $\frac{1}{f^{\alpha}}$ PSD that is non-negligible at high frequency, we observe a $\Gamma_{1}\propto t_{c}^{-2-\alpha}$ relation.

Therefore, if the system $T_{1}$ is only limited by first-order Johnson–Nyquist noise at low $t_{c}$, we should not observe a transition in $B$-field dependence from $\propto B^2$ into $\propto B$ as $t_{c}$ increases because both $1/f$ charge noise and Johnson-Nyquist noise are suppressed as $t_{c}^2$. Therefore, a plausible explanation for the trends in both Fig. \ref{fig:fig5} ($\textbf{d}$) and Fig. \ref{fig:fig5} ($\textbf{e}$) is that a significant second-order relaxation process is involved. 

\begin{figure}[htbp]
    \centering
    \includegraphics[width=1\linewidth]{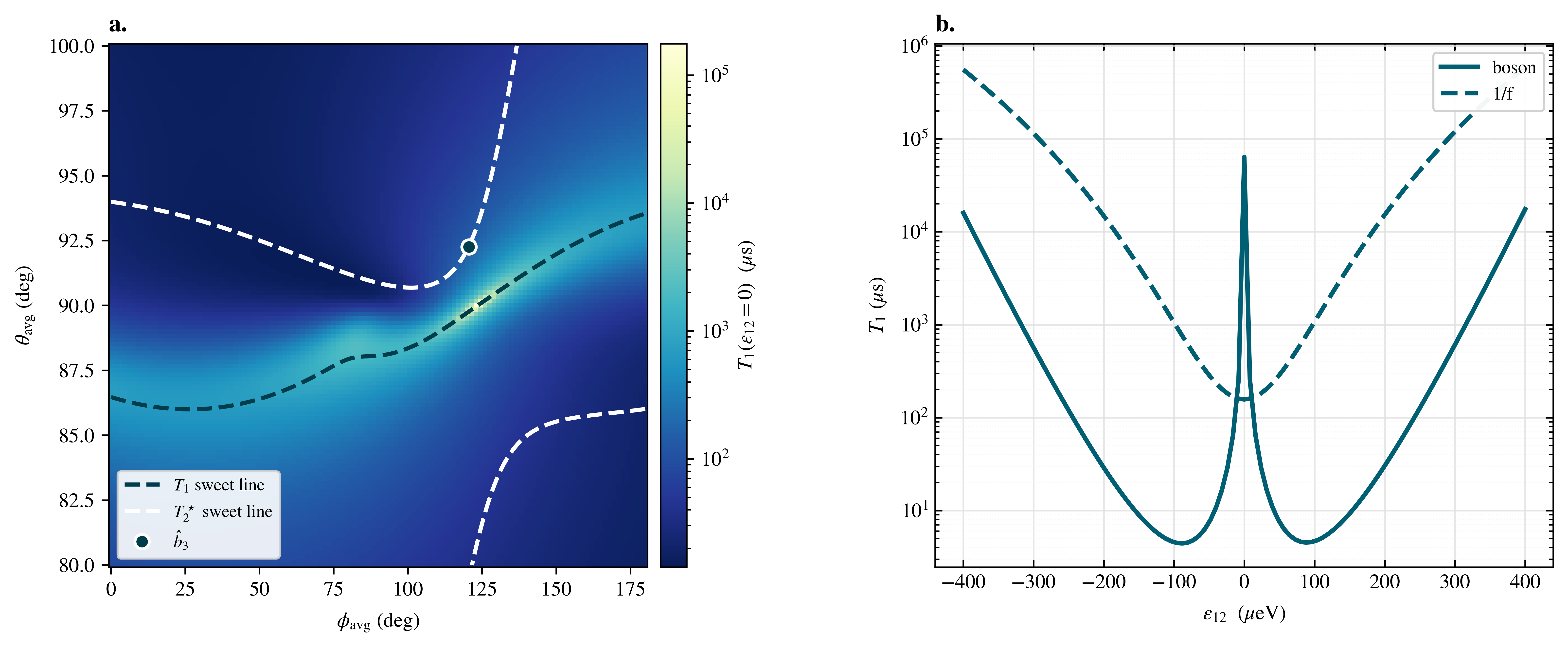}
    \caption{$\textbf{a.}$ The FM qubit relaxation time $T_{1}$ at $\epsilon_{12} = 0$ as a function of $\phi_\mathrm{avg}$ and $\theta_\mathrm{avg}$. $T_1$ sweet line (black dashed line) for both first order Eq.~\eqref{eq:T1_sweetspot_1st_order} and second order relaxation process ~\eqref{eq:Mtildet_2nd_order} is calculated by finding the minimization angle of $\sin(\theta_{12})$ at a given $\phi_{avg}$. The $T_{2}^*$ sweet lines (white dashed lines) are calculated using Eq.~\eqref{eq:T2_sweetspot_1st_order}. 
    $\textbf{b.}$ $T_1$ as a function of $\FMdetuning$ at the magnetic field corresponding to $\hat{\textbf{b}}_{3}$ (green point in Fig.~\ref{fig:fig3}a), calculated using Eq.~\eqref{eq:T1_first_second_order} with energy level transition rate estimated from Eq.~\eqref{eq:1f_level_transition_rate} (dashed curve, 1/f noise case),  Eq.~\eqref{eq:phonon_level_transition_rate_gammae}, ~\eqref{eq:phonon_level_transition_rate_gammat} and ~\eqref{eq:johnson_level_transition_rate} respectively (solid curve, bosonic bath case). 
    }
    \label{fig:S7_sweet_line}
\end{figure}

Further evidence supporting the second-order relaxation process is that $T_{1}$ exhibits a peak-and-dip structure as a function of detuning, as shown in Fig. \ref{fig:S7_sweet_line} (\textbf{b.}), Fig. \ref{fig:fig5} (\textbf{f.}) and Fig.~\ref{fig:Mepsilon_Mt_Mg_modulation}. In the range $\epsilon_{12}< 2t_{c}$, from Eq.~\eqref{eq:Mtilde_t} and Eq.~\eqref{eq:Mtilde_e}, we find that the two qubit subspace transition matrix elements $$|\tilde{M}_{\epsilon,12}| = \frac{2t_{c}^2 |\Delta \textbf{l}_{\perp}|} {\Omega^3} \qquad  \qquad |\tilde{M}_{t,12}| = \frac{2t_{c}|\epsilon_{12}| |\Delta \textbf{l}_{\perp}|} {\Omega^3}$$ vary by less than an order of magnitude over this range due to the suppression given by the factor ${1}/{ \Omega^3}$. Therefore, if the first-order process dominates at the zero-detuning point, we do not expect $\Gamma_{1}^{(1)}\propto \frac{(2t_{c}^2+t_{c}\epsilon_{12}) |\Delta \textbf{l}_{\perp}|} {\Omega^3}$ to drop by an order of magnitude at a slightly detuned point. For the second order matrix element, 
$$|\tilde{M}_{\epsilon,14}|=\frac{t_{c}|\epsilon_{12}|}{2\Omega^{2}}
\frac{E_{z,1}E_{z,2}\sin\theta_{12}}{\Delta E\,\Delta\bar E}
\qquad
|\tilde{M}_{t,14}|=\frac{\epsilon_{12}^{2}}{2\Omega^{2}}
\frac{E_{z,1}E_{z,2}\sin\theta_{12}}{\Delta E\,\Delta\bar E}+\frac{4 t_{c}^2 |\Delta \textbf{l}_{\perp}|}{ \Omega^3} \ .$$ 
We first recall that only the tunneling channel allows the second order relaxation process at $\epsilon_{12} = 0$. Further, we observe that in the denominator, the second-order matrix element is only suppressed as $\propto {1}/{\Omega^2}$, which means that slight detuning will activate the second-order relaxation channel, leading to the double dips observed in Fig.~\ref{fig:fig5}, Fig. ~\ref{fig:S7_sweet_line} (\textbf{b}) and Fig. \ref{fig:S17}.

We show $T_1$ calculated by taking into account both the first-order and second-order relaxation processes described earlier, in Fig.~\ref{fig:S7_sweet_line}a.
The $T_{1}$ and $T_{2}^{\star}$ sweet lines are represented by black and white dashed lines in Fig.~\ref{fig:S7_sweet_line}a.
$T_1$ sweet line corresponds to the magnetic field directions where the tilt $\theta_{12}$ between the Larmor vectors of the two LD qubits in each QD ($\textbf{l}_{i}$) becomes minimized. 
We emphasize that this is the global sweet-line for both first-order and second-order relaxation processes. 
On the other hand, $T_2^\star$ sweet line denotes the collection of magnetic field directions where the size of the two Larmor vectors, i.e. the Zeeman energy, becomes identical.
For the studied system in this work, there is no magnetic field direction where $T_1$ and $T_2^\star$ sweet lines overlap. 
Nevertheless, for the low-magnetic field operation regime, we find that $T_1$ is not a main limiting factor for $T_2^\star$ and $T_2^\mathrm{Hahn}$ when operated on the $T_2^\star$ sweet line.

Furthermore, we present the calculated $T_1$ as a function of $\FMdetuning$ at the magnetic field direction $\hat{\mathbf{\textit{b}}}_3$ (see Fig.~\ref{fig:fig3}a) in Fig.~\ref{fig:S7_sweet_line}b. 
The solid (dashed) curve corresponds to $T_1$ calculated taking into account the bosonic bath noise (1/f charge noise).
Here, we use the charge noise amplitude extracted from the PSD measurements shown in Supplementary~\autoref{sec:CPMG Noise spectroscopy}.
While the calculation suggests that 1/f charge noise dominates the relaxation at $\FMdetuning = 0$, the observed $1/B^2$ scaling of $T_1$ in Fig.~\ref{fig:fig5}d indicates that the relaxation is governed by second-order process rather than the 1/f charge noise. 
We attribute this discrepancy to the uncertainty in the charge-noise amplitude extracted in Supplementary~\autoref{sec:CPMG Noise spectroscopy}, which calls for detailed investigation.

\end{document}

\end{document}